\documentclass{article}
\usepackage{graphicx}
\usepackage{longtable}
\DeclareSymbolFont{AMSa}{U}{msa}{m}{n}
\DeclareMathDelimiter\ulcorner{\mathopen} {AMSa}{"70}{AMSa}{"70}
\DeclareMathDelimiter\urcorner{\mathclose}{AMSa}{"71}{AMSa}{"71}
\makeatletter
\def\uufill{$\m@th\mathopen\ulcorner\mkern-7mu%
  \cleaders\hbox{\rule[6pt]{1dd}{1dd}}\hfill
  \mkern-7mu\mathclose\urcorner$}
\def\overbrack#1{\vbox{\m@th\ialign{##\crcr
      \uufill\crcr\noalign{\kern-\p@\nointerlineskip}%
      $\hfil\displaystyle{#1}\hfil$\crcr}}}
\makeatother

\title{On a Knot Model of Mesons and Baryons}

\author{Sze Kui Ng
\\ {\small Department of Mathematics,
Hong Kong Baptist University, Hong Kong}
}

\begin{document}
\date{}
\maketitle
\begin{abstract}

A quantum knot model of mesons and baryons is established. This quantum knot model is derived from a quantum gauge model which is similar to the QCD gauge model.
We first establish a knot model of the $\pi$ mesons.
We show that the $\pi$ mesons are modeled by the prime knot ${\bf4_1}$ (which is assigned with the prime number $3$). We show that the knot model of the $\pi^+$ meson gives the strong interaction for the formation of the $\pi^+$ meson and the weak decay $\pi^+ \to \mu^+ +\nu_{\mu}$. 
We give a mass machanism for the generation of the masses of the $\pi^0$ meson, $\pi^+$ meson, the muon $\mu^+$ and the muon neutrino $\nu_{\mu}$.
We then extend the knot model to the $K^0$ and $K^+$ mesons.
We show that the $K^0$ and $K^+$ mesons are modeled by the prime knot ${\bf6_1}$ (which is assigned with the prime number $11$) while the anti-particles $\overline{K^0}$ and $K^-$ are modeled by the mirror image of ${\bf6_1}$. We show that the strange degree of freedom of the strange quark is from the linking of knots of the $K$ mesons. This linking gives the strong interaction for the associated production of elementary particles with strangeness. 
We then extend the knot model to the basic mesons $ \eta, \rho, \omega, K^*, \eta^{\prime},$ and $\phi$. These mesons are modeled by the prime knots ${\bf 6_2, 6_3, 6_3, 7_1, 7_2, 7_2}$ respectively where ${\bf 6_2, 6_3, 7_1, 7_2}$ are assigned with the prime numbers $13,17, 19, 23$ respectively. 
 We then extend the knot model to the basic baryons $p,n, \Lambda$, $ \Sigma, \Xi$; $\Delta, \Sigma^*, \Xi^*$, $\Omega^-$ and $\Lambda^*$. We show that these baryons are modeled by the prime knots ${\bf 6_2, 6_2, 7_1, 7_1,7_1}$; ${\bf7_1, 7_1,7_1, 7_2}$ and $ {\bf7_1}$ respectively where ${\bf 6_2, 7_1, 7_2}$ are assigned with the prime numbers $13, 19, 23$ respectively.
 
 This knot model gives a unification of the electromagnetic, the weak and the strong interactions. We show that a meson or a baryon is modeled by a representation of the form $W(K)Z$ where $W(K)$ denotes a quantum knot and the components of the complex vector $Z$ represent quarks. This is as the strong interaction for forming  the meson or the baryon. Then two components $z_i, i=1,2$ of $Z$ may be separated such that they are still attached to $W(K)$ in the form that they are attached to $W(K)$ at  two different positions of the quantum knot $W(K)$. This process of separation is identified as the weak interaction of the two quarks (or leptons) represented by $z_i, i=1,2$ of $Z$. 
 We show that the existing quark model is as a part of this knot model. Thus this knot model is consistent with the existing quark model.



\end{abstract}

\section{Introduction}\label{sec00}

The discovery of the $\pi^+$ meson and its decay may be considered as one of the events for the birth of particle physics \cite{Pow}. The $\pi^+$ meson is interesting in that its structure is simpler than other hadrons but its formation and decay has already involved both the strong and weak interactions. In this paper we first propose a quantum knot model of the $\pi$ mesons. In this knot model we show the formation of the $\pi$ mesons by strong interaction and and the weak decay $\pi^+ \to \mu^+ +\nu_{\mu}$. 
We show that the $\pi$ mesons are modeled by the prime knot ${\bf4_1}$ which is assigned with the prime number $3$. This quantum knot model is derived from a quantum gauge model which is similar to the QCD gauge model.

In this knot model of $\pi^+$ we give a mass machanism for the generation of the masses of the $\pi^+$ meson, the muon $\mu^+$ and the muon neutrino $\nu_{\mu}$.
This mechanism of generating mass supersedes the conventional mechanism of generating mass through the Higgs particles and makes hypothesizing the existence of the Higgs particles unnecessary. This perhaps explains why we cannot physically find such Higgs particles.
The computed masses of $\pi^0$, $\pi^+$ and $\mu^+$ are 135 Mev, 140 Mev and 105 Mev
respectively; while the corresponding observed masses are 135 Mev, 140 Mev
and 105.7 Mev respectively. 
It was a mystery how the mass of the $\mu ^{+}$ particle
comes out in the weak interaction of the $\pi ^{+}$ meson. Our computation
of the mass of the $\mu ^{+}$ particle using the mathematical structure of
the knot model of the $\pi ^{+}$ meson is an evidence that mesons are modeled as knots.

We then extend the knot model to the $K^0$ and $K^+$ mesons.
We give a mass mechanism for the generation of the masses of $K^0$ and $K^+$ mesons and the masses of the anti-particles $\overline{K^0}$ and $K^-$. 
 We show that the $K^0$ and $K^+$ mesons are modeled by the prime knot ${\bf6_1}$ (which is assigned with the prime number $11$) while the anti-particles $\overline{K^0}$ and $K^-$ are modeled by the mirror image of ${\bf6_1}$. We show that the strange degree of freedom of the strange quark is from the linking of knots of the $K$ mesons. This linking gives the strong interaction for the associated production of elementary particles with strangeness.

We then extend the knot model 
to the two nonets of the pseudoscalar and vector mesons which include the mesons $\eta, \rho, \omega, K^*, \eta^{\prime},$ and $\phi(1020)$ which are modeled by the prime knots ${\bf 6_2, 6_3, 6_3, 7_1, 7_2, 7_2}$ respectively and these prime knots ${\bf 6_2, 6_3, 7_1, 7_2}$ are assigned with the prime numbers $13,17, 19, 23$ respectively. Then we also give a knot model to the scalar mesons $a_0(980)$ and $f_0(980)$ and we show that the four mesons $\eta^{\prime},\phi(1020), a_0(980)$ and $f_0(980)$ are closedly related in that they are modeled with the same prime knot ${\bf 7_2}$ which is assigned with the prime number $23$.
Thus we have that the basic pseudoscalar and vector mesons: $\pi^0, K^0$, $\eta, \rho^0$, 
$ K^*, \eta^{\prime}$ are modeled by the prime knots: ${\bf 4_1}$, ${\bf 6_1, 6_2}$, ${\bf 6_3}$, ${\bf7_1, 7_2}$ which are indexed by the consecutative prime numbers: $3, 11, 13, 17, 19, 23 $ appearing in the mass formula of these mesons (We shall show from the quantum knot theory in this paper that the prime numbers $2, 5, 7$ and the corresponding prime knots ${\bf 3_1, 5_1, 5_2}$ are not used for the modeling of mesons). This list of consecutative prime numbers for the modeling of the basic pseudoscalar and vector mesons is thus an evidence for that mesons are modeled as knots. 

From this knot model we have a  mass formula for computing the masses of the basic pseudoscalar and vector mesons. The result of computation is given by the following table of classification of mesons (This  mass formula is different from the Gell-Mann-Okubo mass formula while there are some relations with the Gell-Mann-Okubo mass formula \cite{Gel2}-\cite{Bur2}):
\begin{longtable}{|c|c|c|c|}\hline 
$ I=1 $&$ I=0$&$ \mbox{Strange mesons}\quad I=\frac12$&$ I=0$ \\ \hline 
$\pi(135) $&$ \eta(549)$&$ K(498)$&$ \eta^{\prime}(958)$ \\ $
{\bf 3}\times 45=135$&$ {\bf 13}\times 42=546 $&$ {\bf 11}\times 43+24=497$&${\bf  23}\times 42=966  $ \\ \hline 
$\rho(770)$&$ \omega(783)$&$K^{*}(892)  $&$\phi(1020) $ \\
$ {\bf 17} \times 45=765$&${\bf 17}\times 46=782 $&${\bf 19}\times 46+16=890$&$
{\bf 23}\times 43+32=1021 $ \\ \hline 
\end{longtable}
In the above table the quantum numbers $42,43,45,46$ are determined by a quantum condition which is derived from the structure of the group $SU(2)$ and the winding number of the knot model of mesons; and the quantum numbers $16, 24,32$ are as quantum states of the strange quarks.
This classification of mesons is between mesons and prime knots and is established by the corresponding prime numbers where prime knots are classified by prime numbers by the following table \cite{Ng}:
\begin{displaymath}
\begin{array}{|c|c|c|c|c|c|c|c|c|c|c|c|c|c|c|} \hline
\mbox{prime knot} & {\bf 3_1}& {\bf 4_1}& {\bf 5_1} & {\bf 5_2}& {\bf 6_1}
&{\bf 6_2}&{\bf 6_3}&{\bf 7_1}&{\bf 7_2}&{\bf 7_3}&{\bf 7_4} & {\bf 7_5}&{\bf 7_6} &{\bf 7_7}\\ \hline
 \mbox{prime number}& &3 & 5& 7&11  &13 & 17&19&23&29 &31 &37&41 & 43\\\hline
\end{array}
\end{displaymath}
where the prime knot ${\bf 3_1}$ is assigned with the number $1$ while it is similar to the prime number $2$ \cite{Ng}.

From the above classification of mesons we have the following relation for the two nonets of pseudoscalar and vector mesons:
\begin{equation}
\pi(135) \leftarrow\rightarrow \rho(770) \leftarrow\rightarrow \omega(783) \leftarrow\rightarrow K^{*}(892)\leftarrow\rightarrow K(498) \leftarrow\rightarrow \phi(1020)\leftarrow\rightarrow \eta^{\prime}(958) \leftarrow\rightarrow \eta(549)
\label{relation}
\end{equation}
where $\pi(135)$ and $\rho(770)$ are related by the quantum number $45$; $\rho(770)$ and $\omega(783)$ are related by the prime knot ${\bf 6_3}$ and prime number $17$;
$\omega(783)$ and $K^{*}(892)$ are related by the quantum number $46$;
$K^{*}(892)$ and $K(498) $ are related by the strangeness;
$K(498)$ and $\phi(1020)$ are related by the quantum number $43$;
$\phi(1020)$ and $\eta^{\prime}(958)$ are related by the prime knot ${\bf 7_2}$ and prime number $23$; and
$\eta^{\prime}(958)$ and $\eta(549)$ are related by the quantum number $42$.

From this knot model of mesons we also show that the phenomeon of the generations of quarks is derived from the property of knots. We show that the $u\overline{u}$-mesons  and $d\overline{d}$-mesons (where $u$ and $d$ denote the up and down quarks) such as the $\pi^0$ and $\rho^0$ mesons are modeled by prime knots from the two families ${\bf4_{(\cdot)}}$ and ${\bf6_{(\cdot)}}$ of prime knots with four and six crossings respectively. Thus the generation of the $u$ and $d$ quarks is from the two families ${\bf4_{(\cdot)}}$ and ${\bf6_{(\cdot)}}$. Then we show that the $s\overline{s}$-mesons (where $s$ denotes the strange quark) such as the $\phi(1020)$ meson is modeled by prime knots from the family ${\bf7_{(\cdot)}}$ of prime knots with seven crossings. Then we show that the $c\overline{c}$-mesons (where $c$ denotes the charm quark) such as the $J/\psi$ meson is modeled by prime knots from the family ${\bf8_{(\cdot)}}$ of prime knots with eight crossings. Thus the generation of the $s$ and $c$ quarks is from the families ${\bf7_{(\cdot)}}$ and ${\bf8_{(\cdot)}}$ of prime knots with seven and eight crossings respectively. Continuing in this way we have that the generations of quarks is derived from the property of knots. This is thus an evidence for that mesons are modeled as knots.

We the extend the knot model to the basic octet, decuplet and singlet of baryons which include the baryons $p,n, \Lambda$, $\Sigma, \Xi$; $\Delta, \Sigma^*, \Xi^*$, $\Omega^-$ and $\Lambda^*$. We show that these baryons are modeled by the prime knots ${\bf 6_2, 6_2, 7_1, 7_1,7_1}$; ${\bf7_1, 7_1,7_1, 7_2}$ and $ {\bf7_1}$ respectively where ${\bf 6_2, 7_1, 7_2}$ are assigned with the prime numbers $13, 19, 23$ respectively. These prime numbers 
$13, 19, 23 $ appear in the mass formula of these baryons. We notice that the sequence $13, 19, 23 $ can be regarded as a sequence of consecutive prime numbers for modeling baryons because the missing prime number $17$ is assigned to the prime knot ${\bf 6_3}$ which is an amphichiral knot with the property that it is identified to its mirror image and thus is not suitable to model baryons which are not identified to their anti-particles. This is thus an evidence that baryons are modeled as knots.

For the  baryons $\Lambda, \Sigma, \Xi, \Lambda^*, \Sigma^*, \Xi^*, \Omega^-$  we  introduce a pure strange degree of freedom which is from a $R_s$ matrix of representing knots on $SU(3)$ and this $R_s$ matrix is a linking effect of knots. This linking effect gives the strong interaction between two elementary particles and gives the associated production property for strong interactions producing particles with the strange quark $s$ such as the interaction 
$\pi^- +p\to K^0 + \Lambda$ \cite{Pai}-\cite{Fow}. 

We show that the weak interaction can also be described by this quantum knot model. When the gauge group is $SU(2)\otimes U(1)$ this quantum knot model and the corresponding quantum gauge model is analogous to the usual electroweak theory \cite{Wei}\cite{Gla}\cite{Sal}\cite{Mar}. From this knot modeling of strong and weak interactions we show that the strong and weak interactions are closedly related. The basic relation of the strong and weak interactions can be stated as follows. We have that a meson (or a baryon) is modeled by a representation of the form $W(K)Z$. This is as the strong interaction for forming  the meson (or the baryon). Then two components $z_i, i=1,2$ of $Z$ may be separated such that they are still attached to $W(K)$ in the form that they are attached to $W(K)$ at
two different positions of the quantum knot $W(K)$. This process of separation is identified as the weak interaction of the two quarks (or leptons) represented by the two components $z_i, i=1,2$ of $Z$. This is as the basic relation of the strong and weak interactions. We shall also show that the strength of the strong and weak interactions are closedly related.

During the process of the weak interaction  we have that the components $z_i, i=1,2$ representing the up and down quarks are separated but are still attached to the quantum knot for representing the $\pi^{\pm}$ meson. At this intermediate state with the up and down quarks being separated the quantum knot and the attached components $z_i,i=1,2$ no longer represent the $\pi^{\pm}$ meson but is as a new particle which is identified as the gauge particle $W^{\pm}$. Similarly we also have the knot modeling of the gauge particle $Z^0$ for the weak interaction of neutral currents.

In this knot modeling of strong and weak interactions we can also model the parity $P$ operation and the charge conjugation $C$.
We have that the $\pi$ mesons are modeled by the prime knot ${\bf4_1}$. 
 This prime knot ${\bf 4_1}$ is an amphichiral knot which equals to its mirror image. We show that this property of knot gives the spontaneous symmetry breaking for forming the $\pi$ mesons. We show that this spontaneous symmetry breaking gives the parity $P$ violation, the charge conjugation $C$ violation, and the $CP$ violation of weak interaction. 

This quantum knot model also explains quarks' fractional charges and the phenomenon of quark confinement and asymptotic freedom of strong interaction.
 Thus we conclude that the strong interaction of quarks for forming mesons and baryons is just the quantum knot modeling of mesons and baryons. 

We remark that the quantum gauge model for deriving this quantum knot model of elementary particles is not based on the four dimensional space-time but is based on the one dimensional proper time (We shall describe this quantum gauge model in the next section). Then since knots are of three dimensional nature though this quantum gauge model is not based on the three dimensional space we have that the three dimensional nature of matters (and thus the three dimensional space) naturally comes out from this quantum gauge model. This also shows that why the physical space is three dimensional.

This paper is organized as follows. In section 2 we give a brief description of a quantum gauge model of electrodynamics  and its nonabelian generalization. We shall first consider a nonabelian generalization with a $SU(2)\otimes U(1)$ gauge symmetry which will be for the strong and weak interaction of mesons. With this quantum  model we introduce the generalized Wilson loop for the construction of the  mesons. To investigate the properties of the Wilson loop in section 3 we derive a chiral symmetry from the gauge symmetry of this quantum  model. From this chiral symmetry in section 4 we derive a conformal field theory which includes an affine Kac-Moody algebra and a quantum Knizhnik-Zamolodchikov (KZ) equation. 
A main point of our model on the quantum KZ equation is that we can derive two KZ equations
which are dual to each other. This duality is the main point for the Wilson loop to be exactly solvable and to have a winding property which gives properties of the elementary particles. 
Since this quantum KZ equation is derived from the quantum gauge model by calculus of variation we can regard it as a quantum Euler-Lagrange equation or a quantum Yang-Mills equation.
We then in section 5 to section 9 represent knots by the generalized Wilson loops which are constructed from the quantum KZ equation. 

Then from section 10 to section 12 we use the generalized Wilson loop which is as a quantum knot to give the properties of the $\pi$ mesons and to give a mass mechanism for generating masses of the $\pi$ mesons. 
We give a mass mechanism for generating the masses of $\pi^+$, $\mu^+$ and $\nu_{\mu}$ where we show that the neutrino $\nu_{\mu}$ is without charge and without mass.

Then from section 13 to section 17
we use the generalized Wilson loop which is as a quantum knot to give the properties of the pseudoscalar and vector mesons.
Then in section 18 we show that the phenomeon of the generations of quarks is derived from the property of knots. 
 

Then in section 19 we show that the $\pi$ mesons modeled by the prime knot ${\bf 4_1}$ and that ${\bf 4_1}$ being an amphichiral knot give the $P$, $C$  and $CP$ violation.

 Then in section 20 we show in more detail that this quantum knot modeling is just the required strong and weak interactions and that  strong and weak interactions are closedly related. This gives the unification of the electromagnetic, the weak and the strong interactions. In section 21 we give the knot model of the gauge particles $W^{\pm}$ and $Z^0$. 

Then in section 22 and section 23  we use the gauge symmetry $SU(3)\otimes U(1)$ and the generalized Wilson loop which is as a quantum knot to give the knot models of baryons. 

 \section{A Quantum Gauge Model}\label{sec2}

Let us  
construct a quantum gauge model, as follows. In
probability theory we have the Wiener measure $\nu$ which is a
measure on the space $C[t_0,t_1]$ of continuous functions
\cite{Jaf}. This measure is a well defined mathematical theory for
the Brownian motion and it may be symbolically written in the
following form:
\begin{equation}
d\nu =e^{-L_0}dx
\label{wiener}
\end{equation}
where $L_0 :=
\frac12\int_{t_0}^{t_1}\left(\frac{dx}{dt}\right)^2dt$ is the
energy integral of the Brownian particle and $dx =
\frac{1}{N}\prod_{t}dx(t)$ is symbolically a product of Lebesgue
measures $dx(t)$ and $N$ is a normalized constant.
Once the Wiener measure is defined we may then define other
measures on $C[t_0,t_1]$ as follows\cite{Jaf}. Let a potential
term $\frac12\int_{t_0}^{t_1}Vdt$ be added to $L_0$. Then we have
a measure $\nu_1$ on $C[t_0,t_1]$ defined by:
\begin{equation}
d\nu_1 =e^{-\frac12\int_{t_0}^{t_1}Vdt}d \nu
\label{wiener2}
\end{equation}
Under some condition on $V$ we have that $\nu_1$ is well defined
on $C[t_0,t_1]$. Let us call (\ref{wiener2}) as the Feymann-Kac
formula \cite{Jaf}.
Let us then follow this formula to construct a quantum  model of
electrodynamics, as follows. 
Similar to the formula (\ref{wiener2}) we construct a quantum
model of electrodynamics from the following energy
integral:
\begin{equation}
 \frac12\int_{s_0}^{s_1}[
\frac12\left(\frac{\partial A_1}{\partial x^2}-\frac{\partial
A_2}{\partial x^1}\right)^* \left(\frac{\partial A_1}{\partial
x^2}-\frac{\partial A_2}{\partial x^1}\right)
 +\sum_{j=1}^2
\left(\frac{\partial Z^*}{\partial
x^j}+ie_0A_jZ^*\right)\left(\frac{\partial Z}{\partial
x^j}-ie_0A_jZ\right)]ds \label{1.1}
\end{equation}
where the complex variable $Z=Z(z(s))$ and the real variables
$A_1=A_1(z(s))$ and $A_2=A_2(z(s))$ are continuous functions in a form that they are in terms of an arbitrary (continuously differentiable) closed curve $z(s)=C(s)=(x^1(s),x^2(s)), s_0\leq
s\leq s_1, z(s_0)=z(s_1)$ in the complex plane where
$s$ is a parameter representing the proper time in relativity (We shall also write $z(s)$ in the complex variable form $C(s)=z(s)=x^1(s)+ix^2(s),s_0\leq s\leq s_1$). The complex variable $Z=Z(z(s))$ represents a field of matter( such as the electron)
($Z^*$ denotes its complex conjugate) and the real variables
$A_1=A_1(z(s))$ and $A_2=A_2(z(s))$ represent a connection (or the gauge field of the photon) and $e_0$
denotes the (bare) electric charge.
The integral (\ref{1.1}) has the following gauge symmetry:
\begin{equation}
\begin{array}{rl}
Z^{\prime}(z(s)) & := Z(z(s))e^{ie_0a(z(s))} \\
A'_j(z(s)) & := A_j(z(s))+\frac{\partial a}{\partial x^j} \quad
j=1,2
\end{array}
\label{1.2}
\end{equation}
where $a=a(z)$ is a continuously differentiable real-valued
function of $z$.

We remark that this QED model is similar to the usual Yang-Mills gauge models. 
A  feature of (\ref{1.1}) is that it is not
formulated with the four-dimensional space-time but is formulated
with the one dimensional proper time. This one dimensional nature
let this QED model avoid the usual utraviolet divergence difficulty of
quantum fields. 
As most of the theories in physics are formulated with the space-time let us give reasons of this formulation. We know that with the concept of space-time we have a convenient way to understand
physical phenomena and to formulate theories such as the Newton equation, the Schroedinger
equation , e.t.c. to describe these physical phenomena. However we also know that there
are fundamental difficulties related to space-time such as the utraviolet divergence difficulty of
quantum field theory. To solve these difficulties let us reexamine the concept of space-time.
We propose that the space-time is a statistical concept which is not as basic as the proper
time in relativity. Because a statistical theory is usually a convenient but incomplete
description of a more basic theory this means that some difficulties may appear if we formulate
a physical theory with the space-time. This also means that a way to formulate a basic theory of physics is to formulate it not with the space-time but with the proper time only as the parameter for
evolution. This is a reason that we use (\ref{1.1}) to formulate a QED theory. In this formulation
we regard the proper time as an independent parameter for evolution. From (\ref{1.1}) we may obtain
the conventional results in terms of space-time by introducing the space-time as a statistical method.

Let us explain in more detail how the space-time comes out as a statistics. For statistical purpose when many electrons (or many photons) present we introduce space-time $(t,x)$ as a statistical method
to write $ds^2$ in the form
\begin{equation}
ds^2=dt^2-dx^2
\label{lorentz}
\end{equation}
We notice that for a given $ds$ there may have many $dt$ and $dx$ which correspond to many electrons (or photons) such that (\ref{lorentz}) holds. In this way the space-time is introduced as a statistics.
 By (\ref{lorentz}) we may (and we shall in elsewhere) derive statistical formulas for many electrons (or photons) from
 formulas obtained from (\ref{1.1}). In this way we may obtain the Dirac equation as a statistical equation for electrons and the Maxwell equation as a statistical equation for photons. In this way we may regard the usual QED theory as the statistical extension of this QED model. This statistical interpretation of the usual QED theory is thus an explanation of the mystery that the usual QED theory is so successful in the computation of quantum effects of electromagnetic interaction while it has the difficulty of ultraviolet divergence.
 
 We notice that the relation (\ref{lorentz}) is the famous Lorentz metric (We may generalize it to other metric in general relativity). Here our understanding of the Lorentz metric is that it is a statistical formula
 where the proper time $s$ is more fundamental than the space-time $(t,x)$ in the sense that we first have the proper time and the space-time is introduced via the Lorentz metric only for the
 purpose of statistics. This reverses the order of appearance of the proper time and the 
 space-time in the history of relativity in which we first have the concept of space-time
 and then we have the concept of proper time which is introduced via the Lorentz metric.
  Once we understand that the space-time is a statistical concept from (\ref{1.1}) we can give a solution
  to the quantum measurement problem in the debate about quantum mechanics between Bohr and Einstein. In this debate Bohr insisted
  that with the probability interpretation quantum mechanics is very successful. On the other
  hand Einstein insisted that quantum mechanics is incomplete because of probability interpretation.
  Here we may solve this debate by constructing the above QED model which is a quantum theory as the quantum mechanics
  and unlike quantum mechanics which needs probability interpretation we have that this QED model is deterministic since it is not formulated with the space-time.

Similar to the usual Yang-Mills gauge theory we can generalize this gauge model with $U(1)$ gauge symmetry to
nonabelian gauge models. As an illustration let us consider
$SU(2)\otimes U(1)$ gauge symmetry where 
$SU(2)\otimes U(1)$ denotes the  direct product of the groups $SU(2)$ and $U(1)$. This $SU(2)\otimes U(1)$ gauge symmetry is for the knot model of mesons. We shall later consider the $SU(3)\otimes U(1)$ gauge symmetry which is for the knot model of baryons.
Similar to (\ref{1.1}) we consider the
following energy integral:
\begin{equation}
L := \frac12\int_{s_0}^{s_1}
[\frac12 Tr (D_1A_2-D_2A_1)^{*}(D_1A_2-D_2A_1) +
(D_1^*Z^*)(D_1Z)+(D_2^*Z^*)(D_2Z)]ds
\label{n1}
\end{equation}
where $Z= (z_1, z_2)^{T}$ is a two dimensional complex vector;
$A_j =\sum_{k=0}^{3}A_j^k t^k $ $(j=1,2)$ where $A_j^k$ denotes a
component of a gauge field $A^k$; $t^k=ie_0 T^k$ denotes a generator of $SU(2)\otimes U(1)$ where $T^k$ denotes
a self-adjoint generator of $SU(2)\otimes U(1)$ and $e_0$ denotes the bare electric charge for general interactions including the strong and weak interaction (Here for simplicity we choose a convention that the
complex $i$ is absorbed by $t^k$);
and
$D_j=\frac{\partial}{\partial x^j}-e_0A_j$, $(j=1,2)$. 
From (\ref{n1}) we can develop a nonabelian gauge model as similar
to that for the above abelian gauge model.
We have that (\ref{n1}) is invariant under the following
gauge transformation:
\begin{equation}
\begin{array}{rl}
Z^{\prime}(z(s)) & :=U(a(z(s)))Z(z(s)) \\
A_j^{\prime}(z(s)) & := U(a(z(s)))A_j(z(s))U^{-1}(a(z(s)))+
 U(a(z(s)))\frac{\partial U^{-1}}{\partial x^j}(a(z(s))),
j =1,2
\end{array}
\label{n2}
\end{equation}
where $U(a(z(s)))=e^{a(z(s))}$ and $a(z(s))=\sum_k a^k
(z(s))t^k$. 
We shall mainly consider the case that $a$ is a function of the form $a(z(s))
=\sum_k \mbox{Re}\,
\omega^k(z(s))t^k$ where $\omega^k$ are 
analytic functions of $z$ (We let $\omega(z(s)):=\sum_k\omega^k(z(s))t^k$ and we write $a(z)=\mbox{Re}\,\omega(z)$).

The above gauge model is based on the Banach space $X$ of continuous functions $Z(z(s)), A_j(z(s)), j=1,2, s_0\leq s\leq s_1$
on the one dimensional interval $\lbrack s_0, s_1 \rbrack$. 
Since $L$ is positive and the model is one dimensional (and thus is simpler than the usual two dimensional Yang-Mills gauge model) we have that this gauge model is similar to the Wiener measure except that this gauge model has a gauge symmetry. This gauge symmetry gives a degenerate degree of freedom. In the physics literature the usual way to treat the degenerate degree of freedom of gauge symmetry is to introduce a gauge fixing condition to eliminate the degenerate degree of freedom where each gauge fixing will give equivalent physical results \cite{Fad}. There are various gauge fixing conditions such as the Lorentz gauge condition, the Feynman gauge condition, etc. We shall later in
section \ref{sec6} (on the Kac-Moody algebra) adopt a gauge fixing
condition for the above gauge model. This gauge fixing condition
will also be used to derive the quantum KZ equation in dual form which will be regarded as a quantum Yang-Mill equation since its role will be similar to the classical Yang-Mill equation derived from the classical Yang-Mill gauge model.

\section{Classical Dirac-Wilson Loop } \label{sec4}

Similar to the Wilson loop in quantum field theory \cite{Wit} from our
quantum model we introduce an analogue of Wilson loop, as follows (We shall also call a Wilson loop as a Dirac-Wilson loop).

{\bf Definition}.
 A classical Wilson loop $W_R(C)$ is defined by :
\begin{equation}
W_R(C):= W(z_0, z_1):= Pe^{\int_C A_jdx^j} \label{n4}
\end{equation}
where $R$ denotes a representation of $SU(2)$; $C(\cdot)=z(\cdot)$
is a fixed curve where the quantum gauge models are based on it
as specified in the above section.
As
usual the notation $P$ in the definition of $W_R(C)$ denotes a
path-ordered product \cite{Wit}\cite{Kau}\cite{Baez}.

Let us give some remarks on the above definition of Wilson loop,
as follows.

1) We use the notation $W(z_0, z_1)$ to mean the Wilson loop
$W_R(C)$ which is based on the whole closed curve $z(\cdot)$. Here for
convenience we  use only the end points $z_0$ and $z_1$ of the
curve $z(\cdot)$ to denote this Wilson loop 
(We keep in mind that the definition of $W(z_0, z_1)$ depends on the whole curve $z(\cdot)$ connecting $z_0$ and $z_1$).

Then we  extend the definition of $W_R(C)$ to the case that
$z(\cdot)$ is not a closed curve with $z_0\neq z_1$. When
$z(\cdot)$ is not a closed curve we shall call $W(z_0, z_1)$ as a
Wilson line.

2) We use the above Wilson loop $W(C)$ to represent the unknot (Also called the trivial knot). We shall later generalize this Wilson loop $W(C)$ to generalized Wilson loops which will be shown to be able to represent nontrivial knots. Thus we shall call the generalized Wilson loops as quantum knots. We shall model the mesons by using these quantum knots. Then since knots are of three dimensional nature we have that though the quantum gauge model in this paper is not based on the space-time the three dimensional nature of the world naturally comes out from the quantum gauge model in this paper.

3) In constructing the Wilson loop we need to choose a
representation $R$ of the $SU(2)$ group. We shall see that because a Wilson line 
$W(z_0, z_1)$ is with two variables $z_0$ and $z_1$ a natural representation of a Wilson line or a Wilson loop is the tensor product of the usual two dimensional representation of the $SU(2)$ for constructing the Wilson loop. $\diamond$

A basic property of a Wilson line $W(z_0,z_1)$ is that for a given continuous path $A_i, i=1,2$  on $[s_0, s_1]$ 
the Wilson line $W(z_0,z_1)$ exists on this path and has the
following transition property:
\begin{equation}
W(z_0,z_1)=W(z_0,z)W(z,z_1)
 \label{df2}
\end{equation}
where $W(z_0,z_1)$ denotes the Wilson line of a
curve $z(\cdot)$ which is with $z_0$ as the starting
point and $z_1$ as the ending point and $z$ is a
point on $z(\cdot)$ between $z_0$ and $z_1$.

This property can be prove as follows. We have that $W(z_0,z_1)$ is a limit
(whenever exists)
of ordered product of $e^{A_i\triangle x^i}$ and thus can be
written in the following form:
\begin{equation}
\begin{array}{rl}
W(z_0,z_1)= & I +
\int_{s^{\prime}}^{s^{\prime\prime}}
A_i(z(s))\frac{dx^i(s)}{ds}ds \\
 & + \int_{s^{\prime}}^{s^{\prime\prime}}
[\int_{s^{\prime}}^{s_1} A_i(z(s_1))\frac{dx^i(s_1)}{ds}ds_1]
A_i(z(s_2))\frac{dx^i(s_2)}{ds}ds_2 +\cdot\cdot\cdot
\end{array}
\label{df3}
\end{equation}
where $z(s^{\prime})=z_0$ and $z(s^{\prime\prime})=z_1$. Then
since $A_i$ are continuous on $[s^{\prime}, s^{\prime\prime}]$ and
$x^i(z(\cdot))$ are continuously differentiable on $[s^{\prime},
s^{\prime\prime}]$ we have that the series in (\ref{df3}) is
absolutely convergent. Thus the Wilson line $W(z_0,z_1)$ exists.
Then since $W(z_0,z_1)$ is the limit of ordered
product  we can write $W(z_0,z_1)$ in the form $W(z_0,z)W(z,z_1)$
by dividing $z(\cdot)$ into two parts at $z$. This proves the basic property of Wilson line. $\diamond$

{\bf Remark (Classical version and quantum version of Wilson loop)}. From the above property we have that the Wilson line $W(z_0,z_1)$ exists in the classical pathwise sense where $A_i$ are as classical paths on $[s_0, s_1]$. This pathwise version of the Wilson line $W(z_0,z_1)$; from the Feymann path integral point of view; is as a partial description of the quantum version of the Wilson line $W(z_0,z_1)$ which is as an operator when $A_i$ are as operators. 
We shall in the next section derive and define a quantum generator $J$ of $W(z_0,z_1)$ from the quantum gauge model. Then by using this generator $J$ we shall compute the quantum version of the Wilson line $W(z_0,z_1)$.

We shall denote both the classical version and quantum version of Wilson line by the same notation  $W(z_0,z_1)$ when there is no confusion. 
$\diamond$

By following the usual approach of deriving a chiral symmetry from a gauge transformation of a gauge field
we have the following chiral symmetry
which is derived by applying an analytic gauge transformation with an analytic function $\omega$
for the transformation:
\begin{equation}
W(z_0, z_1) \mapsto W^{\prime}(z_0, z_1)=U(\omega(z_1))
W(z_0, z_1)U^{-1}(\omega(z_0))
\label{n5}
\end{equation}
where $W^{\prime}(z_0, z_1)$ is a Wilson line with gauge field $A_{\mu}^{\prime} =  \frac{\partial U(z)}{\partial x^{\mu}}U^{-1}(z) + U(z)A_{\mu}U^{-1}(z)$.

This chiral symmetry is analogous to the chiral symmetry
of the usual guage theory where $U$ denotes an element of the gauge group \cite{Kau}.
Let us derive (\ref{n5}) as follows.
Let $U(z):=
U(\omega(z(s)))$ and $U(z+dz)\approx U(z)+\frac{\partial
U(z)}{\partial x^{\mu}}dx^{\mu}$ where $dz=(dx^1,dx^2)$. Following
 \cite{Kau} we have
\begin{equation}
\begin{array}{rl}
& U(z+ dz)(1+ dx^{\mu}A_{\mu})U^{-1}(z)\\
=& U(z+ dz)U^{-1}(z)
+ dx^{\mu}U(z+dz)A_{\mu}U^{-1}(s) \\
\approx & 1+ \frac{\partial U(z)}{\partial
x^{\mu}}U^{-1}(z)dx^{\mu}
  + dx^{\mu}U(z+ dz)A_{\mu}U^{-1}(s) \\
\approx & 1+ \frac{\partial U(z)}{\partial
x^{\mu}}U^{-1}(z)dx^{\mu}
+ dx^{\mu}U(z)A_{\mu}U^{-1}(z) \\
=: & 1+  \frac{\partial U(z)}{\partial x^{\mu}}U^{-1}(z)dx^{\mu}
+ dx^{\mu}U(z)A_{\mu}U^{-1}(z)\\
=:& 1 + dx^{\mu}A_{\mu}^{\prime}
\end{array}
\label{n5b}
\end{equation}

From (\ref{n5b}) we have that (\ref{n5}) holds. 

As analogous to  
the WZW model in
conformal field theory \cite{Fra}\cite{Fuc}
from the above symmetry  we have the following formulas for the
variations $\delta_{\omega}W$ and $\delta_{\omega^{\prime}}W$ with
respect to this symmetry (\cite{Fra} p.621):
\begin{equation}
\delta_{\omega}W(z,z')=W(z,z')\omega(z)
\label{k1}
\end{equation}
and
\begin{equation}
\delta_{\omega^{\prime}}W(z,z')=-\omega^{\prime}(z')W(z,z')
\label{k2}
\end{equation}
where $z$ and $z'$ are independent variables and
$\omega^{\prime}(z')=\omega(z)$ when $z'=z$. In (\ref{k1}) the
variation is with respect to the $z$ variable while in (\ref{k2})
the variation is with respect to the $z'$ variable. This
two-side-variations when $z\neq z'$ can be derived as follows. For
the left variation we may let $\omega$ be analytic in a
neighborhood of $z$ and continuously differentiably extended to a
neighborhood of $z'$ such that $\omega(z')=0$ in this neighborhood
of $z'$. Then from (\ref{n5}) we have that (\ref{k1}) holds.
Similarly we may let $\omega^{\prime}$ be analytic in a
neighborhood of $z'$ and continuously differentiably extended to a
neighborhood of $z$ such that $\omega^{\prime}(z)=0$ in this
neighborhood of $z$. Then we have that (\ref{k2}) holds.

\section{A Gauge Fixing Condition and Affine Kac-Moody Algebra} \label{sec6}

This section has two related purposes. One purpose is to find a
gauge fixing condition for eliminating the degenerate degree of
freedom from the gauge invariance of the above quantum gauge model
in section 2. Then another purpose is to find an equation for
defining a generator $J$ of the Wilson line $W(z,z')$. This
defining equation of $J$ can then be used as a gauge fixing
condition. Thus with this defining equation of $J$ the
construction of the quantum gauge model in section 2 is then
completed.

 We shall derive a quantum loop algebra (or the
affine Kac-Moody algebra) structure from the Wilson line $W(z,z')$
for the generator $J$ of $W(z,z')$. To this end let us first
consider the classical case. Since $W(z,z')$ is constructed from $
SU(2)$ we have that the mapping $z \to W(z,z')$ (We consider
$W(z,z')$ as a function of $z$ with $z'$ being fixed) has a loop
group structure \cite{Lus}\cite{Seg}. For a loop group we have the
following generators:
\begin{equation}
J_n^a = t^a z^n \qquad n=0, \pm 1, \pm 2, ...
\label{km1}
\end{equation}
These generators satisfy the following algebra:
\begin{equation}
[J_m^a, J_n^b] =
if_{abc}J_{m+n}^c
\label{km2}
\end{equation}
This is  the so called loop algebra \cite{Lus}\cite{Seg}. Let us
then introduce the following generating function $J$:
\begin{equation}
J(w) = \sum_a J^a(w)=\sum_a j^a(w) t^a
\label{km3}
\end{equation}
where we define
\begin{equation}
J^a(w)= j^a(w) t^a :=
\sum_{n=-\infty}^{\infty}J_n^a(z) (w-z)^{-n-1}
\label{km3a}
\end{equation}

From $J$ we have
\begin{equation}
J_n^a=  \frac{1}{2\pi i}\oint_z dw (w-z)^{n}J^a(w)
\label{km4}
\end{equation}
where $\oint_z$ denotes a closed contour integral  with center $z$. This formula
can be interpreted as that
$J$ is the generator of the loop group and that
$J_n^a$ is the directional generator in the direction
$\omega^a(w)= (w-z)^n$. We may generalize $(\ref{km4})$
to the following  directional generator:
\begin{equation}
  \frac{1}{2\pi i}\oint_z dw \omega(w)J(w)
\label{km5}
\end{equation}
where the analytic function
$\omega(w)=\sum_a \omega^a(w)t^a$ is regarded
as a direction and we define
\begin{equation}
 \omega(w)J(w):= \sum_a \omega^a(w)J^a
\label{km5a}
\end{equation}

Then since $W(z,z')\in SU(2)$, from the variational formula
(\ref{km5}) for the loop algebra of the loop group of $SU(2)$ we
have that the variation of $W(z,z')$ in the direction $\omega(w)$
is given by
\begin{equation}
W(z,z')
  \frac{1}{2\pi i}\oint_z dw \omega(w)J(w)
\label{km6}
\end{equation}

Now let us consider the quantum case which is based on the quantum
gauge model in section 2. For this quantum case we shall define a
quantum generator $J$ which is analogous to the $J$ in
(\ref{km3}). We shall choose the equations (\ref{n8b}) and
(\ref{n6}) as  the equations for defining the quantum generator
$J$. Let us first give a
formal derivation of the equation (\ref{n8b}), as follows.
 Let us consider the
following formal functional integration:
\begin{equation}
\langle W(z,z')A(z) \rangle := \int dA_1dA_2dZ^{*}dZ  e^{-L}
W(z,z')A(z) \label{n8a}
\end{equation}
where $A(z)$ denotes a field from the quantum gauge model (We
first let $z'$ be fixed as a parameter).

Let us  do a calculus of variation on this integral to derive a variational
equation by applying a gauge transformation on (\ref{n8a}) as follows
(We remark that such variational equations are usually called the
Ward identity in the  physics literature).

Let $(A_1,A_2,Z)$ be regarded as a coordinate system of the integral
(\ref{n8a}).
Under a gauge transformation (regarded as
a change of coordinate) with gauge function $a(z(s))$ this coordinate
is changed to another coordinate denoted by
$(A_1^{\prime}, A_2^{\prime}, Z^{\prime})$.
As similar to the usual change of variable for integration we have that
the integral  (\ref{n8a}) is unchanged
under a change of variable and we have the following
equality:
\begin{equation}
\begin{array}{rl}
& \int dA_1^{\prime}
 dA_2^{\prime}dZ^{\prime *}dZ^{\prime}
 e^{-L^{\prime}} W^{\prime}(z,z')A^{\prime}(z) \\
= & \int dA_1dA_2dZ^{*}dZ  e^{-L} W(z,z')A(z)
\end{array}
\label{int}
\end{equation}
where $W^{\prime}(z,z')$ denotes the Wilson line based on
$A_1^{\prime}$ and $A_2^{\prime}$ and similarly $A^{\prime}(z)$
denotes  the field obtained from $A(z)$ with $(A_1, A_2,Z)$
replaced by $(A_1^{\prime}, A_2^{\prime},Z^{\prime})$.

Then it can be shown that the differential is unchanged under a
gauge transformation \cite{Fad}:
\begin{equation}
dA_1^{\prime}
dA_2^{\prime}dZ^{\prime *}dZ^{\prime}
= dA_1dA_2dZ^{*}dZ
\label{int2}
\end{equation}
Also by the gauge invariance property the factor $e^{-L}$ is
unchanged under a gauge transformation. Thus from (\ref{int}) we
have
\begin{equation}
0 = \langle W^{\prime}(z,z')A^{\prime}(z)\rangle -
  \langle W(z,z')A(z)\rangle
\label{w1}
\end{equation}
where the correlation notation
$\langle \cdot\rangle$ denotes the integral with
respect to the differential
\begin{equation}
e^{-L}dA_1dA_2dZ^{*}dZ
\label{w1a}
\end{equation}

We can now carry out the calculus of variation. From the gauge
transformation we have the formula
$W^{\prime}(z,z')=U(a(z))W(z,z')U^{-1}(a(z'))$ 
where $a(z)=\mbox{Re}\,\omega(z)$. This
gauge transformation gives a variation of $W(z,z')$ with the
gauge function $a(z)$  
as the variational direction $a$  
in the variational formulas (\ref{km5}) and  (\ref{km6}). Thus analogous
to the variational formula (\ref{km6}) we have that the variation
of $W(z,z')$ under this gauge transformation is given by
\begin{equation}
W(z,z')
  \frac{1}{2\pi i}\oint_z dw a(w)J(w) 
\label{int3}
\end{equation}
where the generator $J$ for this variation is to
be specified. This $J$ will be a quantum generator
which generalizes the classical generator $J$ in
(\ref{km6}).

Thus under a gauge transformation with gauge function $a(z)$ from (\ref{w1}) we have the
following variational equation:
\begin{equation}
0= \langle W(z,z')[\delta_{a}A(z)+\frac{1}{2\pi i}\oint_z
dw a(w)J(w)A(z)]\rangle 
\label{w2}
\end{equation}
where $\delta_{a}A(z) $ 
denotes the variation of the field
$A(z)$ in the direction $a(z)$.  
From this equation an ansatz of
$J$ is that $J$ satisfies the following equation:
\begin{equation}
W(z,z')[\delta_{a}A(z)+\frac{1}{2\pi i}\oint_z
dw a(w)J(w)A(z)] =0 \label{n8bb}
\end{equation}
From this equation we have the following variational equation:
\begin{equation}
\delta_{a}A(z)=\frac{-1}{2\pi i}\oint_z dw a(w)J(w)A(z)
\label{n8bre}
\end{equation}
This completes the formal calculus of variation. Now (with the
above derivation as a guide) we choose the following equation (\ref{n8b}) as one of the
equation for defining the generator $J$:
\begin{equation}
\delta_{\omega}A(z)=\frac{-1}{2\pi i}\oint_z dw\omega(w)J(w)A(z)
\label{n8b}
\end{equation}
where we generalize the direction $a(z)=\mbox{Re}\,\omega(z)$ to the analytic direction $\omega(z)$
(This generalization has the effect of extending the real measure to include the complex Feymann path integral). 

Let us now choose one more equation for determine the generator
$J$ in (\ref{n8b}). This choice will be as  
a gauge fixing
condition. As analogous to the WZW model in conformal field
theory \cite{Fra}\cite{Fuc} \cite{Kni}  let us consider a $J$
given by:
\begin{equation}
J(z) := -k W^{-1}(z, z')\partial_z W(z, z') \label{n6}
\end{equation}
where we define $\partial_z=\partial_{x^1} +i\partial_{x^2} $ and we set $z'=z$ after the differentiation with respect to $z$; $ k>0 $ is a constant which is fixed when the $J$ is determined to be of the form (\ref{n6}) and the minus sign is chosen by
convention. In the WZW model \cite{Fra}\cite{Kni}
 the $J$ of the form (\ref{n6})
is the  generator  of the chiral symmetry of the WZW model. We can
write the $J$ in (\ref{n6}) in the following form:
\begin{equation}
 J(w) = \sum_a J^a(w) =
\sum_a j^a(w) t^a  
\label{km8}
\end{equation}
We see that the generators $t^a$ of $SU(2)$ appear in this form of
$J$ and  this form is analogous to the classical $J$ in
(\ref{km3}). This shows that
 this $J$ is a possible candidate for the generator
$J$ in (\ref{n8b}).

Since $W(z,z')$ is constructed by gauge field we need to have a
gauge fixing for the computations related to $W(z,z')$. Then since
the $J$ in (\ref{n8b}) and (\ref{n6}) is constructed from
$W(z,z')$ we have that in defining this $J$ as the generator $J$
of $W(z,z')$ we have chosen a condition for the gauge fixing. In
this paper we shall always choose this defining equations
(\ref{n8b}) and (\ref{n6}) for $J$ as the gauge fixing condition.

In summary we introduce the following definition.

{\bf Definition}. The generator $J$ of the quantum Wilson line $W(z,z')$ whose classical version is defined by (\ref{n4}), is an operator defined by the two conditions (\ref{n8b}) and (\ref{n6}).
$\diamond$

{\bf Remark}. We remark that the condition (\ref{n6}) first defines $J$ classically. Then the condition (\ref{n8b}) raises this classical $J$ to the quantum generator $J$. $\diamond$

Now we want to show that this generator $J$ in (\ref{n8b}) and
(\ref{n6}) can be uniquely solved (This means that the gauge
fixing condition has already fixed the gauge that the degenerate
degree of freedom of gauge invariance has been eliminated so that
we can carry out computation). 

Let us now solve $J$.
From (\ref{n5}) and (\ref{n6}) we
have that the variation $\delta_{\omega}J$ of the generator $J$ in
(\ref{n6}) is given by \cite{Fra}(p.622) \cite{Kni}:
\begin{equation}
\delta_{\omega}J= \lbrack J, \omega\rbrack -k\partial_z \omega
\label{n8c}
\end{equation}

From (\ref{n8b}) and (\ref{n8c}) we have that $J$ satisfies the
following relation of current algebra
\cite{Fra}\cite{Fuc}\cite{Kni}:
\begin{equation}
J^a(w)J^b(z)=\frac{k\delta_{ab}}{(w-z)^2}
+\sum_{c}if_{abc}\frac{J^c(z)}{(w-z)} \label{n8d}
\end{equation}
where as a convention the regular term of the product
$J^a(w)J^b(z)$ is omitted. Then by following
\cite{Fra}\cite{Fuc}\cite{Kni} from (\ref{n8d}) and (\ref{km8})
we can show that the $J_n^a$ in (\ref{km3})  for the corresponding Laurent series of the quantum generator $J$  
satisfy the
following  Kac-Moody algebra:
\begin{equation}
[J_m^a, J_n^b] = if_{abc}J_{m+n}^c + km\delta_{ab}\delta_{m+n, 0}
\label{n8}
\end{equation}
where $k$ is  usually called the central extension or the level of
the Kac-Moody algebra.

{\bf Remark}. Let us also consider the other side of the chiral 
symmetry. 
Similar to the $J$ in (\ref{n6}) we define a generator
$J^{\prime}$ by:
\begin{equation}
J^{\prime}(z')= k\partial_{z'}W(z, z')W^{-1}(z, z') \label{d1}
\end{equation}
where after differentiation with respect to $z'$ we set $z=z'$.
Let us then consider
 the following formal correlation:
\begin{equation}
\langle A(z')W(z,z') \rangle := \int
dA_1dA_2dZ^{*}dZ
  A(z')W(z,z')e^{-L}
\label{n8aa}
\end{equation}
where $z$ is fixed. By an approach similar to the above derivation
of (\ref{n8b}) we have the following  variational equation:
\begin{equation}
\delta_{\omega^{\prime}}A(z') =\frac{-1}{2\pi i}\oint_{z^{\prime}}
dwA(z')J^{\prime}(w) \omega^{\prime}(w) \label{n8b1}
\end{equation}
where as a gauge fixing we choose the $J^{\prime}$ in (\ref{n8b1})
be the $J^{\prime}$ in (\ref{d1}). Then similar to (\ref{n8c}) we
also have
\begin{equation}
\delta_{\omega^{\prime}}J^{\prime}= \lbrack  J^{\prime},
\omega^{\prime}\rbrack -k\partial_{z'} \omega^{\prime}
\label{n8c1}
\end{equation}
Then from (\ref{n8b1}) and (\ref{n8c1}) we can derive the current
algebra and the Kac-Moody algebra for $J^{\prime}$ which are of
the same form of (\ref{n8d}) and (\ref{n8}). From this we  have
$J^{\prime}=J$. $\diamond$

Now with the above current algebra $J$ and the formula (\ref{n8b}) we can
follow the usual approach
in conformal field theory to derive a quantum
Knizhnik-Zamolodchikov (KZ) equation for the product of
primary fields in a conformal field theory \cite{Fra}\cite{Fuc}\cite{Kni}.
Here we derive the KZ equation for the product of $n$ Wilson lines $W(z, z')$.
An important point is that from the two sides of
$W(z, z')$  we can derive two quantum KZ equations which are
dual to each other. These two quantum KZ equations are different from the usual KZ equation in that they are equations for the quantum operators $W(z, z')$ while the usual KZ equation is for the correlations of quantum operators.
With this difference we can follow the usual approach
in conformal field theory to derive the following quantum
Knizhnik-Zamolodchikov equation \cite{Ng}\cite{Fra} \cite{Fuc}:
\begin{equation}
\partial_{z_i}
 W(z_1, z_1^{\prime})\cdot\cdot\cdot 
W(z_n, z_n^{\prime})
=\frac{-1}{k+g_0}
\sum_{j\neq i}^{n}
\frac{\sum_a t_i^a \otimes t_j^a}{z_i-z_j}
 W(z_1, z_1^{\prime})\cdot\cdot\cdot 
W(z_n, z_n^{\prime})
\label{n9}
\end{equation}
for $i=1, ..., n$
where $g_0$ denotes the dual Coxeter number of a group multiplying with $e_0^2$ and we have $g_0=2e_0^2$ for the group $SU(2)$ (When the gauge group is $U(1)$ we have $g_0=0$).

We remark that in (\ref{n9}) we have
defined $t_i^a:= t^a$ and
\begin{equation}
\begin{array}{rl}
 & t_i^a \otimes t_j^a W(z_1, z_1^{\prime})\cdot\cdot\cdot
W(z_n, z_n^{\prime}) \\
:=& W(z_1, z_1^{\prime})\cdot\cdot\cdot
 [t^aW(z_i, z_i^{\prime})]\cdot\cdot\cdot
[t^aW(z_j, z_j^{\prime})]\cdot\cdot\cdot
W(z_n, z_n^{\prime})
\end{array}
\label{n9a}
\end{equation}

It is interesting and important that we also have
the following quantum Knizhnik-Zamolodchikov equation with repect to
the $z_i^{\prime}$ variables which is dual to (\ref{n9}):
\begin{equation}
\partial_{z_i^{\prime}}
 W(z_1,z_1^{\prime})\cdot\cdot\cdot W(z_n,z_n^{\prime})
= \frac{-1}{k+g_0}\sum_{j\neq i}^{n}
 W(z_1, z_1^{\prime})\cdot\cdot\cdot 
W(z_n, z_n^{\prime})
\frac{\sum_a t_i^a\otimes t_j^a}{z_j^{\prime}-z_i^{\prime}}
\label{d8}
\end{equation}
for $i=1, ..., n$
where we have defined:
\begin{equation}
\begin{array}{rl}
 &  W(z_1, z_1^{\prime})\cdot\cdot\cdot
W(z_n, z_n^{\prime})t_i^a \otimes t_j^a \\
:=& W(z_1, z_1^{\prime})\cdot\cdot\cdot
 [W(z_i, z_i^{\prime})t^a]\cdot\cdot\cdot
[W(z_j, z_j^{\prime})t^a]\cdot\cdot\cdot
W(z_n, z_n^{\prime})
\end{array}
\label{d8a}
\end{equation}

{\bf Remark}. From the quantum gauge model we derive the above quantum KZ equation in dual form by calculus of variation. This quantum KZ equation in dual form may be considered as a quantum Euler-Lagrange equation or as a quantum Yang-Mills equation since it is analogous to the classical Yang-Mills equation which is derived from the classical Yang-Mills gauge model by calculus of variation. 
 $\diamond$

\section{Solving Quantum KZ Equation In Dual Form}\label{sec8a}

Let us consider the following product of two quantum
Wilson lines:
\begin{equation}
G(z_1,z_2, z_3, z_4):=
 W(z_1, z_2)W(z_3, z_4)
\label{m1}
\end{equation}
where the two quantum Wilson lines $W(z_1, z_2)$ and
$W(z_3, z_4)$ represent two pieces
of curves starting at $z_1$ and $z_3$ and ending at
$z_2$ and $z_4$ respectively.

We have that this product $G$ satisfies the KZ equation for the
variables $z_1$, $z_3$ and satisfies the dual KZ equation
for the variables $z_2$ and $z_4$.
Then
by solving the two-variables-KZ equation in (\ref{n9}) we have that a form of $G$ is
given by \cite{Chari}\cite{Koh}\cite{Dri}:
\begin{equation}
e^{-t\log [\pm (z_1-z_3)]}C_1
\label{m2}
\end{equation}
where $t:=\frac{1}{k+g_0}\sum_a t^a \otimes t^a$
and $C_1$ denotes a constant matrix which is independent
of the variable $z_1-z_3$.

We see that $G$ is a multivalued analytic function
where the determination of the $\pm$ sign depended on the choice of the
branch.

Similarly by solving the dual two-variable-KZ equation
 in (\ref{d8}) we have that
$G$ is of the form:
\begin{equation}
C_2e^{t\log [\pm (z_4-z_2)]}
\label{m3}
\end{equation}
where $C_2$ denotes a constant matrix which is independent
of the variable $z_4-z_2$.

From (\ref{m2}), (\ref{m3}) and we let
$C_1=Ae^{t\log[\pm (z_4-z_2)]}$,
$C_2= e^{-t\log[\pm (z_1-z_3)]}A$ where $A$ is a constant matrix  we have that
$G$ is given by:
\begin{equation}
G(z_1, z_2, z_3, z_4)=
e^{-t\log [\pm (z_1-z_3)]}Ae^{t\log [\pm (z_4-z_2)]}
\label{m4}
\end{equation}
where at the singular case that $z_1=z_3$ we simply define $\log [\pm (z_1-z_3)]=0$. Similarly for $z_2=z_4$.

Let us find a form of the initial operator $A$. We notice that there are two operators $\Phi_{\pm}(z_1-z_2):=e^{-t\log [\pm (z_1-z_3)]}$ and $\Psi_{\pm}(z_i^{\prime}-z_j^{\prime})$ acting on the two sides of $A$ respectively where  the two independent variables  $z_1, z_3$ of $\Phi_{\pm}$ are mixedly from the two quantum Wilson lines $W(z_1, z_2)$ and
$W(z_3, z_4)$ respectively and the the two independent variables  $z_2, z_4$ of $\Psi_{\pm}$ are mixedly from the two quantum Wilson lines $W(z_1, z_2)$ and $W(z_3, z_4)$ respectively. From this we determine the form of $A$ as follows.

Let $D$ denote a representation of $SU(2)$. Let $D(g)$ represent an element $g$ of  $SU(2)$
and let $D(g)\otimes D(g)$ denote the tensor product representation of $SU(2)$. Then
in the KZ equation we define:
\begin{equation}
[t^a\otimes t^a] [D(g_1)\otimes D(g_1)]\otimes
[D(g_2)\otimes D(g_2)]
:=[t^aD(g_1)\otimes D(g_1)]\otimes
[t^aD(g_2)\otimes D(g_2)]
\label{tensorproduct}
\end{equation}
and
\begin{equation}
[D(g_1)\otimes D(g_1)]\otimes
[D(g_2)\otimes D(g_2)][t^a\otimes t^a]
:=[D(g_1)\otimes D(g_1)t^a]\otimes
[D(g_2)\otimes D(g_2)t^a]
\label{tensorproduct2}
\end{equation}

Then we let $U({\bf a})$  
denote the universal
enveloping algebra 
where ${\bf a}$ denotes an algebra which is formed by the Lie
algebra $su(2)$ and the identity matrix. 

Now let the initial operator $A$ be of the form $A_1\otimes A_2\otimes A_3\otimes A_4$ with $A_i,i=1,...,4$
taking values in $U({\bf a})$. 
In
this case we have that in (\ref{m4}) the operator
$\Phi_{\pm}(z_1-z_2):=e^{-t\log [\pm (z_1-z_3)]}$ acts on $A$ from
the left via the following formula:
\begin{equation}
t^a\otimes t^a A=
[t^a A_1]\otimes A_2\otimes [t^a A_3]\otimes A_4
\label{ini2}
\end{equation}

Similarly the operator
$\Psi_{\pm}(z_1-z_2):=e^{t\log [\pm (z_1-z_3)]}$
in (\ref{m4}) acts on $A$ from the right via the following formula:
\begin{equation}
A t^a\otimes t^a =
A_1\otimes [A_2 t^a]\otimes A_3\otimes[A_4 t^a]
\label{ini3}
\end{equation}

We may generalize the above tensor product of
two quantum Wilson lines as follows.
Let us consider a tensor product of $n$ quantum Wilson lines:
$W(z_1, z_1^{\prime})\cdot\cdot\cdot W(z_n, z_n^{\prime})$
where the variables $z_i$, $z_i^{\prime}$
are all independent. By solving the two KZ equations
we have that this tensor product is given by:
\begin{equation}
W(z_1, z_1^{\prime})\cdot\cdot\cdot W(z_n, z_n^{\prime})
=\prod_{ij} \Phi_{\pm}(z_i-z_j)
A\prod_{ij}
\Psi_{\pm}(z_i^{\prime}-z_j^{\prime})
\label{tensor}
\end{equation}
where $\prod_{ij}$ denotes a product of
$\Phi_{\pm}(z_i-z_j)$ or
$\Psi_{\pm}(z_i^{\prime}-z_j^{\prime})$
for $i,j=1,...,n$ where $i\neq j$.
In (\ref{tensor}) the initial operator
$A$ is represented as a tensor product of operators $A_{iji^{\prime}j^{\prime}}, i,j,i^{\prime}, j^{\prime}=1,...,n$ where each $A_{iji^{\prime}j^{\prime}}$ is of the form of the initial operator $A$ in the above tensor product of two-Wilson-lines case and is acted by $\Phi_{\pm}(z_i-z_j)$ or
$\Psi_{\pm}(z_i^{\prime}-z_j^{\prime})$ on its two sides respectively.

\section{Computation of Quantum Wilson Lines}\label{sec 8aa}

Let us consider the following product of two quantum
Wilson lines:
\begin{equation}
G(z_1,z_2, z_3, z_4):=
W(z_1, z_2)W(z_3, z_4)
\label{h1}
\end{equation}
where the two quantum Wilson lines $W(z_1, z_2)$ and
$W(z_3, z_4)$ represent two pieces
of curves starting at $z_1$ and $z_3$ and ending at
$z_2$ and $z_4$ respectively.
As shown in the above section we have that $G$
is given by the following formula:
\begin{equation}
G(z_1, z_2, z_3, z_4)=
e^{-t\log [\pm (z_1-z_3)]}Ae^{t\log [\pm (z_4-z_2)]}
\label{m4a}
\end{equation}
where the product is  
a 4-tensor.

Let us set $z_2=z_3$. Then   
the 4-tensor $W(z_1, z_2)W(z_3, z_4)$ is reduced to the 2-tensor 
$W(z_1, z_2)W(z_2, z_4)$. 
By using (\ref{m4a}) the 2-tensor 
$W(z_1, z_2)W(z_2, z_4)$
is given by:
\begin{equation}
W(z_1, z_2)W(z_2, z_4)
=e^{-t\log [\pm (z_1-z_2)]}A_{14}e^{t\log [\pm (z_4-z_2)]}
\label{closed1}
\end{equation}
where $A_{14}=A_1\otimes A_4$ is a 2-tensor reduced from the 4-tensor 
$A=A_1\otimes A_2\otimes A_3\otimes A_4$ in (\ref{m4a}). In this reduction the $t$ operator of $\Phi=e^{-t\log [\pm (z_1-z_2)]}$ acting on the left side of $A_1$ and $A_3$ in $A$ is reduced to acting on the left side of $A_1$ and $A_4$ in $A_{14}$. Similarly  the $t$ operator of $\Psi=e^{-t\log [\pm (z_4-z_2)]}$ acting on the right side of $A_2$ and $A_4$ in $A$ is reduced to acting on the right side of $A_1$ and $A_4$ in $A_{14}$.

Then since $t$ is a 2-tensor operator we have that $t$ is as a matrix acting on the two sides of the 2-tensor $A_{14}$ which is also as a matrix with the same dimension as $t$.
Thus $\Phi$ and $\Psi$ are as matrices of the same dimension as the matrix
$A_{14}$  acting on $A_{14}$ by the usual matrix operation.
Then since $t$ is a Casimir operator for the 2-tensor group representation of $SU(2)$ we have that
$\Phi $  and $\Psi $ commute
with $A_{14}$ since  $\Phi $  and $\Psi$ are exponentials
of $t$ (We remark that $\Phi $  and $\Psi $ are in general not commute with the 4-tensor initial operator $A$).
Thus we have:
\begin{equation}
e^{-t\log [\pm (z_1-z_2)]}A_{14}e^{t\log[\pm (z_4-z_2)]}
=e^{-t\log [\pm (z_1-z_2)]}e^{t\log[\pm (z_4-z_2)]}A_{14}
\label{closed1a}
\end{equation}

We let $W(z_1, z_2)W(z_2, z_4)$ be as a representation of the quantum Wilson line $W(z_1,z_4)$ and we write $W(z_1,z_4)=W(z_1, z_2)W(z_2, z_4)$. 
Then we have the following representation of  
$W(z_1,z_4)$:
\begin{equation}
W(z_1,z_4)=W(z_1,w_1)W(w_1,z_4)=e^{-t\log [\pm (z_1-w_1)]}e^{t\log[\pm (z_4-w_1)]}A_{14}
\label{closed1a1}
\end{equation}
This representation of the quantum Wilson line $W(z_1,z_4)$ means that the line (or path) with end points $z_1$ and $z_4$ is specified that it passes the intermediate point $w_1=z_2$. This representation shows the quantum nature that the path is  not specified at other intermediate points except the intermediate point $w_1=z_2$. This unspecification of the path is of the same quantum nature of the Feymann path description of quantum mechanics.

Then let us consider another representation of the quantum Wilson line $W(z_1,z_4)$. We consider $W(z_1,w_1)W(w_1,w_2)W(w_2,z_4)$ which is obtained from the tensor $W(z_1,w_1)W(u_1,w_2)W(u_2,z_4)$ by two reductions where $z_j$, $w_j$, $u_j$, $j=1,2$ are independent variables. For this representation we have: 
\begin{equation}
W(z_1,w_1)W(w_1,w_2)W(w_2,z_4)
=e^{-t\log [\pm (z_1-w_1)]}e^{-t\log [\pm (z_1-w_2)]}
e^{t\log[\pm (z_4-w_1)]}e^{t\log[\pm (z_4-w_2)]}A_{14}
\label{closed1a2}
\end{equation}
This representation of the quantum Wilson line $W(z_1,z_4)$ means that the line (or path) with end points $z_1$ and $z_4$ is specified that it passes the intermediate points $w_1$ and $w_2$. This representation shows the quantum nature that the path is  not specified at other intermediate points except the intermediate points $w_1$ and $w_2$. This unspecification of the path is of the same quantum nature of the Feymann path description of quantum mechanics.

Similarly we may represent the quantum Wilson line $W(z_1,z_4)$ by path with end points $z_1$ and $z_4$ and is specified only to pass at finitely many intermediate points. Then we let the quantum Wilson line $W(z_1,z_4)$ as an equivalent class of all these representations. Thus we may write $W(z_1,z_4)=W(z_1,w_1)W(w_1,z_4)=W(z_1,w_1)W(w_1,w_2)W(w_2,z_4)=\cdot\cdot\cdot$.

{\bf Remark}. Since $A_{14}$ is a 2-tensor  
we have that a natural group representation for the Wilson line $W(z_1,z_4)$ is the 2-tensor group representation.

\section{Representing Braiding of Curves by Quantum Wilson Lines}\label{sec 9aa}

Consider again the product $G(z_1, z_2, z_3, z_4)=W(z_1,z_2)W(z_3,z_4)$.
We have that $G$ is a multivalued analytic function
where the determination of the $\pm$ sign depended on the choice of the
branch.

Let the two pieces of curves be crossing at $w$. Then we have $W(z_1,z_2)=W(z_1,w)W(w,z_2)$ and
 $W(z_3,z_4)=W(z_3,w)W(w,z_4)$. Thus we have
\begin{equation}
W(z_1,z_2)W(z_3,z_4)=
W(z_1,w)W(w,z_2)W(z_3,w)W(w,z_4)
\label{h2}
\end{equation}

If we interchange $z_1$ and $z_3$, then from
(\ref{h2}) we have the following ordering:
\begin{equation}
 W(z_3,w)W(w, z_2)W(z_1,w)W(w,z_4)
\label{h3}
\end{equation}

Now let us choose a  branch. Suppose that
these two curves are cut from a knot and that
following the orientation of a knot the
curve represented by  $W(z_1,z_2)$ is before the
curve represented by  $W(z_3,z_4)$. Then we fix a branch such that the  product in (\ref{m4a}) is
with two positive signs:
\begin{equation}
W(z_1,z_2)W(z_3,z_4)=
e^{-t\log(z_1-z_3)}Ae^{t\log(z_4-z_2)}
\label{h4}
\end{equation}

Then if we interchange $z_1$ and $z_3$ we have:
\begin{equation}
W(z_3,w)W(w, z_2)W(z_1,w)W(w,z_4) =
e^{-t\log[-(z_1-z_3)]}Ae^{t\log(z_4-z_2)}
\label{h5}
\end{equation}
From (\ref{h4}) and (\ref{h5}) as a choice of branch we have:
\begin{equation}
W(z_3,w)W(w, z_2)W(z_1,w)W(w,z_4) =
R W(z_1,w)W(w,z_2)W(z_3,w)W(w,z_4)
\label{m7a}
\end{equation}
where $R=e^{-i\pi t}$ is the monodromy of the KZ equation.
In (\ref{m7a}) $z_1$ and $z_3$ denote two points on a closed curve
such that along the direction of the curve the point
$z_1$ is before the point $z_3$ and in this case we choose
a branch such that the angle of $z_3-z_1$ minus the angle
of $z_1-z_3$ is equal to $\pi$.

{\bf Remark}. We may use other representations of $W(z_1,z_2)W(z_3,z_4)$. For example we may use the following representation:
\begin{equation}
\begin{array}{rl}
 &W(z_1,w)W(w, z_2)W(z_3,w)W(w,z_4)\\
= &e^{-t\log(z_1-z_3)}e^{-2t\log(z_1-w)}e^{-t2\log(z_3-w)}Ae^{t\log(z_4-z_2)}e^{2t\log(z_4-w)}e^{2t\log(z_2-w)}
\end{array}
\label{h4a}
\end{equation}
Then the interchange of $z_1$ and $z_3$ changes only $z_1-z_3$ to $z_3-z_1$. Thus the formula (\ref{m7a}) holds. Similarly
all other representations of $W(z_1,z_2)W(z_3,z_4)$ will give the same result. $\diamond$

Now from (\ref{m7a}) we can take a convention that the ordering (\ref{h3}) represents that
the curve represented by  $W(z_1,z_2)$ is upcrossing
the curve represented by  $W(z_3,z_4)$ while
(\ref{h2}) represents zero crossing of these two
curves.

Similarly from the dual KZ equation as a choice of branch which
is consistent with the above formula we have:
\begin{equation}
W(z_1,w)W(w,z_4)W(z_3,w)W(w,z_2)=
W(z_1,w)W(w,z_2)W(z_3,w)W(w,z_4)R^{-1}
\label{m8a}
\end{equation}
where $z_2$ is before $z_4$. We take a convention that the ordering in (\ref{m8a}) represents that
the curve represented by $W(z_1,z_2)$ is undercrossing the curve represented by $W(z_3,z_4)$.
Here along the orientation of a closed curve the piece of curve
represented by $W(z_1,z_2)$ is before the piece of curve represented by
$W(z_3,z_4)$. In this case since the angle of $z_3-z_1$ minus the angle
of $z_1-z_3$ is equal to $\pi$ we have that the
angle of $z_4-z_2$ minus the angle of $z_2-z_4$ is
also equal to $\pi$ and this gives the $R^{-1}$ in this formula
(\ref{m8a}).

From (\ref{m7a}) and (\ref{m8a}) we have:
\begin{equation}
 W(z_3,z_4)W(z_1,z_2)=
RW(z_1,z_2)W(z_3,z_4)R^{-1} \label{m9}
\end{equation}
where $z_1$ and $z_2$ denote the end points of a curve which is before a curve with end points $z_3$ and $z_4$.
From (\ref{m9}) we see that the algebraic structure of these
quantum Wilson lines $W(z,z')$
is analogous to the quasi-triangular quantum
group \cite{Fuc}\cite{Chari}.

\section{Computation of Quantum Dirac-Wilson Loop}\label{sec10a}

Let us consider again the quantum Wilson line $W(z_1,z_4)=W(z_1, z_2)W(z_2, z_4)$.
Let us set $z_1=z_4$. In this case the quantum Wilson line forms a closed loop.
Now in (\ref{closed1a}) with $z_1=z_4$ we have that $e^{-t\log  \pm (z_1-z_2)}$
and $e^{t\log \pm (z_1-z_2)}$ which come from the two-side KZ
equations cancel each other and from the multivalued property of
the $\log$ function we have:
\begin{equation}
W(z_1, z_1) =R^{n}A_{14} \quad\quad n=0, \pm 1, \pm 2, ...
\label{closed2}
\end{equation}
where $R=e^{-i\pi t}$ is the monodromy of the KZ equation where the integer $n$ is as a winding number \cite{Chari}. 

{\bf Remark}. It is clear that if we use other representation of the quantum Wilson loop $W(z_1,z_1)$ (such as the representation $W(z_1,z_1)=W(z_1,w_1)W(w_1,w_2)W(w_2,z_1)$) then we will get the same result as (\ref{closed2}). $\diamond$

{\bf Remark}. For simplicity we shall drop the subscript of $A_{14}$ in (\ref{closed2}) and simply write $A_{14}=A$.
 $\diamond$
 
{\bf Remark}. In the case that the gauge group is $SU(2)\otimes U(1)$ since the gauge field $A^0$ for $ U(1)$ is independent of the gauge fields $A^j, j=1,2,3$ for $SU(2)$ when the winding of the Wilson loop on $ U(1)$ is independent of the winding of the Wilson loop on $SU(2)$ we have the following more general expression:
\begin{equation}
W(z_1, z_1)
=R_{U(1)}^{n_1} R_{SU(2)}^{n_2}A \qquad  n_1,n_2 =0, \pm 1, \pm 2, ...
\label{closed21}
\end{equation}
where $R_{SU(2)}$ 
denotes  the monodromy of KZ equation for $SU(2)$ and $R_{U(1)}$ denotes the monodromy of the KZ equation for $U(1)$.
$\diamond$

\section{Generalized Dirac-Wilson Loop As Quantum Knot}\label{sec8aa}

Now
we have that the Wilson loop $W(z_1, z_1)$ corresponds to a closed
curve in the complex plane with starting and ending
point $z_1$.
Let this Wilson loop $W(z_1, z_1)$ represent the unknot. 
Then from (\ref{closed2}) we have the following invariant for the unknot:
\begin{equation}
Tr W(z_1, z_1)= Tr R_{U(1)}^{n_1} R_{SU(2)}^{n_2}A \qquad  n_1,n_2 =0, \pm 1, \pm 2, ...
\label{m6}
\end{equation}

In the following let us generalize the Wilson loop $W(z_1, z_1)$ and the definition (\ref{m6}) 
to nontrivial knots.
Let $W(z_i,z_j)$ represent a piece of curve
with starting point $z_i$ and ending point $z_j$.
Then we let:
\begin{equation}
W(z_1,z_2)W(z_3,z_4)
\label{m11}
\end{equation}
represent two pieces of uncrossing curve.
Then by interchanging $z_1$ and $z_3$ we have:
\begin{equation}
W(z_3,w)W(w,z_2)W(z_1,w)W(w,z_4)
\label{m12}
\end{equation}
represent the curve specified by $W(z_1,z_2)$ upcrossing the
curve specified by $W(z_3,z_4)$.

Now for a given knot diagram we may cut it into a sum of
parts which are formed by two pieces of curves crossing  each other. Each of these parts is represented
by  (\ref{m12})( For a knot diagram of the unknot
with zero crossings we simply do not need to cut the
knot diagram).
Then we define the trace of a knot with a
given knot diagram by the following form:
\begin{equation}
 Tr \cdot\cdot\cdot 
 W(z_3,w)W(w,z_2)W(z_1,w)W(w,z_4)
\cdot\cdot\cdot 
 \label{m14}
\end{equation}
where we use (\ref{m12})  to represent the state of the
two pieces of curves specified by 
 $W(z_1,z_2)$ and
$W(z_3,z_4)$. The 
 $\cdot\cdot\cdot$ means the product
of a sequence of parts represented by 
(\ref{m12}) according to the state of
each part. The ordering of the sequence in (\ref{m14})
 follows the ordering of the parts given by the orientation of the
knot diagram. We shall call the sequence of crossings in
the trace (\ref{m14}) as the generalized Wilson
loop of the knot diagram. For the knot diagram of the unknot with zero crossings we simply
let it be $W(z,z)$ and call it the Wilson loop.

{\bf Examples of generalized Wilson loops.}
As an example of generalized Wilson loops let us
consider the trefoil knots in Fig.2 \cite{Mur}-\cite{Rol}. Starting at $z_1$ let the crossings be denoted by 1, 2 and 3.
Then we have the following circling property of the trefoil knots and
the generalized Wilson loops representing the trefoil knots:
\begin{equation}
123 = 123(1) =231 = 231(2)=312 = 312(3)=123 = \cdot\cdot\cdot
\label{cr1}
\end{equation}

As one more example let us consider the figure-eight
knot denoted by ${\bf 4_1}$ in Fig.3. 
The knot diagram of this knot has
four crossings. 
\begin{figure}[hbt]
\centering
\includegraphics[scale=0.6]{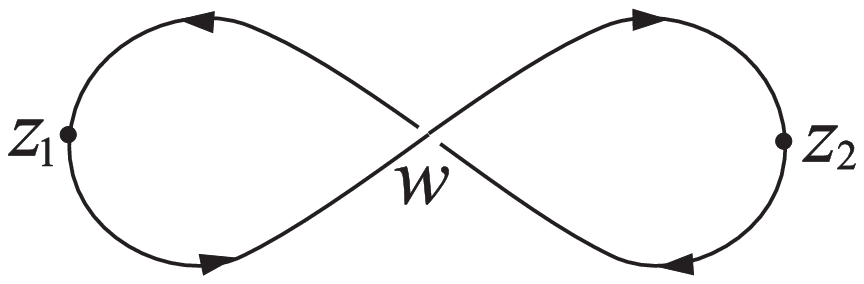}

                Fig.1
\end{figure}

\begin{figure}[hbt]
\centering
\includegraphics[scale=0.5]{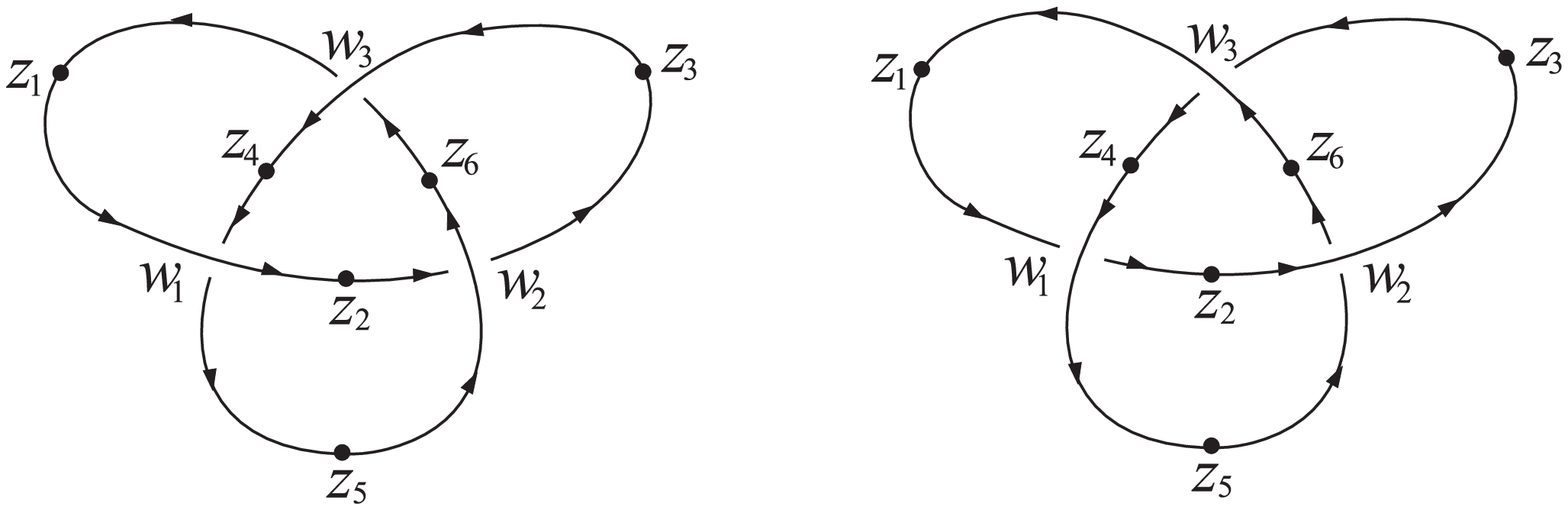}

             Fig.2a  \hspace*{5cm}  Fig.2b
\end{figure}

\begin{figure}[hbt]
\centering
\includegraphics[scale=0.5]{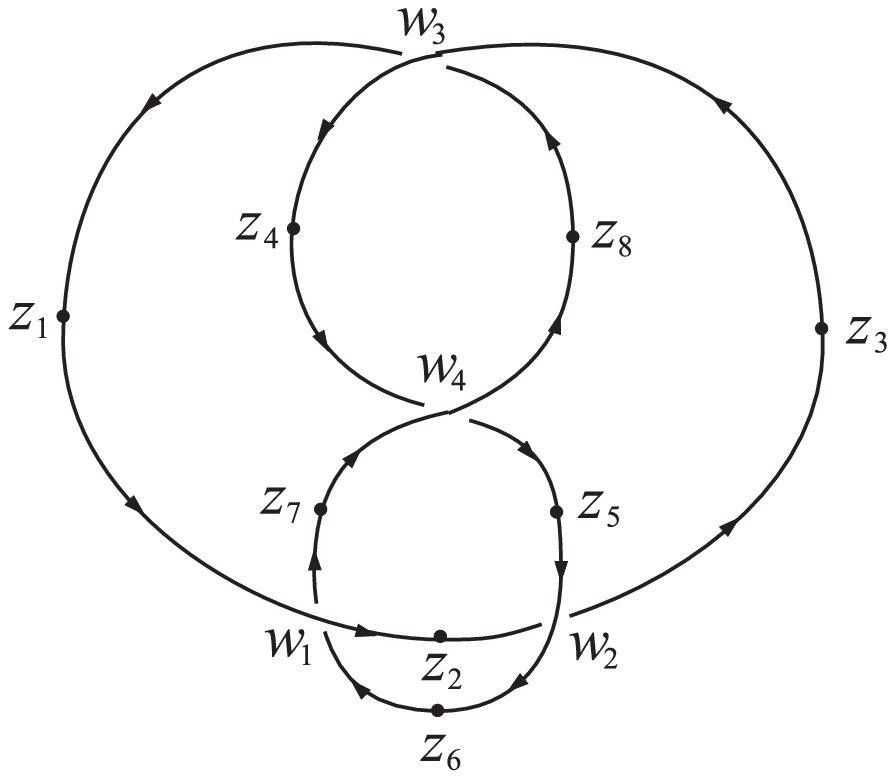}

Fig.3
\end{figure}
Starting at $z_1$ let us denote these crossings by
1, 2, 3 and 4. 
Then we have the following circling property of the figure-eight knot and
the generalized Wilson loop representing the figure-eight knot:
\begin{equation}
\begin{array}{rl}
& 1234 = 1234(2) =1342 = 1342(1)=3421 = 3421(4)=3214 =
3214(3)=2143 \\
& = 2143(1)=2431 = 2431(2)=4312
= 4312(3)=4123 = 4123(4)=1234 = \cdot\cdot\cdot
\end{array}
\label{cr2}
\end{equation}

Let us give an illustration of the above generalized Wilson loop and the corresponding knot invariant.
Let us consider the knot in Fig.1.
For this knot we have that (\ref{m14}) is given by:
\begin{equation}
Tr W(z_2,w)W(w,z_2)W(z_1,w)W(w,z_1)
\label{m15a}
\end{equation}
where the product of Wilson lines  is from the definition (\ref{m12})
representing a crossing at $w$.
In applying (\ref{m12}) we let $z_1$ be the
starting and the ending point. 

Then  we have that (\ref{m15a}) is equal to:
\begin{equation}
\begin{array}{rl}
&Tr W(w,z_2)W(z_1,w)W(w,z_1)W(z_2,w) \\
=&Tr RW(z_1,w)W(w,z_2)R^{-1}
RW(z_2,w)W(w,z_1)R^{-1} \\
=&Tr W(z_1,z_2)W(z_2,z_1) \\
=&Tr W(z_1,z_1) 
\end{array}
\label{m16}
\end{equation}
where we have used (\ref{m9}).
We see that (\ref{m16}) is just the knot invariant (\ref{m6}) of 
the unknot.
Thus the knot in Fig.1 is with the same knot invariant of the unknot and this agrees with the fact that this knot is topologically equivalent
to the unknot.

Let us then consider the trefoil knot in Fig.2a.
By (\ref{m12}) and similar to the above examples
we have that the definition (\ref{m14})
for this knot is given by:
\begin{equation}
\begin{array}{rl}
&Tr W(z_4,w_1)W(w_1,z_2)W(z_1,w_1)W(w_1,z_5)\cdot
W(z_2,w_2)W(w_2,z_6) \\
&W(z_5,w_2)W(w_2,z_3)\cdot 
W(z_6,w_3)W(w_3,z_4)W(z_3,w_3)W(w_3,z_1) \\
=&Tr W(z_4,w_1)RW(z_1,w_1)W(w_1,z_2)
R^{-1}W(w_1,z_5)\cdot
W(z_2,w_2)RW(z_5,w_2) \\
&W(w_2,z_6)R^{-1}W(w_2,z_3)\cdot 
W(z_6,w_3)RW(z_3,w_3)W(w_3,z_4)R^{-1}W(w_3,z_1) \\
\end{array}
\label{trefol}
\end{equation}
By braidings we have that (\ref{trefol}) is equal to
$Tr RW(z_3,z_3)$ \cite{Ng}.
This is as a knot invariant for the trefoil knot in Fig.2a.

Similarly we can show that 
 the trefoil knot in Fig. 2b which
is the mirror image of the trefoil knot in Fig.2a is with knot invariant
$Tr W(z_5,z_5)R^{-1}$ \cite{Ng}.
We notice that the knot invariants for the two
trefoil knots are different. This shows that these two
trefoil knots are not topologically equivalent.

Let $W(K)$ denote the generalized Wilson loop of
a knot $K$. We can show that the generalized Wilson loop of a knot diagram has all the properties of the knot diagram  and that
(\ref{m14}) is  a knot invariant \cite{Ng}. From this we can regard a generalized Wilson loop as a quantum knot. 
Then similar to the case of the trefoil knot we may show that we can write $W(K)$ in the form $W(K)=R^{-j} W(z,z)$
where  $W(z,z)$ denotes the Wilson loop for
the unknot and $j$ is an integer \cite{Ng}. We call $j$ as the power index for the knot $K$. 
More generally we may write $W(K)$ in the form $W(K)=R^{-n_2 j} W(z,z)$ for some fixed winding number $n_2$ for all knots where we replace $R$ with $R^{-n_2 }$.

\section{Knot Model of $\pi$ Mesons}\label{sec12aa}

In the above sections we have introduced a quantum gauge theory from which a quantum knot theory is derived. In the following sections we want to show that this gauge theory and the derived quantum knot theory can give a knot model of mesons.
Before giving the details of this knot model let us first explain the idea of this knot model. The basic idea of this knot model is that when the gauge field $A$ forms a generalized Wilson loop $W(K)$ for a knot $K$ we have that the correseponding matter field $Z$ acted by this generalized Wilson loop is in a strong interaction. This matter field $Z$ is then considered as a field with components as quarks and the gauge field $A$ may be considered as the gluon as similar to the QCD theory (It may be more appropriate to consider the generalized Wilson loop $W(K)$ as the gloun). For the modeling of mesons we use the $SU(2)$ group to model the strong interaction for the formation of mesons. We shall show that for the modeling of baryons we use the $SU(3)$ group to model the strong interaction for the formation of baryons.

In the following let us write out the details of this knot model.
Let us first consider the $\pi^0$ meson. 
The $\pi^0$ meson is composed with $u\overline{u}$ or $d\overline{d}$ 
where $u$ and $d$ represent the up and down quarks respectively. 
Let us represent $\pi^0$ with the following observable:
\begin{equation}
 Z^* W_{SU(2)}(K)Z
\label{sw12}
\end{equation}
where $Z=(z_1,z_2,z_1,z_2)^T =: (u,\overline{u},u,\overline{u})^T$ is a four dimensional complex vector where $z_1, z_2$ are complex variables,  The complex variables $u=z_1 $, $\overline{u}=z_2 $ will represent the up and 
antiup quarks respectively. 

On the other hand we may also write $Z=(z_1,z_2,z_1,z_2)^T =: (d,\overline{d},d,\overline{d})^T$  where the complex variables $d=z_1 $, $\overline{d}=z_2 $ will represent the down and antidown quarks respectively.

In (\ref{sw1}) the generalized Wilson loop $W_{SU(2)}(K)$ 
is as an interaction matrix formed by the gauge field $A$ acting on $Z$ and that we regard $u$ and $\overline{u}$ are in strong interaction via $W_{SU(2)}(K)$ which represents a knot $K$. For the $\pi$ mesons the knot $K$ will be chosen to be the knot ${\bf 4_1}$. 

We remark that for defining the parity of $\pi^0$ we shall extend the above model (\ref{sw12}) of $\pi^0$ to the following form:
\begin{equation}
 \frac12 Z^* [W_{SU(2)}(K)-\overline{W_{SU(2)}(K)}]Z
\label{sw121}
\end{equation}
where $\overline{W_{SU(2)}(K)}$ denotes the conjugate of $W(K)$ obtained by changing the sign of the winding number $n_2$ of $W(K)$.
Here without loss of generality we shall choose the model (\ref{sw12}) of $\pi^0$ to investigate the properties of $\pi^0$ except the parity of $\pi^0$.

Let us then consider the $\pi^+$ meson. We represent $\pi^+$ with the following operator:
\begin{equation}
 Z_1^* W_{SU(2)\otimes U(1)}(K) Z
\label{sw32}
\end{equation}
where we let $Z_1=(z_1,z_1,z_1,z_2) =: (u,u,u,\overline d)$ be a four dimensional complex vector where $z_1, z_2$ are complex variables and 
$Z=(z_1,z_2,z_1,z_2) =: (u,\overline d,u,\overline d)$ where the complex variables $u=z_1 $, $\overline d=z_2 $  represent the up and down quarks respectively. The Wilson loop $W_{SU(2)\otimes U(1)}(K)$ 
is on the group $SU(2)\otimes U(1)$ where the self-adjoint generator
of $U(1)$ is of the form $\frac{e_0}{3}I$ where $e_0$ denotes the bare
electric charge and $I$ denotes the usual self-adjoint generator of $U(1)$. We shall show that this $U(1)$ group gives $\frac{2e}{3}$ charge to the up quark $u$ and gives $\frac{e}{3}$ charge to the down quark $\overline d$ respectively (The observed electric charge $e=e_0 n_e$ for some winding number $n_e$).

We remark that for the parity of $\pi^+$ we shall extend the above model (\ref{sw32}) of $\pi^+$ to the following form:
\begin{equation}
 \frac12 Z_1^* [W_{SU(2)\otimes U(1)}(K)-\overline{W_{SU(2)\otimes U(1)}(K)}]Z
\label{sw122}
\end{equation}
where $\overline{W_{SU(2)}(K)}$ denotes the conjugate of $W(K)$ obtained by changing the sign of the winding number $n_2$ for $SU(2)$ of $W(K)$.
Here without loss of generality we shall choose the model (\ref{sw32}) of $\pi^+$ to investigate the properties of $\pi^+$ except the parity of $\pi^+$.

Similarly we may construct the $\pi^-$ meson. 

\section{Masses of $\pi$ Mesons}

Let us introduce a mass mechanism for generating masses to elementary particles. Let us first consider the positron $e^+$ (or the electron $e^-$).
Let $W_{e^+}(C):=W_{U(1)}(C)$ denote a Wilson loop which will be used for the construction of the positron $e^+$ where $C$ denotes an unknot. Let $W_{e^+}(C)=R_{U(1)}^{n_1}A_{e^+}$ (for some winding number $n_1$) where $R_{U(1)}$ denotes the $R$ matrix for $U(1)$ and $A_{e^+}$ denotes an initial operator for the Wilson loop $W_{e^+}(C)$. By choosing the winding number $n_1$ we may write $R_{U(1)}^{n_1}A_{e^+}= e^{imc^2}A_{e^+}$ where $m$ denotes the mass of positron and $c$ denotes the speed of light and we set  $c=1$ (We have that $m$ is of the form $m=eq$ where $e=e_0 n_e$ ($n_e$ is a winding number) denotes the observed electric charge and $q=q_{min}$ is regarded as the magnetic charge given by $q_{min}=\frac{n_m e_0 \pi}{k}$ for some winding number $n_m$ and the constant $k$ is from the Wilson loop and the current operator $J$ \cite{Ng2}). We set $m=0.5$ Mev. Then we construct the positron as the observable $W_{e^+}(C)z$ where $z$ is the complex number appearing in the quantum model in section 2 and is used to represent the positron (This means that a positron is represented by the whole observable $W_{e^+}(C)z$ while the complex number $z$ is a part for the construction of a positron). Then we construct the observable operator $z^*W_{e^+}(C)z$. This observable is regarded as the mass mechanism for giving mass to the positron $W_{e^+}(C)z$ where the mass of positron is given by the value $m$ in this observable.

This mass mechanism for generating mass can be extended to other elementary particles. Here let us first consider the $\pi$ mesons.
Let $W_{\pi^0}(K)=R_{SU(2)}^{-n_2 j}A_{\pi^0}$ for some fixed winding number $n_2$
where $R_{SU(2)}$ denotes the $R$ matrix for $SU(2)$; $j$ is the power index of the knot $K$ for the $\pi$ mesons which will be determined to be the knot ${\bf 4_1}$.

Now let us consider the structure of $W_{\pi^0}(K)$ and $R_{SU(2)}$.
Let us consider the knot $K$ for the $\pi^0$ meson. Let us choose 
$K= {\bf 4_1}$. 
 For this knot we have $j=3$ (as shown in \cite{Ng}).  
Let us then choose the winding number $n_2$ and the constant for $R_{SU(2)}$ 
such that  $R_{SU(2)}^{n_2}=e^{-i 30m T}$ where $T=\sum_{a=1}^{3} T^a \otimes T^a$ is a  Casimir opeartor for $SU(2)$ and that $T^a$ are self-adjoint generators of $SU(2)$ given by:
\begin{equation}
T^1= \left ( \begin{array}{cc}
             0 & 1 \\
             1 & 0 \end{array}\right), \qquad
T^2= \left ( \begin{array}{cc}
             0 & i \\
            -i & 0 \end{array}\right), \qquad
T^3= \left ( \begin{array}{cc}
             1 & 0 \\
             0 & -1 \end{array}\right)
\end{equation}
 Let us call this Casimir opeartor $T$ as the mass operator for the $\pi$ mesons.
We have
\begin{equation}
 T=\sum_{a=1}^{3} T^a \otimes T^a =
\left ( \begin{array}{cccc}
             1 &  0 &  0 & 0 \\
             0 & -1 &  2 & 0 \\
             0 &  2 & -1 & 0 \\
             0 &  0 &  0 & 1  \end{array}\right)
\label{casimor11}
\end{equation}
This Casimir operator has eigenvalues $1$ and $-3$ where $1$ is with
multiplicity $3$. Since $1$ is positive it is considered as an eigenvalue for the mass of the $\pi$ mesons. On the other hand the negative eigenvalue $-3$ is not considered as an eigenvalue for the mass of the $\pi$ mesons. It can be checked  that
\begin{equation}
v_1=\left ( \begin{array}{cccc}
             1 & 0 & 0 & 0 \\
             0 & 0 & 0 & 0 \\
             0 & 0 & 0 & 0 \\
             0 & 0 & 0 & 0  \end{array}\right),
\quad
v_2=\left ( \begin{array}{cccc}
             0 & 0 & 0 & 0 \\
             0 & 0 & 0 & 0 \\
             0 & 0 & 0 & 0 \\
             0 & 0 & 0 & 1  \end{array}\right),
\quad
v_3= \left ( \begin{array}{cccc}
             0 &       0 &       0 & 0 \\
             0 & \frac12 & \frac12 & 0 \\
             0 & \frac12 & \frac12 & 0 \\
             0 &       0 &       0 & 0  \end{array}\right),
\label{casimir1a}
\end{equation} 
are  three independent eigenvectors for the eigenvalue $1$. 
Let us set the initial operator $A_{\pi^0} =\sum_1^3 v_a b_a$  where $b_a$ denote operators constructed from the gauge fields. With this operator $A_{\pi^0}$ the part from the eigenvalue $-3$ will be eliminated. Then we have
\begin{equation}
W_{\pi^0}(K)=R_{SU(2)}^{-n_2 j}A_{\pi^0}=e^{i 3\times 30 m}A_{\pi^0}
=e^{i 90 m}A_{\pi^0}
\label{Casimor211}
\end{equation}
where we have $j=3$ for the knot ${\bf 4_1}$. Then the $\pi^0$ meson is represented by the observable
\begin{equation}
Z^*W_{\pi^0}(K)Z=Z^*e^{i 90 m}A_{\pi^0}Z 
\label{casimor31}
\end{equation}
where $Z=(z_1,z_2,z_1,z_2)^T$.  Write out this observable we have  
\begin{equation}
Z^*W_{\pi^0}(K)Z= e^{i 90 m}b_1 z_1^*z_1 +  
e^{i 90 m}b_2 z_2^*z_2+
\frac12 e^{i 90 m}b_3 [z_1^*(z_1+z_2)+ z_2^*(z_1+z_2)]             
\label{casimor41}
\end{equation}
where the terms with the factors $z_i^*z_i, i=1,2$ or $\frac12(z_1^*z_1+z_2^*z_2)$ are
mass mechanisms analogous to the positron (or electron) case. Since there are three such  mass mechanisms (each eigenvector of 1 corresponds to a mass mechanism) we have that the mass $m_{\pi^0}$ of $\pi^0$ is given by:
\begin{equation}
m_{\pi^0}=3\cdot 90 m =270 m = 135 Mev
\label{casimor51}
\end{equation}
This agrees with the experimental mass $135 Mev$ of the $\pi^0$ meson. We may write this formula in the form $3\times 45 Mev=135 Mev$ where the prime number $3$ is for the prime knot ${\bf 4_1}$ and $45$ is as a winding number (or proportional to a winding number) of the monodromy $R_{SU(2)}$.

Let us then consider the $\pi^+$ meson. Let $W_{\pi^+}(K)$ denote the Wilson loop for $\pi^+$ where we also choose the knot $K={\bf 4_1}$ for the $\pi^+$ meson. We have $W_{\pi^+}(K)= W_{U(1)\otimes SU(2)}(K)$. For the $\pi^+$ meson in addition to the generators $e_0 T^a$ of $SU(2)$ for the $\pi^0$ we have  one more generator $\frac{e_0}{3}I$ for the $U(1)$ group where $I$ denotes an identity matrix. Thus we have:
\begin{equation}
W_{\pi^+}(K)=R_{U(1)}^{-n_2}R_{SU(2)}^{-n_2 3}A_{\pi^0}=
e^{i \frac19 30 m}e^{i 90 m}A_{\pi^0}
\label{Casimor61}
\end{equation}
where since $U(1)$ is an abelian group we have that the knot $K$ on
$U(1)$ is the same as an unknot and thus $R_{U(1)}$
does not have the index $3$ for $K$; and $R_{U(1)}$ and $R_{SU(2)}$ are with the same winding number $n_2$ since they are from the same knot $K$; and 
the factor $\frac19$ is from the $\frac13$ of the
generator $\frac{e_0}{3}I$. Then similar to the $\pi^0$ meson we have
the following observable representing the $\pi^+$ meson:
\begin{equation}
Z_1^*W_{\pi^+}(K)Z=Z_1^*e^{i \frac19 30 m}e^{i 90 m}A_{\pi^0}Z 
\label{casimor7a1}
\end{equation}
Write out this observable we have:
\begin{equation}
 Z_1^*W_{\pi^+}(K)Z
= e^{i \frac19 30 m}e^{i 90 m}b_1 z_1^*z_1 +  
e^{i \frac19 30 m}e^{i 90 m}b_2 z_2^*z_2+
\frac12 e^{i \frac19 30 m}e^{i 90 m}b_3 [z_1^*(z_1+z_2)+ z_1^*(z_1+z_2)]             
\label{casimor71}
\end{equation}
As similar to the $\pi^0$ meson since there are two $z_1^*z_1$ terms and one $z_2^*z_2$ term  we have that the mass $m_{\pi^+}$ of $\pi^+$ is given by:
\begin{equation}
m_{\pi^+}=3\cdot(\frac{10}{3} m + 90 m) =280 m = 140 Mev
\label{casimor81}
\end{equation}
This agrees with the experimental mass $140 Mev$ of the $\pi^+$ meson.

In (\ref{casimor71}) and (\ref{casimor81}) we have that the electric charges of $u$ and $\overline d$ are from $\frac{e_0}{3}I$ and the two $z_1^*z_1$ and one  $z_2^*z_2$ terms respectively. Because of the electromagnetic generator
$\frac{e_0}{3}I$ we have that each $z_1$ or $z_2$ is with charge $\frac{e}{3}$ (The observed electric charge $e=e_0 n_e$ for some winding number $n_e$). Then since there are two $z_1^*z_1$ terms for the $u$ quark and one $z_2^*z_2$ term for the $\overline d$ quark we have that the $u$ quark is with electric charge $\frac{2e}{3}$ and the $\overline d$ quark is with electric charge $\frac{e}{3}$. This gives the required charges for the $u$ and $\overline d$ quarks.

\section{Weak Interaction of  $\pi^+$ Meson}\label{sec100}

Let us consider the knot modeling of the following weak interaction:
\begin{equation}
\pi^+ \to \mu^+  + \nu_{\mu}
\label{sw4}
\end{equation}
Let $\pi^+$ be represented by $Z_1^*W_{\pi^+}(K)Z$ where
$K={\bf 4_1}$ and $Z_1=(z_1,z_1,z_1,z_2)^T$, $Z=(z_1,z_2,z_1,z_2)^T$.

Let a gauge fixing be such that $A_3=\frac13 A_0$. This gauge fixing is as 
a cause of the weak interaction $\pi^+ \to \mu^+  + \nu_{\mu}$. Then we have a generator $T_{\pi^+}$ given by:
\begin{equation}
T_{\pi^+}= \frac13 I + \frac13 T^3=\frac23 \left ( \begin{array}{ll}
                  1 & 0 \\
                  0 & 0
             \end{array}\right)
\label{sw5}
\end{equation}
From this generator we have that the antidown quark $\overline d$ represented by $z_2$
becomes a particle without electromagnetic interaction and without mass. 
Then in this case the antidown quark $\overline d$ will deviate from the closed loop interaction formed by the knot $K$. This means that the variable $z_2(s)$ representing $\overline d$ deviates from the proper time $s$ to another proper time $s'$ where $s$ is the value that  
$Z(s)=(z_1(s),z_2(s),z_1(s),z_2(s))^T$ for $z_1(s)$ and  $z_2(s)$ meet together by the closed loop $K$ which represents the strong interaction for the formation of the $\pi^+$ meson. 
Under this deviation this $Z$ variable becomes a vector $(z_1(s),z_2(s'),z_1(s),z_2(s'))^T$ and is still acted by the knot $K$.
This deviation of $z_2$ gives the weak interaction that
the $\pi^+$ meson decays into the muon lepton $\mu^+$ and the neutrino $\nu_{\mu}$ where the antidown quark $\overline d$  becomes the neutrino $\nu_{\mu}$ and the up quark $u$ becomes the muon lepton $\mu^+$. 

From (\ref{sw5}) we have that each $z_1$ is with charge $\frac23 e$ and there are two $z_1$ variables. Thus the muon lepton $\mu^+$ will be with charge $\frac23 e + \frac23 e =\frac43 e$ and we need to reduce this charge by a factor $\frac34$ to the charge $e$
for the conservation of electric charge. 
We shall reduce the winding number for 
the value $90 m=90eq=\frac43 e\cdot\frac34\cdot 90q$ from the generator $T_{\pi^+}$ by a factor $\frac34$ to $e\cdot\frac34\cdot 90q$ to get this effect.  

Let us then compute the mass of the muon $\mu$ which is transformed from $u$ under the above weak interaction. We have that the mass operator of the muon $\mu$ is from the following generator:
\begin{equation}
M_{\pi^+}^{'}= T_{\pi^+}\otimes T_{\pi^+} + T^1\otimes T^1 + T^2\otimes T^2
=\left ( \begin{array}{cccc}
             \frac49 &  0 &  0 & 0 \\
             0 &  0 &  2 & 0 \\
             0 &  2 &  0 & 0 \\
             0 &  0 &  0 & 0  \end{array}\right)
\label{mouna}
\end{equation}
This operator has three non-negative eigenvalues: $\frac49, 2$ and $0$ where the eigenvalues $\frac49, 2$ is for the up quark represented by $z_1$ and the eigenvalue $0$ is for the antidown quark $\overline d$ represented by $z_2$.
From this operator of $\pi^+$ we have that the antidown quark $\overline d$ becomes a particle without electromagnetic charge and without mass and thus becomes the neutrino. On the other hand the up quark $u$ becomes the muon $\mu$ with mass given by
\begin{equation}
m_{\mu}=\frac49 \cdot\frac34\cdot 90m + 2\cdot 90m =210m=105 Mev
\label{mouna1}
\end{equation}
where the factor $\frac34$ is from the reducing of the winding number for the generator $T_{\pi^+}$ and is for the conservation of electric charge and for
$\mu$ to have electric charge $e$.
This approximates well the experimental mass $105.7$ Mev of the muon $\mu$. When the neutrino
deviates from $\mu$ we have the decay of the meson $\pi^+$ under weak interaction. This gives a modeling of the weak interaction $\pi^+\to \mu^+ + \nu_{\mu}$. 

In this decay we have that the $\mu^+$ lepton will reduce the winding numbers of its $R$ matrix until it becomes the positron $e^+$ and the corresponding neutrino $\nu_{\mu}$ will become the neutrino $\nu_{e}$. In this case we then have 
the following weak interaction:
\begin{equation}
\mu^+ +\nu_{\mu} \to e^+ + \nu_{e}
\label{moun1}
\end{equation}

\section{Knot Model of the $K^0$ and $K^+$ Mesons}\label{sec12a}

In this section let us consider the knot model of the $K^0$ and $K^+$ mesons.
Let us first consider the $K^0$ meson. 
The $K^0$ meson is composed in the form $d\overline{s}$ 
where $s$ and $d$ represent the strange and down quarks respectively. We shall show that we can model a strange quark $s$ by introducing a strange degree of freedom to a down quark $d$.
Let us represent $K^0$ with the following observable:
\begin{equation}
 Z^* R_s^{n_s}W_{SU(2)}(K)Z
\label{sw1}
\end{equation}
where $Z=(z_1,z_1,z_2,z_2)^T =: (d,d,\overline{s},\overline{s})^T$ is a four dimensional complex vector where $z_1, z_2$ are complex variables.  The complex variables $d=z_1 $, $\overline{s}=z_2 $ will represent the down and 
antistrange quarks respectively. The $R_s$ in (\ref{sw1}) denotes a $R$ matrix for the $SU(2)$ group and this $R_s$ matrix is independent of the $R$ matrix for $W_{SU(2)}(K)$. This $R_s$ matrix is for the strange degree of freedom of the strange quark. We shall later give a more detail description of this $R_s$ matrix.
Let us here give an explanation for the appearance of this $R_s$ matrix for the strange degree of freedom. In the history of particle physics we have that the strange degree was discovered from some strong interactions such as the following one:
\begin{equation}
\pi^- + p \to K^0 + \Lambda^0
\label{Casimor3s}
\end{equation}
where $K^0$ is with strangeness $+1$ and $\Lambda^0$ is with stangeness 
$-1$. This is known as the associated production where $K^0$ and $\Lambda^0$ is a pair of strange particles with the sum of strangeness equal to zero \cite{Pai}-\cite{Fow}. In our knot model of elementary particles we model this strong interaction by the linking of the two knots for the forming of $K^0$ and $\Lambda^0$. In knot theory there are various linking of knots \cite{Ada}-\cite{Rol}. Let us choose the type of linking which links two knots with the linking number equal to zero for the modeling of the associated production \cite{Ada}-\cite{Rol}. In the quantum knots and links theory in \cite{Ng} this type of linking gives a new degree of freedom in the form of a $R^{n_s}$ matrix (the power index $n_s$ is an integer) and a $R^{-n_s}$ matrix such that these two matrices act respectively on one of the two Wilson loops representing the two knots.

 Let us denote this $R$ matrix by $R_s$. When the $R_s^{n_s}$ matrix (and $R_s^{-n_s}$ matrix) acts on a knot it gives a new degree of freedom to the knot and this new degree of freedom may then be identified with the strange degree of freedom where $R_s^{n_s}$ and $R_s^{-n_s}$ give different signs of the strange degree of freedom. Then since 
 the sum of the two power indexes of  $R_s$ is equal to zero
 we have that the sum of the strangeness of $K^0$ and $\Lambda^0$ is equal to zero. Thus this type of linking is just the modeling of the strong interaction (\ref{Casimor3s}) giving the $K^0$ and $\Lambda^0$ mesons which are modeled as two knots with this  type of linking.

In (\ref{sw1}) the generalized Wilson loop $W_{SU(2)}(K)$ is on a representation of the $SU(2)$ group and is as an interaction matrix formed by the gauge field $A$ acting on $Z$ and that we regard $d$ and $\overline{s}$ are in strong interaction via the quantum knot $W_{SU(2)}(K)$. For the 
$K^0$ and $K^+$ mesons in the pseudoscalar representation of $SU(2)$ the knot $K$ will be chosen to be the knot ${\bf 6_1}$. 

Let us then consider the $K^+$ meson. We represent $K^+$ with the following operator:
\begin{equation}
 Z_1^* R_s^{n_s}W_{SU(2)\otimes U(1)}(K) Z
\label{sw3}
\end{equation}
where we let $Z_1=(z_1,z_1,z_1,z_2) =: (u,u,u,\overline{s})$ be a four dimensional complex vector where $z_1, z_2$ are complex variables and 
$Z=(z_1,z_1,z_2,z_2) =: (u,u,\overline{s},\overline{s}$ where the complex variables $u=z_1 $, $\overline d=z_2 $  represent the up and strange quarks respectively. The Wilson loop $W_{SU(2)\otimes U(1)}(K)$ 
is on the group $SU(2)\otimes U(1)$ where the self-adjoint generator
of $U(1)$ is of the form $\frac{e}{3}I$ where $e$ denotes the
electric charge and $I$ denotes the usual self-adjoint generator of $U(1)$. We shall show that this $U(1)$ group gives $\frac{2e}{3}$ charge to the up quark $u$ and gives $\frac{e}{3}$ charge to the down quark $\overline{s}$ respectively.

We remark that the generalized Wilson loop $R_s^{n_s}W_{SU(2)\otimes U(1)}(K)$ for the $K^+$ meson when the $U(1)$ is omitted will be
slightly different from the generalized Wilson loop $R_s^{n_s}W_{SU(2)}(K)$ for the $K^0$ meson where the $R_s^{n_s}$ matrix will act on different vectors for the $K^+$ and $K^0$ mesons.

Similarly we may construct the $\overline{K^0}$ meson and the $K^-$ meson by the mirror image of the knot ${\bf 6_1}$. We have that the  mirror image of the knot ${\bf 6_1}$ and the knot ${\bf 6_1}$ are 
 dual to each other and this gives a way to model the $K^0$ meson and the $K^+$ meson and their anti-particles $\overline{K^0}$ and  $K^-$.

\section{Masses of the $K^0$ and $K^+$ Mesons}

Similar to the $\pi$ mesons we may give masses to the $K^0$ and $K^+$ mesons, as follows.
Let $W_{K^0}(K)=R_{SU(2)}^{-n_2 j}A_{K^0}$
where $R_{SU(2)}$ denotes the $R$ matrix for $SU(2)$; $j$ is the power index of the knot for the $K^0$ and $K^+$ mesons which will be determinated to be the knot ${\bf 6_1}$; and $n_2$ is a winding number as that for the $\pi$ mesons.

Now let us consider the structure of $W_{K^0}(K)$ and $R_{SU(2)}$.
Let us consider the knot $K$ for the $K^0$ meson. Let us choose 
$K= {\bf 6_1}$. 
This knot is assigned with the prime number $j=11$ \cite{Ng}.

For the $K^0$ meson we then have the product of two $R$ matrices $R_s^{n_s}R_{SU(2)}^{-n_2 j}$ where $R_s^{n_s}$ is independent of $R_{SU(2)}$. This independece will give the degree of freedom for the strange quark. We define this $R_s$ matrix such that it is only act on the component of  the initial operator $A_{K^0}$ for the strange quark.

Let us then choose the winding number and the constant for $R_{SU(2)}$ (as that for the $\pi$ mesons) such that $R_{SU(2)}^{-n_2 j}$ is of the form $R_{SU(2)}^{-n_2 j}=e^{-i 30m T}$ where $T=c^2\sum_{a=1}^{2} T^a \otimes T^a +T^3 \otimes T^3$ 
where $c$ is a constant such that $0<c^2<1$ and is closed to $1$ and is to be determined.
 Let us call this Casimir opeartor $T$ as the mass operator for the $K^0$ and $K^+$ mesons mesons.
We have:
\begin{equation}
 T=c^2\sum_{a=1}^{2} T^a \otimes T^a +T^3 \otimes T^3=
\left ( \begin{array}{cccc}
             1 &   0  &  0   & 0 \\
             0 &  -1  & 2c^2 & 0 \\
             0 & 2c^2 & -1   & 0 \\
             0 &   0  &  0   & 1  \end{array}\right)
\label{casimor}
\end{equation}
This Casimir operator has eigenvalues $1$, $2c^2-1$ and $-2c^2-1$ where $1$ is with
multiplicity $2$ and $2c^2-1$ is with multiplicity $1$. We choose $c$ closed to $1$ such that $2c^2-1>0$. This $c$ closed to $1$ is as a deviation from $1$.
Comparing to $c^2= 1 $ for the $\pi$ mesons this constant $c^2\neq 1 $ represents the existence of the structure matrix $\sum_{a=1}^{2} T^a \otimes T^a$ of $SU(2)$ and thus represents the $K^0$ and $K^+$ mesons.

Since $1$ and $2c^2-1$ are positive they are considered as eigenvalues for mass of the $K^0$ and $K^+$ mesonsmesons. On the other hand the negative eigenvalue $-2c^2-1$ is not considered as eigenvalue for mass of the $K^0$ and $K^+$ mesonsmesons. It can be checked  that the eigenmatrices $v_i, i=1,2,3$ in (\ref{casimir1a}) 
are  three independent eigenmatrices for the eigenvalues $1$ and $2c^2-1$. 
Let us set the initial operator $A_{K^0} =\sum_1^3 v_a b_a$  where $b_a$ denote operators constructed from the gauge fields. With this operator $A_{K^0}$ the part from the eigenvalue $-2c^2-1$ will be eliminated. Then we have:
\begin{equation}
R_{SU(2)}^{-n_2 j}A_{K^0}=
e^{i 11\times 30 m}(v_1b_1+v_2b_2)+e^{i 11\times (2c^2-1)30 m}v_3b_3
\label{Casimor2}
\end{equation}
where we have $j=11$ for the knot ${\bf 6_1}$.

Then the $R_s^{n_s}$ matrix which is for the strange degree of freedom is defined to act on the component of  $A_{K^0}$ for the strange quark which is given by $v_2b_2 + v_{32}b_3$ where
$v_{32}$ is given by:
\begin{equation}
v_{32}:= \left ( \begin{array}{cccc}
             0 &  0 &       0 & 0 \\
             0 &  0 & \frac12 & 0 \\
             0 &  0 & \frac12 & 0 \\
             0 &  0 &       0 & 0  \end{array}\right),
\label{casimir1as1}
\end{equation} 
where we write $v_3$ in the form $v_3=v_{31}+v_{32}$.
Then we have: 
\begin{equation}
R_s^{n_s}A_{K^0}:= (v_1b_1 + v_{31}b_3)+ R_s^{n_s}(v_2b_2 + v_{32}b_3)
\label{casimir1as2}
\end{equation}
where we let $v_2b_2 + v_{32}b_3$ for the strange quark $s$ and $v_1b_1 + v_{31}b_3$ for the anti-strange quark 
$\overline{s}$.
From this we then have:
 \begin{equation}
 \begin{array}{rl}
W_{K^0}(K)=&
R_s^{n_s}R_{SU(2)}^{-n_2 j}A_{K^0}=
R_s^{n_s}[e^{i 11\cdot 30 m}(v_1b_1+v_2b_2)+e^{i 11\cdot (2c^2-1)30 m}v_3b_3]\\
=&[e^{i 11\cdot 30 m}v_1b_1 + 
e^{i 11\cdot (2c^2-1)30 m}v_{31}b_3]+ 
R_s^{n_s}[e^{i 11\cdot 30 m}v_2b_2 + 
 e^{i 11\cdot (2c^2-1)30 m}v_{32}b_3] \\
=&[e^{i 11\cdot 30 m}v_1b_1 + 
 e^{i 11\cdot (2c^2-1)30 m}v_{31}b_3]+ 
e^{i\alpha m}
[e^{i 11\cdot 30 m}v_2b_2 + 
 e^{i 11\cdot (2c^2-1)30 m}v_{32}b_3]
 \end{array}
\label{Casimor2s}
\end{equation}
where $e^{i\alpha m}$ is from $R_s^{n_s}$ and the constant $\alpha$ is to be determined.
Then we have that the $K^0$ meson is  represented by the observable:
\begin{equation}
Z^*R_s^{n_s}W_{K^0}(K)Z
\label{casimor3}
\end{equation}
where $Z=(z_1,z_1,z_2,z_2)^T$.  Write out this observable we have:  
\begin{equation}
\begin{array}{rl}
 & Z^*R_s^{n_s}W_{\pi^0}(K)Z \\
= &
e^{i 11\cdot 30 m} b_1 z_1^*z_1 +  
\frac12 e^{i 11\cdot (2c^2-1)30 m}b_{31} [z_1^*(z_1+z_2)] \\
 & +e^{i\alpha m}e^{i 11\cdot 30 m}b_2 z_2^*z_2+
\frac12 e^{i\alpha m}e^{i 11\cdot (2c^2-1)30 m}b_{32}[ z_2^*(z_1+z_2)]
\end{array}            
\label{casimor4}
\end{equation}
where the terms with the factors $z_i^*z_i, i=1,2$  are
mass mechanisms analogous to the positron (or electron) case and the terms with factor $z_2^*z_2$ are for the strange quark. 

Then we have that the mass of the $K^0$ meson which are from the nonstrange degree of freedom is given by:
\begin{equation}
11[1\cdot 2 +(2c^2-1)\cdot1]2\cdot m\cdot15 = 11\cdot 43 Mev=473 Mev
\label{casimor5s1}
\end{equation}
where we choose $c^2$ by requiring that $c^2 <1$ is as the smallest deviation from $1$ and is a rational number
such that the expression $[1\cdot 2 +(2c^2-1)\cdot1]\cdot15=[1+2c^2]\cdot 15 $ is an integer. From this integer condition we have that $c^2$ is determined by the equation $[1+2c^2]\cdot 15=43$ which gives $c^2=\frac{14}{15}$. We remark that this integer condition is from the property of quantum knots that the total winding of a quantum knot is an integer which is as the quantum phenomeon of energy and that the number $15$ is as the original winding number of the $R$-matrix of the quantum knot. Thus this integer condition can be interpreted as a quantum condition.

Let us then choose $\alpha =32$. Then we have that the total mass of $K^0$ meson which is from the strange degree of freedom is given by:
\begin{equation}
32 m + \frac12 32m=(16+8) Mev =24 Mev
\label{casimor5s2}
\end{equation}
Thus we have that the total mass $m_{K^0}$ of the $K^0$ meson is given by:
\begin{equation}
m_{K^0}= 473 Mev + (16+8) Mev = 497 Mev
\label{casimor5s3}
\end{equation}
This agrees with the experimental mass $497.7 Mev$ of the $K^0$ meson. We may write this formula in the form $11\times 43 + 24 Mev=497 Mev$ where the prime number $11$ is for the prime knot ${\bf 6_1}$ and $43$ is as a winding number (or proportional to a winding number) of the monodromy $R_{SU(2)}$.

Let us then consider the $K^+$ meson. Let $W_{K^+}(K)$ denote the Wilson loop for $K^+$ where we also choose the knot ${\bf 6_1}$ for the $K^+$ meson. For the $K^+$ meson in addition to the generators $e_0 T^a$ of $SU(2)$ for the $K^0$ we have  one more generator $\frac{e_0}{3}I$ for the $U(1)$ group where $I$ denotes an identity matrix. Thus we have:
\begin{equation}
\begin{array}{rl}
R_s^{n_s}W_{K^+}(K)=&
R_s^{n_s}R_{U(1)}^{-n_2}R_{SU(2)}^{-n_2 11}A_{K^+}=
e^{i \frac19 30 m}
R_s^{n_s}[e^{i 11\cdot 30 m}(v_1b_1+v_2b_2)+e^{i 11\cdot (2c^2-1)30 m}v_3b_3] \\
=& e^{i \frac19 30 m}
[e^{i 11\cdot 30 m}v_1b_1 + e^{i 11\cdot (2c^2-1)30 m}v_3b_3)]+ 
 e^{i \frac19 30 m}R_s^{n_s}[e^{i 11\cdot 30 m}v_2b_2 ] \\
=& e^{i \frac19 30 m}
[e^{i 11\cdot 30 m}v_1b_1 + e^{i 11\cdot (2c^2-1)30 m}v_3b_3)]+ 
e^{i \frac19 30 m}e^{i\alpha m}e^{i 11\cdot 30 m}v_2b_2
\end{array}
\label{Casimor6}
\end{equation}
where as for the $\pi^+$ meson since $U(1)$ is an abelian group we have that the knot $K$ on
$U(1)$ is the same as an unknot and thus the $R$-matrix  $R_{U(1)}$
does not have the index $11$ for $K$; the factor $\frac19$ is from the $\frac13$ of the
generator $\frac{e_0}{3}I$; and the winding number $n_2$ is as that for the $\pi^+$ meson. 

For the $K^+$ meson we have that $A_{K^+}=A_{K^0}$ where the component of  $A_{K^+}$ for the strange quark is $v_2b_2$ which is different from the strange component of $A_{K^0}$. Thus we have that the $R_s^{n_s}$ matrix for $K^+$ is defined by:
\begin{equation}
R_s^{n_s}A_{K^+}:= (v_1b_1 + v_3b_3)+ R_s^{n_s} v_2b_2 
\label{casimir1as3}
\end{equation}

Then similar to the $K^0$ meson we have
the following observable representing the $K^+$ meson:
\begin{equation}
Z_1^*R_s^{n_s}W_{K^+}(K)Z
\label{casimor7a}
\end{equation}
Write out this observable we have:
\begin{equation}
\begin{array}{rl}
  & Z_1^*R_s^{n_s}W_{K^+}(K)Z \\
=& e^{i \frac19 30 m}e^{i 11\cdot 30 m}b_1 z_1^*z_1 +  
e^{i \frac19 30 m}e^{i\alpha m}e^{i 11\cdot 30 m}b_2 z_2^*z_2+ \\
 & \frac12 e^{i \frac19 30 m}e^{i 11\cdot(2c^2-1)30 m}b_3 [z_1^*(z_1+z_2)+ z_1^*(z_1+z_2)]
\end{array}             
\label{casimor7}
\end{equation}
As similar to the $K^0$ meson since there are two $z_1^*z_1$ terms and one $z_2^*z_2$ term  we have that the mass of $K^+$ which is from the nonstrange degree of freedom is given by:
\begin{equation}
m_{K^+}=
2\cdot [\frac19 30m + 11\cdot 30m]+ [\frac19 30m +11\cdot(2c^2-1)30m]
=10m +11\cdot 86m = (5+ 473) Mev = 478 Mev
\label{casimor8s1}
\end{equation}
Then the mass of $K^+$ which is from the strange degree of freedom is given by:
\begin{equation}
\alpha m =32 m =16 Mev
\label{casimor8s2}
\end{equation}
Thus we have that the total mass of $K^+$ is given by:
\begin{equation}
(478 +16) Mev =494 Mev
\label{casimor8s3}
\end{equation}
This agrees with the experimental mass $493.7 Mev$ of the $K^+$ meson.
We may write the formula for computing the mass of $K^+$ as 
$11\times 43+16+5=494$ where the prime number $11$ is from the prime knot
$6_1$ and $43$ is as a winding number.
We notice that this knot model of $K$ mesons explains why the mass of the $K^0$ meson is greater than the mass of the $K^+$ meson even though the 
$K^+$ meson is with charge. 

In (\ref{casimor7})  we have that the electric charges of $u$ and $\overline s$ are from $\frac{e_0}{3}I$ and the two $z_1^*z_1$ and one  $z_2^*z_2$ terms respectively. Because of the electromagnetic generator
$\frac{e_0}{3}I$ we have that each $z_1$ or $z_2$ is with charge $\frac{e}{3}$ (The observed charge $e=e_0 n_e$ for some winding number $n_e$ \cite{Ng2}). Then since there are two $z_1^*z_1$ terms for the $u$ quark and one $z_2^*z_2$ term for the $\overline s$ quark we have that the $u$ quark is with electric charge $\frac{2e}{3}$ and the $\overline s$ quark is with electric charge $\frac{e}{3}$. This gives the required charges for the $u$ and $\overline s$ quarks.

Similar to the models of $K^{0}$ and $K^{+}$ we model their anti-particles, the $\overline{K^{0}}$ and $K^{-}$ mesons, by the mirror image $\overline{\bf 6_1}$ of the prime knot ${\bf 6_1}$ which is not an amphichiral knot \cite{Ada}-\cite{Rol}. We have that $\overline{\bf 6_1}$ is assigned with the number $-11$ while ${\bf 6_1}$ is assigned with the prime number $11$ since $\overline{\bf 6_1}$ is the  mirror image of ${\bf 6_1}$ \cite{Ng}. For simplicity we omit the details.

\section{Parity and Conjugate Operations }\label{sec15a}

Let us in this section consider the model of the parity operator $P$, the conjugate operator $C$ and the G-parity operator in this knot model. 

Let $W(K)$ be the generalized Wilson loop of a knot $K$. Then we define the parity operator $P$ on
$W(K)$ by:
\begin{equation}
PW(K) := \overline{W(K)}
\label{P1}
\end{equation}
where $\overline{W(K)}$ denotes the conjugate of $W(K)$ obtained by changing the sign of the winding number $n_2$ of $W(K)$. This means that $P$ changes left spinning to right spinning, and vice versa; and thus can be used to model the parity operation.

Then we define the charge conjugatation operator $C$ on $Z_1^* W(K)Z_2$ by: 
\begin{equation}
CZ_1^* W(K)Z_2:= Z_2^* W(K)Z_1
\label{P2}
\end{equation}

Then we define $C$ on $W(K)\frac{i}{\sqrt2}[Z+\overline{Z}]$ by:
\begin{equation}
CW(K)\frac{i}{\sqrt2}[Z+\overline{Z}]:=-W(K)\frac{i}{\sqrt2}[Z+\overline{Z}]
\label{P2a}
\end{equation}
This means that the charge conjugatation operator $C$ maps the vector $\frac{i}{\sqrt2}(Z+\overline{Z})$ to its conjugate $-\frac{i}{\sqrt2}(Z+\overline{Z})$.

Then we define $C$ on $W(K)\frac1{\sqrt2}[Z-\overline{Z}]$ by:
\begin{equation}
CW(K)\frac{1}{\sqrt2}[Z-\overline{Z}]:=W(K)\frac{1}{\sqrt2}[\overline{Z}-\overline{\overline{Z}}]
=-W(K)\frac{1}{\sqrt2}[Z-\overline{Z}]
\label{P2b}
\end{equation}
This means that the charge conjugatation operator $C$ maps the vector $\frac{1}{\sqrt2}(Z-\overline{Z})$ to its conjugate $-\frac{1}{\sqrt2}(Z-\overline{Z})$.

When a $U(1)$ gauge group is presented  giving a factor $R_{U(1)}^{n_1}$ multiplied to $W(K)$ we define 
the charge conjugatation operator $C$ on $Z_1^*R_{U(1)}^{n_1}W(K)Z_2$ by:
\begin{equation}
CZ_1^*R_{U(1)}^{n_1}W(K)Z_2 := Z_2^* R_{U(1)}^{-n_1}W(K)Z_1
\label{P2a1}
\end{equation}
where the sign of the winding number $n_1$ is changed. This  represents that the sign of the electric charge is changed under the charge conjugatation operator $C$. On the other hand when the parity $P$ is on $R_{U(1)}^{n_1}W(K)$ we define $P$ such that the sign of $n_1$ is unchanged:
\begin{equation}
PZ_2^* R_{U(1)}^{n_1}W(K)Z_1 :=Z_1^* R_{U(1)}^{n_1}\overline{W(K)}Z_2
\label{P1b}
\end{equation}

As an example let us consider the parity and charge conjugation of $\pi$ mesons.  As in the above section on the knot model of $\pi^0$ we have that the knot model of $\pi^0$ is given by:
\begin{equation}
Z^* \frac12[W({\bf 4_1})- \overline{W({\bf 4_1})}]Z 
\label{P1c}
\end{equation}
where $\pi^0$ is modeled by the figure-eight knot ${\bf 4_1}$ which is a prime knot indexed by the prime number $3$ and is an amphichiral knot equivalent to its mirror image \cite{Ng}\cite{Ada}-\cite{Rol}.

Under $P$ we have that the model of $\pi^0$ is changed by:
\begin{equation}
PZ^* \frac12[W({\bf 4_1})- \overline{W({\bf 4_1})}]Z 
=-Z^* \frac12[W({\bf 4_1})-\overline{W({\bf 4_1})}]Z 
\label{P1c1}
\end{equation}
Thus the parity of $\pi^0$ is $-1$. From (\ref{P1c1}) we also have that $\pi^0$ is with two states: $\pm Z^* \frac12[W({\bf 4_1})- \overline{W({\bf 4_1})}]Z$. 

On the other hand under $C$ we have that the model of $\pi^0$ is given by:
\begin{equation}
CZ^* \frac12[W({\bf 4_1})- \overline{W({\bf 4_1})}]Z 
=Z^* \frac12[W({\bf 4_1})- \overline{W({\bf 4_1})}]Z 
\label{P1c2}
\end{equation}
Thus the charge conjugation $C$ of $\pi^0$ is $+1$.

Similarly  
$CP$ changes $\pi^+$ to $\pi^-$ where we model $\pi^+$ and $\pi^-$ respectively by the following models: 
\begin{equation}
Z_1^*  R_{U(1)}^{n_1}\frac12[W({\bf 4_1})- \overline{W({\bf 4_1})}]Z  
\label{P1c3}
\end{equation}
and
\begin{equation}
Z^*  R_{U(1)}^{-n_1}\frac12[\overline{W({\bf 4_1})}-W({\bf 4_1})]Z_1  
\label{P1c4}
\end{equation}

From the property that the prime knot ${\bf 4_1}$ is an amphichiral knot that it is equal to its mirror image
we can derive the nonconservation of parity in the weak interaction  \cite{Lee}\cite{Wu}. We shall in elsewhere give a detail derivation that the prime knot ${\bf 4_1}$ being an amphichiral knot gives the nonconservation of parity of the  weak interaction.

We remark that the above definition of charge conjugation $C$ is defined to model the usual charge conjugation  in the literature of particle physics. Here we note that while the above definition of the $CP$ transformation is just the $CP$ transformation in the literature of particle physics we have that the above definition of charge conjugation $C$ is different from the usual charge conjugation in that the above definition of charge conjugation $C$ is complementary to the parity $P$ such that the transformation $CP$ maps a particle to its anti-particle.
 Here the point is that we found that the parity  $P$ which maps a particle to its mirror image is actually a part of the usual charge conjugation which maps a particle to its anti-particle. This means that the mirror image of a particle is as a part of the anti-particle of this particle. As an example we consider the $K^0$ particle. In the above section we have shown that the $K^0$ meson is modeled by the quantum knot $W({\bf 6_1})$. Then the parity $P$ maps $W({\bf 6_1})$ to 
$\overline{W({\bf 6_1})}=W(\overline{\bf 6_1})$ where $\overline{\bf 6_1}$ denotes the mirror image of the prime knot ${\bf 6_1}$. Then since the anti-particle of $K^0$, denoted by $\overline{K^0}$, is modeled by the quantum prime knot $W(\overline{\bf 6_1})$ we have that the parity  $P$ actually maps the $K^0$ meson to its anti-particle $\overline{K^0}$. In this case the parity  $P$ is actually equal to the usual charge conjugation which maps a particle to its anti-particle. As another example we have shown in the above that the parity  $P$ is a part of the transformation $CP$ of mapping the $\pi^+$ meson to its anti-particle $\pi^-$. Thus the parity  $P$ is a part of the usual charge conjugation which maps a particle to its anti-particle.

Let us then consider the internal $G$-parity. Let us define the $G$-parity operator $G$ on $Z_1^*W(K)Z_2$ by:
\begin{equation}
GZ_1^*W(K)Z_2 :=Z_1^* \overline{W(K)}Z_2
\label{P3}
\end{equation}

Then we define $G$ on $[Z-\overline{Z}]^*W(K)\frac{1}{2}[Z-\overline{Z}]$ by:
\begin{equation}
G[Z-\overline{Z}]^* W(K)\frac{1}{2}[Z-\overline{Z}]:= [Z-\overline{Z}]^*\overline{W(K)}\frac{1}{2}[\overline{Z}-\overline{\overline{Z}}]
= -[Z-\overline{Z}]^*\overline{W(K)}\frac{1}{2}[Z-\overline{Z}]
\label{P3c}
\end{equation}

Similarly we define $G$ on $W(K)\frac{i}{\sqrt2}[Z+\overline{Z}]$ by:
\begin{equation}
GW(K)\frac{i}{\sqrt2}[Z+\overline{Z}]:= \overline{W(K)}\frac{i}{\sqrt2}[\overline{Z}+\overline{\overline{Z}}]
= \overline{W(K)}\frac{i}{\sqrt2}[Z+\overline{Z}]
\label{P3a}
\end{equation}
and we define $G$ on $W(K)\frac{1}{\sqrt2}[Z-\overline{Z}]$ by:
\begin{equation}
GW(K)\frac{1}{\sqrt2}[Z-\overline{Z}]:= \overline{W(K)}\frac{1}{\sqrt2}[\overline{Z}-\overline{\overline{Z}}]
= -\overline{W(K)}\frac{1}{\sqrt2}[Z-\overline{Z}]
\label{P3b}
\end{equation}

As an example we have that the $G$-parity of $\pi^0$ is given by:
\begin{equation}
PZ^* \frac12[W({\bf 4_1})- \overline{W({\bf 4_1})}]Z 
=-Z^* \frac12[W({\bf 4_1})- \overline{W({\bf 4_1})}]Z 
\label{P1c5}
\end{equation}
Thus the $G$-parity of $\pi^0$ is $-1$. Similarly the $G$-parity of $\pi^+$ (or $\pi^-$) is $-1$.

\section{Quantum Condition for Knot Modeling of Mesons }\label{sec151}

In the above knot modeling of $K^0$ and $K^+$ mesons we have determined a quantum condition which gives $c^2=\frac{14}{15}$ and the number $43$ for the $K^0$ and $K^+$ mesons. This $c^2$ is for a multiplet which contains a meson with a charge matrix $\frac13 I$ ($I$ denotes the two dimensional identity matrix) such as the $K^+$ meson. This is because of the number $15=3\cdot5$ where the number $3$ is for electric charge.
Let us then consider another $c^2$ multiplied to the matrix $T^3\otimes T^3$. Then we consider a mass operator of the following form:
\begin{equation}
 T=\sum_{a=1}^{2} T^a \otimes T^a +c^2T^3 \otimes T^3=
\left ( \begin{array}{cccc}
             c^2&   0 &  0  & 0 \\
             0  & -c^2&  2  & 0 \\
             0  &   2 & -c^2& 0 \\
             0  &   0 &  0  & c^2  \end{array}\right)
\label{casim1}
\end{equation}
When $c^2<1$ we let this operator be an operator for mesons without the charge matrix $\frac13 I$.
This Casimir operator has eigenvalues $c^2$, $2-c^2$ and $-2-c^2$ where $1$ is with
multiplicity $2$ and $2-c^2$ is with multiplicity $1$. We choose $c$ closed to $1$ such that $2-c^2>0$. This $c$ closed to $1$ is as a deviation from $1$.
Comparing to $c^2= 1 $ for the $\pi$ mesons this constant $c^2\neq 1 $ represents the existence of the structure matrix $T^3 \otimes T^3$ of $SU(2)$.

Then similar to the mass operator (\ref{casimor}) for the $K^0$ and $K^+$ mesons for the mass operator (\ref{casim1}) we choose $c^2$ by requiring that $c^2 <1$ is as the smallest deviation from $1$ and is a rational number of the form $\frac{a}{5}$
such that the expression:
 \begin{equation}
[c^2\cdot 2 +(2-c^2)\cdot1]\cdot5=[c^2+2]\cdot 5  
\label{integer}
\end{equation}
is an integer.
Here $c^2$  is of the form $\frac{a}{5}$ because the factor $3$ is omitted from $15$ for mesons without the charge matrix $\frac13 I$.
 
We remark that similar to the case of the $K^0$ and $K^+$ mesons
this integer condition is from the property of quantum knots that the total winding of a quantum knot is an integer which is as the quantum phenomeon of energy and that the number $15$ is as the original winding number of the $R$-matrix of the quantum knot. Thus this integer condition can be interpreted as a quantum condition.

From (\ref{integer}) we determine that $c^2=\frac{4}{5}=\frac{12}{15}$. 
Thus from (\ref{integer}) we have the following quantum number:
 \begin{equation}
[c^2\cdot 2 +(2-c^2)\cdot1]\cdot15=[c^2+2]\cdot 15=42  
\label{integer2}
\end{equation}
Thus from the 
two quantum steps:  
$c^2=\frac{12}{15}, \frac{14}{15}<1$ we have two quantum numbers $42, 43$ for modeling mesons where the quantum step $c^2=\frac{12}{15}$ for (\ref{integer}) gives the quantum number $42$ and the quantum step $c^2= \frac{14}{15}$ for (\ref{casimor}) gives the quantum number $43$.

On the other hand for $c^2>1$ and when the mass operator (\ref{integer}) is also for mesons with the charge matrix $\frac13 I$ then from (\ref{integer}) we have the first quantum step $c^2=1+\frac{1}{15}$. For this quantum step from (\ref{integer}) we have the following quantum number:
 \begin{equation}
[c^2\cdot 2 +(2-c^2)\cdot1]\cdot15=[c^2+2]\cdot 15=46  
\label{integer3}
\end{equation}

Thus from the quantum condition we have four quantum numbers $42, 43, 45, 46$ for modeling mesons (The $\pi$ mesons are modeled by the basic quantum number $45$). We shall see in the next section that these quantum numbers are suitable to model mesons which are with properties consistent with the properties of these quantum numbers.

\section{Knot Model of Pseudoscalar and Vector Mesons }\label{sec15}

We have established knot models for the $\pi$ mesons and the $K^0$ and $K^+$ mesons. By similar constructions we may establish knot models for other mesons. Let us here briefly consider the mesons of the nonets of the pseudoscalar and vector mesons, as follows.

We shall model the $\eta,  \rho, \omega, K^*, \eta'$ and $\phi$ mesons by the generalized Wilson loops representing the prime knots ${\bf 6_2},{\bf 6_3},{\bf 6_3},{\bf 7_1},{\bf 7_2}$ and ${\bf 7_2}$ respectively. These prime knots ${\bf 6_2},{\bf 6_3},{\bf 7_1}$ and ${\bf 7_2}$ are assigned with the prime numbers $13, 17, 19$ and $ 23$ respectively \cite{Ng}. 

Let us first consider the $\eta$ meson. Let us model $\eta$ by the following observable:
\begin{equation}
(Z-\overline{Z})^*\frac12[W({\bf 6_2})-W(\overline{\bf 6_2})]\frac12(Z-\overline{Z})
\label{phi11}
\end{equation}
where $\overline{\bf 6_2}$ denotes the mirror image of ${\bf 6_2}$; and since the $\eta$ meson is without charge from the above quantum condition we have that $W({\bf 6_2})$ is with the mass operator (\ref{integer}) with $c^2=\frac{12}{15}$ which gives quantum number $42$.
  Then we have that the mass of $\eta$ is given by:
\begin{equation}
\frac12[13\cdot 42 + 13\cdot 42 ]
Mev =546 Mev
\label{phi12}
\end{equation}
where the prime number $13$ is for the prime knot ${\bf 6_2}$ and the two numbers $13\cdot 42$ are computed from $W({\bf 6_2})$ and $W(\overline{\bf 6_2})$ respectively where the minus sign of $W(\overline{\bf 6_2})$ does not affect the computation of the mass of $\eta$.
This approximates quite well the experimental mass $549 Mev$ of $\eta$.

From the above modeling of the parity $P$ and conjuate operation $C$ we have that the $PC$ of $\eta$ is $-+$ and the $G$ of $\eta$ is $+$.

Let us then consider the $\eta'$ and the $\phi$ meson which are related to the strange degree of freedom. 
For these two mesons we shall also consider the scalar mesons $a_0(980)$ and $f_0(980)$ which will be shown to be related to these two mesons that these four mesons are modeled by the prime knot ${\bf 7_2}$ with the prime number $23$.

Let us consider the $\phi$ meson which is a meson of the form $\overline{s} s$ where $s$ denotes a strange quark. By the reason for the $K^0$ and $K^+$ mesons (and since $\phi$ is a vector) we consider the following model for $\phi$:
\begin{equation}
R_s^{n_s}W(K_{\phi})\frac{i}{\sqrt2}[Z+\overline{Z}]
\label{phi1}
\end{equation}
where $Z=(z_1,z_1,z_2,z_2)^T$ with $z_1$ for $\overline{s}$ and $z_2$ for strange quark $s$; and $W(K_{\phi})$ denotes the generalized Wilson loop for $\phi$ where $K_{\phi}$ denotes a knot for $\phi$ (We shall choose $K_{\phi}={\bf 7_2}$). In (\ref{phi1}) we have only one side of $Z$ and we replace $Z$ with the vector
$\frac{i}2[Z+\overline{Z}]$. 
This represents that $\phi$ is a vector.

Then since $\phi$ is a particle identified with its anti-particle we  model $\phi$ with the following observable:
\begin{equation}
\frac12 R_s^{n_s}[W(K_{\phi})-W(\overline K_{\phi})]
\frac{i}{\sqrt2}[Z+\overline{Z}]
\label{phi1a}
\end{equation}
where $\overline K_{\phi}$ denotes the mirror image of $K_{\phi}$. 

This model of $\phi$ gives that the parity, charge conjugate $C$ and $G$-parity of $\phi$ are equal to $-1,-1$ and $-1$ respectively.

Then the mass observable for $\phi$ is given by:
\begin{equation}
\frac12[Z+\overline{Z}]^*R_s^{n_s}[W({\bf 7_2})-
W(\overline{\bf 7_2})]\frac12[Z+\overline{Z}]
\label{phi2}
\end{equation}

Then we need to specify the mass operator of $\phi$ for $W_{\phi}(K_{\phi})$. We let the mass operator of $\phi$ be the operator  
(\ref{casimor}) with $c^2=\frac{14}{15}$ as that for the $K^0$ and $K^+$ mesons.

Let us then consider the $R_s^{n_s}$ matrix which gives the strange degree of freedom to the $s$ and $\overline{s}$ quarks of $\phi$. Similar to the $K^0$ and $K^+$ mesons we let $R_s^{n_s}$ be given by:
\begin{equation}
R_s^{n_s}A_{\phi}:= v_3b_3 + R_s^{n_s}[v_1b_1+v_2b_2]
\label{phi3}
\end{equation}
where $A_{\phi}=v_1b_1+v_2b_2 +v_3b_3$ and $R_s^{n_s}$ does not 
on $v_3b_3$ where $v_1$ and $v_2$ are for $\overline{s}$ and $s$ quarks respectively (We require that $R_s^{n_s}$ does not on $v_3b_3$ because if $R_s^{n_s}$ also acts on $v_3b_3$ then $R_s$ is the same $R$-matrix as that for the $u$ and $d$ quarks and thus gives no strange degree of freedom). We have 
\begin{equation}
R_s^{n_s}[v_1b_1+v_2b_2]=e^{i\alpha m}[v_1b_1+v_2b_2]
\label{phi4}
\end{equation}
where $\alpha=32$. 

Then similar to the computation of the masses of the $K^0$ and $K^+$ mesons we have that the mass of $\phi$ is given by:
\begin{equation}
\frac12[(23\cdot 43 + 2\cdot 16)+ (23\cdot 43 + 2\cdot 16)]
Mev =[23\cdot 43 + 2\cdot 16] Mev=1021 Mev
\label{phi5}
\end{equation}
where the prime number $23$ is from the prime knot ${\bf 7_2}$ and $2\cdot 16 Mev=2\cdot\alpha m$ is from the strange and anti-strange degree of freedom; and the quantum number $43$ is from the mass operator (\ref{casimor}) with $c^2=\frac{14}{15}$. This agrees with the experimental mass $1020 Mev$ of $\phi$.

Let us then consider the scalar mesons $a_0(980)$ and
$f_0(980)$. We suppose that these two mesons are with similar structure as the $\phi$ meson but they are without the strange and anti-strange degrees of freedom. Thus let us model $a_0(980)$ by the following observable:
\begin{equation}
\frac12[Z-\overline{Z}]^*[W({\bf 7_2})+W(\overline{\bf 7_2})]\frac12[Z-\overline{Z}]
\label{phi6}
\end{equation}
where $W({\bf 7_2})$ is the same generalized Wilson loop as that for $\phi$ and is with quantum number $43$. As we have seen in the construction of the $K^0$ and $K^+$ mesons that for obtaining this number $43$ which is less than $45$ we have the number $c^2<1$ which gives an effect of the strange and anti-strange degrees of freedom that this effect is not the strange and anti-strange degrees of freedom which is from the $R_s$ matrix. Let us call the strange and anti-strange degrees of freedom which is from the $R_s$ matrix as the pure strange and anti-strange degrees of freedom. With the difference of pure and nonpure strange degree of freedom we can then describe the diffeence between $\phi$ meson and mesons such as $a_0(980)$ which has the effect of $s\overline{s}$ but is not a pure $s\overline{s}$ such as the $\phi$ meson. Thus the pure strange and anti-strange degree of freedom is for characterizing the $K^0$ and $K^+$ and $\phi$ mesons while the effect of strange and anti-strange degrees of freedom which is not from the $R_s$ matrix is for mesons such as $a_0(980)$ and $f_0(980)$.
 
We have that (\ref{phi6}) is of the form 
$[Z-\overline{Z}]^*[Z-\overline{Z}]$ which gives the scalar property. Then 
we have that  the $P$, $C$ and $G$ of $a_0(980)$ are equal to $+1, +1$ and $+1$ respectively. 

Then we model $f_0(980)$ by the following observable:
\begin{equation}
\frac12 [Z+\overline{Z}]^*[W({\bf 7_2})+W(\overline{\bf 7_2})]\frac12[Z+\overline{Z}]
\label{phi7}
\end{equation}
Then we have that $f_0(980)$ and $a_0(980)$ are with the same mass which is given by:
\begin{equation}
\frac12[23\cdot 43 + 23\cdot 43 ]
Mev =989 Mev
\label{phi8}
\end{equation}
where the prime number $23$ is from the prime knot ${\bf 7_2}$ ; and the quantum number $43$ is  from the mass operator (\ref{casimor}) with $c^2=\frac{14}{15}$.
This approximates quite well the experimental mass $980 Mev$ of $f_0(980)$ and $a_0(980)$.

Similar to the $a_0(980)$ meson we have that the $PC$ of $f_0(980)$ is $++$ and the $G$ of $f_0(980)$ is $+1$.

Let us then consider the $\eta'$ meson. Let us again suppose that $\eta'$ is with similar structure as the $\phi$ meson but is without the pure strange and anti-strange degrees of freedom from the $R_s$ matrix. Thus the $\eta'$ meson is with the same Wilson loop $W({\bf 7_2})$ as the $\phi$ meson (but may not with the same mass operator).
Then assuming that $\eta'$ is identified with its anti-particle we model $\eta'$ with the following observable:
\begin{equation}
(Z-\overline{Z})^*\frac12[W({\bf 7_2})- W(\overline{\bf 7_2})]\frac12(Z-\overline{Z})
\label{phi9}
\end{equation}
where since the $\eta'$ meson is similar to  the $\eta$ meson we have that $W({\bf 7_2})$ is with the mass operator (\ref{integer}) and with $c^2=\frac{12}{15}$ which gives quantum number $42$ as that for the $\eta$ meson. Then we have that the mass of $\eta'$ is given by:
\begin{equation}
\frac12[23\cdot 42 + 23\cdot 42 ]
Mev =966 Mev
\label{phi10}
\end{equation}
where the prime number $23$ is from the prime knot ${\bf 7_2}$.
This approximates quite well the experimental mass $958 Mev$ of $\eta'$.

We have that the $PC$ of $\eta^{'}$ is $-+$ and the $G$ of $\eta'$ is $+1$.

We remark that we describe $\eta'$ by the nonpure effect of strange and anti-strange degree of freedom which is not from the $R_s$ matrix. This agrees with the usual description of $\eta'$ that $\eta'$ has the effect of
strange and anti-strange degree of freedom  but it is not of the pure $s\overline{s}$ form.


Let us then consider the $K^*$ mesons. We model the $K^{*0}$ meson by the following observable:
\begin{equation}
R_sW({\bf 7_1})]\frac{i}{\sqrt2}[Z+\overline{Z}]
\label{phi13}
\end{equation}
where the prime knot ${\bf 7_1}$ is assigned with the prime number $19$. The mass observable of $K^{*0}$ is given by:
\begin{equation}
[Z+\overline{Z}]^*R_sW({\bf 7_1})\frac12[Z+\overline{Z}]
\label{phi14}
\end{equation}
Let the generalized Wilson loop $W({\bf 7_1})$ be with  quantum number $46$ where the strange degree of freedom of $W({\bf 7_1})$ gives $46$ 
which is different from $45$. Then $46$ is also different from $43$ which also gives strange degree of freedom of $W({\bf 6_1})$ for the $K^0$ meson. This difference between $43$ and $46$ gives a distinction between the $K^0$ and the $K^{*0}$ mesons. Then similar to the $K^0$, $K^+$ and the $\phi$ mesons we have that the mass of $K^{*0}$ is given by:
\begin{equation}
[19\cdot 46 + 16 ]
Mev =890 Mev
\label{phi14a}
\end{equation}
where $16 Mev$ is from the pure strange degree of freedom of the $R_s$ matrix and is the value for the strange degree of freedom of the $K^+$ and the $\phi$ mesons.
This approximates well the experimental mass $892 Mev$ of $K^{*0}$.

Similar to the relation of the $K^0$ and  $K^+$ mesons from the above model of the $K^{*0}$ meson we can model the $K^{*+}$ meson. For simplicity we omit the details.

Then also similar to relation of the $K^0$ and  $K^+$ mesons with their anti-particles  we model the anti-particles $\overline{K^{*0}}$ and $K^{*-}$ of the $K^{*0}$ and $K^{*+}$ mesons by the mirror image $\overline{\bf 7_1}$ of the prime knot ${\bf 7_1}$ which is not an amphichiral knot. For simplicity we omit the details.

Let us then consider the $\omega$ and $\rho$ mesons.
 We model $\omega$ with the following observable:
\begin{equation}
\frac12[W({\bf 6_3})-\overline{W({\bf 6_3})}]\frac{i}{\sqrt2}[Z+\overline{Z}]
\label{phi15}
\end{equation}
where ${\bf 6_3}$ is an amphichiral knot which is equivalent to its mirror image.
Then the mass observable of $\omega$ is given by:
\begin{equation}
[Z+\overline{Z}]^*
\frac12[W({\bf 6_3})-\overline{W({\bf 6_3})}]\frac12[Z+\overline{Z}]
\label{phi16}
\end{equation}

Similar to the $\phi$ meson we have that the $PC$ of $\omega$ is $--$. Then the $G$ of $\omega$ is $-1$.

Then we model $\rho$ by the following observable:
\begin{equation}
\frac12[W({\bf 6_3})-\overline{W({\bf 6_3})}]\frac{1}{\sqrt2}[Z-\overline{Z}]
\label{phi17}
\end{equation}
and the mass observable of $\rho$ is given by:
\begin{equation}
\frac12[Z-\overline{Z}]^*[W({\bf 6_3})-\overline{W({\bf 6_3})}]\frac12[Z-\overline{Z}]
\label{phi18}
\end{equation}

Let us then compute the masses of $\rho$ and $\omega$. For the $\rho$ meson we let the Wilson loop $W({\bf 6_3})$ be with winding number proportional to $45$ as that for the $\pi$ mesons. Then we have that the mass of $\rho$ is given by:
\begin{equation}
 17\cdot 45 
Mev =765 Mev
\label{phi19}
\end{equation}
where the prime number $17$ is for the prime knot ${\bf 6_3}$. This approximates quite well the experimental mass $770 Mev$ of $\rho$.
Similar to the $\phi$ meson we have that the $PC$ of $\rho$ is $--$. Then the $G$ of $\rho$ is $+1$.

Let us then consider mass of the $\omega$ meson. For this meson we let the generalized Wilson loop $W({\bf 6_3})$ be with the quantum number $46$. As the $K^{*0}$ meson this number $46$ gives the strange degree of freedom which is not the pure strange degree of freedom from the $R_s$ matrix. Then we have that the mass of $\omega$ is given by:
\begin{equation}
 17\cdot 46 
Mev =782 Mev
\label{phi19a}
\end{equation}
This approximates quite well the experimental mass $783 Mev$ of $\omega$.

We remark again that with the nonpure strange and anti-strange degree of freedom we can describe that $\omega$ has the effect of strange and anti-strange degree of freedom while it is not of the pure $s\overline{s}$ form.

In summary we have the 
classification table of pseudoscalar and vector mesons in the section of the introduction of this paper.
In this classification table  the prime knot ${\bf 3_1}$ is assigned with the number $1$ while it is similar to the prime number $2$ \cite{Ng}. 
From this classification table it is interesting to note that since the prime number $2$ is not used to represent knots we have that the prime number $2$  is not used to model mesons and baryons (Thus the prime knot ${\bf 3_1}$ which is not assigned with a prime number though it is similar to the prime number $2$ is not used to model mesons and baryons). This agrees with the fact that the prime number $2$ does not appear in the above mass formula for the computation of the masses of mesons.

\section{Generations of Quarks as Knot Phenomena of Mesons}\label{sec16ab}

Since knots are composed with prime knots we have that the form of prime knots is more stable than the form of nonprime knots when knots are used to model elementary particles. Then prime knots are assigned with prime numbers (except the prime knot ${\bf 3_1}$ which is similar to the prime number $2$). Thus the prime numbers appearing in the above mass formula is an evidence for the knot model of elementary particles.

We have that the $\pi^0$ meson is modeled by the prime knot ${\bf 4_1}$ which is an amphichiral knot and is assigned with the prime number $3$. Then ${\bf 4_1}$ is an amphichiral knot with the property that it is equivalent to its mirror image. This property is exactly fit with the property of $\pi^0$ that $\pi^0$ is identified with its anti-particle (We have that $\pi^+$ and $\pi^-$ are also modeled by ${\bf 4_1}$ where we introduce the $U(1)$ gauge group to give positive and negative charge to $\pi^+$ and $\pi^-$ respectively such that $\pi^+$ is the anti-particle of $\pi^-$).

Then the next amphichiral knot is the knot ${\bf 6_3}$ which is assigned with the prime number $17$. This knot then must be used to model a meson similar to the $\pi^0$ meson with the property that it is identified with its anti-particle. Thus this meson must be the $\rho(770)$ meson which is a meson identified with its anti-particle. Further since the mass and the property of the $ \omega(783)$ is similar to that of the $\rho(770)$ meson we have that this $ \omega(783)$ meson must also be modeled by the prime knot ${\bf 6_3}$. Thus the appearing of the prime number $17$ in the mass formula of $\rho(770)$ and $ \omega(783)$ (where the number $46$ is as a deviation from the number $45$ showing the structure of $SU(2)$ for the knot modeling) is an evidence for the knot model of elementary particles.

Furthermore the quark model (which is a part of the knot model) of the $\pi^0$ meson  is of a $u\overline{u}$ or a $d\overline{d}$ form. Thus there are two degrees of freedom for forming the quark model of the $\pi^0$ meson. Thus to represent these two degree of freedom if the $\pi^0$ meson is identified as a meson modeled by an amphichiral knot and of the $u\overline{u}$ form then there must be a meson similar to the $\pi^0$ meson such that it is modeled by an amphichiral knot and is of the $d\overline{d}$ form. Then this meson is exactly the $\rho(770)$ meson which is modeled by the amphichiral knot ${\bf 6_3}$ and can be regarded as of the $d\overline{d}$ form. Thus the family ${\bf 4_{(\cdot)}}$ of prime knots with four crossings (which consists of only the prime knot ${\bf 4_1}$) and the family ${\bf 6_{(\cdot)}}$ of prime knots with six crossings (which consists of prime knots ${\bf 6_1}, {\bf 6_2}, {\bf 6_3}$) together form a region which gives the modeling of basic mesons consists of the up and down quarks where these two families of prime knots; the ${\bf 4_{(\cdot)}}$ and ${\bf 6_{(\cdot)}}$; must be with the property that each of them contains an amphichiral knot.

It then also follows that the family ${\bf 5_{(\cdot)}}$ consisting of prime knots ${\bf 5_1}, {\bf 5_2}$ with five crossings should not be used to model mesons since this family does not contain an amphichiral knot. Now we have that the prime knots ${\bf 5_1}$ and $ {\bf 5_2}$ are assigned with the prime numbers $5$ and $7$ respectively and they are indeed not used to model mesons; as shown in the above computation of the masses of mesons. This agreement is an evidence that mesons are modeled as knots.

Now for the family ${\bf 6_{(\cdot)}}$ we still have two knots: ${\bf 6_1}$ and $ {\bf 6_2}$ which should be used to model mesons in the two multiplets of pseudoscalar and vector mesons. This is indeed the case that the $K^0=K(498)$ meson and the $ \eta(549)$ meson are modeled by ${\bf 6_1}$ and $ {\bf 6_2}$ respectively; as shown by the above computation of masses of these two mesons where the $K^0=K(498)$ meson and the $ \eta(549)$ meson are assigned with the prime numbers $11$ and $13$ respectively. This agreement is an evidence that mesons are modeled as knots.

Beyond the region of ${\bf 4_{(\cdot)}}$ and ${\bf 6_{(\cdot)}}$  we come into a new region ${\bf 7_{(\cdot)}}$ of prime knots with seven crossings. This new region contains no amphichiral knots (It is a fact in knot theory that the family of prime knots of odd number of crossings contains no amphichiral knots while the family of prime knots of even number of crossings contains amphichiral knots \cite{Ada}-\cite{Rol}).
We may then expect that this new region can give additional properties of elementary particles in the knot modeling of elementary particles. Since the previous region gives the modeling of mesons of the $u\overline{u}$ or $d\overline{d}$ form
we expect that this region gives the modeling of mesons of the $s\overline{s}$ form. 
Thus we expect that the knots ${\bf 7_1}$ and ${\bf 7_2}$ can be used to model $K$ mesons; or mesons of the $s\overline{s}$ form; or mesons related to the $s\overline{s}$ form. 
This is indeed the case that the $K^{*}(892)$ meson is modeled by the prime knot ${\bf 7_1}$ assigned with the prime number $19$ and the $\phi(1020)$ meson (which is a $s\overline{s}$-meson ) is modeled by the prime knot ${\bf 7_2}$ assigned with the prime number $23$ , as showned in the above computation of masses of the $K^{*}(892)$ mesons and the $\phi(1020)$ meson. This agreement is an evidence that mesons are modeled as knots.

We remark that the prime knot ${\bf 7_2}$ is not an amphichiral knot but the average 
$\frac12[W({\bf 7_2})-W(\overline{\bf 7_2})]$ can be used to model the $\phi(1020)$ meson which is as a meson identified with its anti-particles and is of the $s\overline{s}$ quark form. This means that mesons of the $s\overline{s}$ form  are not modeled by amphichiral knots but are modeled by the average of  non-amphichiral knots and their mirror images. This is distinct from  mesons of the $u\overline{u}$ or $d\overline{d}$ form (We shall see in the following that mesons of the $c\overline{c}$ form where $c$ denotes a charm quark can be modeled by amphichiral knots; as similar to the $\pi^0$ and $\rho(770)$ mesons of the $u\overline{u}$ or $d\overline{d}$ form modeled by amphichiral knots).

Then we consider the remaining meson $\eta^{\prime}(958)$ of the two nonets of pseudoscalar and vector mesons. This meson is related to meson of the form $s\overline{s}$ and thus it should also be modeled by prime knots in the family ${\bf 7_{(\cdot)}}$. This is indeed the case that $\eta^{\prime}(958)$ is modeled by the prime knot
${\bf 7_2}$ assigned with the prime number $23$; as shown in the above computation of the mass of $\eta^{\prime}(958)$. This agreement is an evidence that mesons are modeled as knots.

Then beyond the region of ${\bf 7_{(\cdot)}}$  we come into a new region ${\bf 8_{(\cdot)}}$ of prime knots with eight crossings. This new region contains many amphichiral knots \cite{Ada}-\cite{Rol}.
We may then expect that this new region can give additional properties of elementary particles in the knot modeling of elementary particles. Since this region is similar to the previous region of ${\bf 4_{(\cdot)}}$ and ${\bf 6_{(\cdot)}}$ giving the modeling of basic mesons of the $u\overline{u}$ or $d\overline{d}$ form
we have  that this region should be used to model mesons of the $c\overline{c}$ form.
This is indeed the case. As examples let us consider the $c\overline{c}$-mesons in the following table \cite{PDA2}:
\begin{displaymath}
\begin{array}{|c|c|c|c|} \hline
\mbox{$c\overline{c}$-meson} & \mbox{computed mass} &\mbox{$c\overline{c}$-meson} & \mbox{computed mass}\\ \hline
 \eta_c(1S)(2979)&{\bf 59}\times 50+24=2974 &
\eta_c(2S)(3594)&{\bf 79}\times 45+36=3591\\ \hline

 J/\psi(1S)(3096)&{\bf 61}\times 50+48=3098 &
 \psi(2S)(3686)&{\bf 73}\times 50+36=3686\\ \hline

 \chi_{c0}(1P)((3415)&{\bf 71}\times 48=3408 &
\psi(3770)&{\bf 83}\times 45+36=3771 \\ \hline

 \chi_{c1}(1P)((3510)&{\bf 73}\times 48=3504 &
\psi(3836)&{\bf 83}\times 45+2\cdot48= 3831\\ \hline

 h_c(1P)(3526)&{\bf 73}\times 48+24=3528 &
 \psi(4040)&{\bf 89}\times 45+36=4041 \\ \hline

\chi_{c2}(1P)((3556)&{\bf 79}\times 45=3555 &
 \psi(4160)&{\bf 83}\times 50+2\cdot36=4162\\ \hline

 & &\psi(4415)&{\bf 97}\times 45+ 48=4413 \\ \hline
\end{array}
\end{displaymath}
where $12, 24, 36$ and $48$ are for the $c\overline{c}$ degree of freedom (and the $24, 36, 48$ are from excited states of the state of $12$), as similar to that for the $s\overline{s}$ degree of freedom in the above computation for mesons with strange quarks. 
From this table we see that the computed masses approximate the experimentl masses. We notice that the computed masses of $\chi$ mesons do not have the term of multiple of $12, 24, 36$ and $48$. This agrees with the property of $\chi$  that they are not a complete $c\overline{c}$ meson. Also we notice that there are many $c\overline{c}$-mesons in this mass region. 

We notice that the mass region of the $c\overline{c}$-mesons corresponds to the region ${\bf 8_{(\cdot)}}$ of prime knots with eight crossings; as shown in the  following table of classification of prime knots in the region ${\bf 8_{(\cdot)}}$ by prime numbers:
\begin{displaymath}
\begin{array}{|c|c|c|c|c|c|c|c|c|c|} \hline
\mbox{prime knot} & {\bf 8_1}&{\bf 8_2} &{\bf 8_3}&{\bf 8_4} &{\bf 8_5} & {\bf 8_6}&{\bf 8_7}&{\bf 8_8}&{\bf 8_9}\\ \hline
\mbox{prime number}&47  &53 &59 &61 & 67&71 &73 &79 &83 \\ \hline
\mbox{prime knot} &{\bf 8_{10}}&{\bf 8_{11}}
&{\bf 8_{12}}&{\bf 8_{13}}&{\bf 8_{14}}&{\bf 8_{15}}&{\bf 8_{16}} &{\bf 8_{17}} &{\bf 8_{18}} \\ \hline
 \mbox{prime number} &89&97&101 &103&107&109&113 &127 &131 \\ \hline
\end{array}
\end{displaymath}
where the prime knots ${\bf 8_3},{\bf 8_9},{\bf 8_{12}}, {\bf 8_{17}}, {\bf 8_{18}}$ are amphichiral knots which correspond to $c\overline{c}$-mesons by the assigned prime numbers $59, 83,101,127, 131$.
 
Then beyond the region ${\bf 8_{(\cdot)}}$ of  prime knots  we come into the region ${\bf 9_{(\cdot)}}$ of  prime knots with nine crossings. This new region contains no amphichiral knots.
We may then expect that this new region can give additional properties of elementary particles in the knot modeling of elementary particles.  Since this region is similar to the previous region  ${\bf 7_{(\cdot)}}$ giving the modeling of basic mesons of the $s\overline{s}$ form
we have  that this region should be used to giving modeling of mesons of the $b\overline{b}$ form where $b$ denotes the bottom quark. 

Then beyond the region ${\bf 9_{(\cdot)}}$ of  prime knots  we come into the region ${\bf 10_{(\cdot)}}$ of  prime knots with ten crossings. This new region contains many amphichiral knots.
We may then expect that this new region can give additional properties of elementary particles in the knot modeling of elementary particles.  Since this region is similar to the previous two regions  
${\bf 4_{(\cdot)}}$, ${\bf 6_{(\cdot)}}$ and ${\bf 8_{(\cdot)}}$ giving the modeling of $u\overline{u}$ and $c\overline{c}$ mesons  we have  that this region should be used to giving the modeling of $t\overline{t}$-mesons  where $t$ denotes the top quark.

Continuing in this way
this shows that the phenomeon of the generations of quarks is derived from the properties of knots. This is an evidence that elementary particles are modeled as knots.

\section{Knot Modeling of Weak Interaction and $CP$ Violation}

Let us first consider the well known $\tau-\theta$ puzzle:
\begin{equation}
\tau^+ \to \pi^+ \pi^- \pi^+, \quad  \pi^0 \pi^0 \pi^+; \qquad \theta^+ \to \pi^+ \pi^0,
\label{cp1}
\end{equation}
where the particles $\tau,\theta$ are identified as the same meson $K^+$ for solving this puzzle. The solving of this puzzle gives the well known parity $P$ violation of weak interaction \cite{Lee}\cite{Wu}.

Let us here give the knot modeling of this $P$ violation of weak interaction and the $CP$ violation of the following weak interaction \cite{Chr}: 
\begin{equation}
K_L^0 \to \pi^+ \pi^- \pi^0, \quad \pi^0 \pi^0 \pi^0; \qquad K_L^0 \to \pi^+ \pi^-, \pi^0 \pi^0
\label{cp2}
\end{equation}

We shall show that since  the figure-eight knot ${\bf 4_1}$ is an amphichiral knot that it equals to its mirror image and that $\pi$ mesons are modeled by ${\bf 4_1}$ we have the $P$ violation and $CP$ violation of weak interaction. In this knot modeling of weak interaction we shall also show that in the weak interaction (\ref{cp1}) the $K^+$ meson for which the $\tau$ and $\theta$ are identified to should be the meson
$K_L^+ :=\frac12(K^+ -K^-)$.

Lut us consider the knot model (\ref{P1c}) of $\pi^0$ meson. Since the figure-eight knot ${\bf 4_1}$ is an amphichiral knot that ${\bf 4_1}=\overline{\bf 4_1}$, in defining the parity of $\pi^0$ we need to use 
$\overline{W({\bf 4_1})}$ instead of $W(\overline{\bf 4_1})$ where $\overline{\bf 4_1}$ denotes the mirror image of ${\bf 4_1}$. Thus in defining the knot model (\ref{P1c}) of $\pi^0$ we have to choose one of the two states: $\pm\frac12[W({\bf 4_1})- \overline{W({\bf 4_1})}]$. This chosen of states is a spontaneous symmetry breaking of the left-right symmetry. Thus we have that the $\pi^0$ meson modeled by the amphichiral knot ${\bf 4_1}$ spontaneously breaks the parity $P$ symmetry. Then since the weak interaction is a slow interaction with respect to the strong interaction we have that it may have enough time for the appearing of the phenomenon of this parity $P$ violation of the $\pi^0$ meson to be observed in a weak decay. Thus the phenomenon of parity $P$ violation may be observed in a waek interaction.

On the other hand it is hard to observe this parity $P$ violation of the $\pi$ mesons in a strong decay because of the very rapid interaction of a strong decay where before the possible appearing of the phenomenon of the parity $P$ violation of the $\pi$ mesons the strong decay has already completed. Thus the phenomenon of parity $P$ violation is hard to be observed in a strong interaction.

As an example let us consider again the weak interaction: $\pi^+ \to \mu^+  +\nu_{\mu}$. In this weak decay the components $z_1$ and $z_2$ (representing the up and down quarks) of the complex vector $Z$ of the model of $\pi^+$ are separated at two different proper times while $Z$ is still acted on by the chosen quantum state $\frac12[W({\bf 4_1})- \overline{W({\bf 4_1})}]$. Then in this weak decay the components $z_1$ and $z_2$ has become to represent the $\mu^+$ and $\nu_{\mu}$ respectively. Thus we have that the left-hand spinning of the chosen quantum state $\frac12[W({\bf 4_1})- \overline{W({\bf 4_1})}]$ has become the left-hand spinning of the $\mu^+$ and $\nu_{\mu}$ respectively. Thus in this weak decay the neutrino $\nu_{\mu}$ always has the left-hand spinning and one cannot find a neutrino $\nu_{\mu}$ with the right-hand spinning. This is thus a phenomenon of the parity $P$ violation of the $\pi^+$ meson where the left-hand spinning of $\nu_{\mu}$ is from the spontaneous symmetry breaking of the parity $P$ symmetry of the chosen quantum state $\frac12[W({\bf 4_1})- \overline{W({\bf 4_1})}]$ of $\pi^+$.

This phenomenon of the parity $P$ violation of the $\pi^+$ meson may be reinterpreted that in the nature we have only the $\pi^+$ meson with the  chosen quantum state $\frac12[W({\bf 4_1})- \overline{W({\bf 4_1})}]$ and there are no 
$\pi^+$ meson with another quantum state $-\frac12[W({\bf 4_1})- \overline{W({\bf 4_1})}]$.

Let us then consider the derived  $C$ violation of the $\pi$ mesons of this example. By a charge conjugation $C$ operation the following model of the $\pi^+$ meson:
\begin{equation}
Z_1^*  R_{U(1)}^{n_1}\frac12[W({\bf 4_1})- \overline{W({\bf 4_1})}]Z  
\label{P1c31}
\end{equation}
is changed to the following state:
\begin{equation}
Z^*  R_{U(1)}^{-n_1}\frac12[W({\bf 4_1})-\overline{W({\bf 4_1})}]Z_1  
\label{P1c41}
\end{equation}
However in the nature this state does not represent the $\pi^-$ meson and does not exist. This is thus a charge conjugation $C$ violation.

On the other hand  if we apply the $CP$ transformation then we have that the model of the $\pi^+$ meson (\ref{P1c31}) is transformed to the model of the $\pi^-$ meson; as shown in the above section on parity. Thus the $CP$ symmetry holds in this example of weak interaction. This is a well known fact in particle physics \cite{Chr}-\cite{Wol}. Here we use the knot modeling of weak interaction to derive this $CP$ symmetry.

As another example of weak interaction let us then consider the knot modeling of the $CP$ violation of the weak interaction (\ref{cp2}) \cite{Chr}-\cite{Wol}.
With the knot modeling we have that the $CP$ of $K_L^0$ is $-1$; the $CP$ of $\pi^+ \pi^- \pi^0$ is $-1$; and the $CP$ of $\pi^+ \pi^- $ is $1$. Thus this knot  modeling gives the $CP$ violation of the weak interaction (\ref{cp2}). This agrees with the existing method  of determining the $CP$ violation of the weak interaction (\ref{P1c}) \cite{Chr}-\cite{Wol}.

We want to show that this  $CP$ violation of the weak interaction (\ref{cp2}) is also due to that the $\pi$ mesons are modeled by the prime knot ${\bf 4_1}$ which is with the special property that it is an amphichiral knot. As analyzed
in the above we have that whenever a $\pi^0$ meson is produced in an interaction the left-right symmetry is spontaneously broken by the chosen of one of the quantum states: $\pm \frac12[W({\bf 4_1})- \overline{W({\bf 4_1})}]$. Further this phenomenon of spontaneous symmetry breaking may be observed in a weak interaction which is a slow interaction such that it may have enough time for the appearing of this phenomenon of spontaneous symmetry breaking. Now we have that the $\pi^0$ meson is produced in the weak interaction (\ref{cp2}). Thus the $CP$ violation of this weak interaction (\ref{cp2}) is  the phenomenon of spontaneous symmetry breaking of the $\pi^0$ meson given by the chosen of one of the quantum states: $\pm \frac12[W({\bf 4_1})- \overline{W({\bf 4_1})}]$ of the $\pi^0$ meson. This shows that the $CP$ violation of the weak interaction (\ref{cp2}) is due to that the $\pi^0$ meson is modeled by the prime knot ${\bf 4_1}$ which is with the special property that it is an amphichiral knot.

We remark that  in the weak interaction (\ref{cp2}) we identify the $K_L^0$ meson as the following meson
 \begin{equation}
K_L^0:=\frac12(K^0- \overline{K^0})
 \label{cp3}
\end{equation}
which in the knot modeling is with $P=-1$, $C=1$ and $CP=-1$. This is more appropriate as a meson with longer life than the meson $\frac{1}{\sqrt{2}}(K^0+\overline{K^0})$ which is regarded as the $K_L^0$ meson in the literature 
\cite{Chr}-\cite{Wol} (In the literature the meson $\frac1{\sqrt{2}}(K^0+\overline{K^0})$ is also defined to with $CP=-1$).
This is because that the $\overline{K^0}$ meson is the anti-particle of $K^0$ and if these two particles combineed in  the symmetric form $ \frac1{\sqrt{2}}(K^0+\overline{K^0})$ then these two particles would quickly annihilate each other and thus the meson 
$\frac1{\sqrt{2}}(K^0+\overline{K^0})$ would be with shorter life.

As another example of weak interaction let us then consider the knot modeling of the $P$ violation of the $\tau-\theta$ puzzle (\ref{cp1}) \cite{Lee}-\cite{Wol}. Let us consider the following meson:
\begin{equation}
K_L^+:=\frac12(K^+ - \overline{K^+})  
\label{cp4}
\end{equation}
In this knot modeling this meson is with $P=-1$ while the $K^+$ is not an eigenstate of $P$. 
We see that this $K_L^+$ meson and the weak interaction (\ref{cp1}) is completely analogous to the above example of the 
$K_L^0$ meson and the weak interaction (\ref{cp2}).
Thus in this knot modeling it is more appropriate to regard $K_L^+$ rather than $K^+$ as the $\tau-\theta$ particle (In the literature the meson $K^+$ is defined to with $P=-1$ \cite{Lee}-\cite{Wol}). 

Again as analyzed
in the above we have that whenever a $\pi^+$ meson is produced in an interaction the left-right symmetry is spontaneously broken by the chosen of one of the quantum states: $\pm \frac12[W({\bf 4_1})- \overline{W({\bf 4_1})}]$. Further this phenomenon of spontaneous symmetry breaking may be observed in a weak interaction which is a slow interaction such that it may have enough time for the appearing of this phenomenon of spontaneous symmetry breaking. Now we have that the $\pi^+$ meson is produced in the weak interaction (\ref{cp1}). Thus the $P$ violation of this weak interaction (\ref{cp1}) is  the phenomenon of spontaneous symmetry breaking of the $\pi^+$ meson given by the chosen of one of the quantum states: $\pm \frac12[W({\bf 4_1})- \overline{W({\bf 4_1})}]$ of the $\pi^+$ meson. This shows that the $P$ violation of the weak interaction (\ref{cp1}) is due to that the $\pi^+$ meson is modeled by the prime knot ${\bf 4_1}$ which is with the special property that it is an amphichiral knot.

In summary from this knot modeling of weak interaction we conclude that the $P$ violation, the $C$ violation 
and the $CP$ violation in weak interactions are all due to the appearing of a meson (such as the $\pi$ mesons) which is modeled by an amphichiral knot (where the $\pi$ mesons are modeled by the prime knot ${\bf 4_1}$) which is with the special property that it is equal to its mirror image.

\section{Unification of Strong and Weak Interactions}

Let us investigate in more detail why the above knot modeling of the interaction (\ref{sw4}) starting from the gauge fixing $A_3=\frac13 A_0$ is the modeling of the weak decay of the $\pi^+$ meson. Let us use this knot modeling to give a rough estimate of the strength of this interaction (\ref{sw4}) to show that the strength of this interaction is consistent with the strength of the weak interaction.
First we have that the strength of QED is the fine structure constant $\alpha =\frac1{137}$ and we have $\alpha \sim e^2$. Let us suppose that the charge for the electromagnetic interaction, the weak interaction and the strong interaction are of the same charge $e$ (This agrees with the idea of the grand unified field theory in particle physics that these three interactions should finally be with the same charge). Then according to the Yukawa theory we have that the residual force of strong interaction between two particles such as proton and neutron is given by the exchange of the $\pi$ mesons. From the Yukawa theory with the electric charge $e$ for strong interaction we have that the strength of this residual force of strong interaction is roughly proportional to:
\begin{equation}
 G_s:=\frac{e^2}{\frac{M^2}{m^2}}
 \label{strength1}
 \end {equation}
where $M$ is the mass of the $\pi^0$ meson and $m $ is the mass of the electron (When $M=m$ we have the case of the electromagnetic interaction and we have $G_s=e^2$ which is just the strength of the electromagnetic interaction). From this we see that the residual force of strong interaction is in fact not strong because $M>100$ and that the charge $e$ is small. This however agrees with the fact that the range of this residual force of strong interaction is very short comparing to the long range of the electromagnetic interaction. Then why the strong interaction is strong? Let us give a more detail description of the knot modeling of the strong interaction in the above sections, as follows.

In the above sections we identify the strong interaction of quarks as a closed loop interaction formed by a knot $K$ of the gauge field and the quarks  interacted by the gauge field  are components of $Z(s)$ acted by the generalized Wilson loop $W(K)$. This form of interaction is different from  the electromagnetic interaction in the form of the exchange of photon which is described by the usual Feynman diagram. This strong interaction is for the formation of one elementary particle such as the $\pi^+$ meson or the proton. Then we need a strong interaction between two elementary particles. The above residual force is an interaction between two elementary particles but it is not strong. Thus there should have another effect of the strong interaction. Let us model this effect of strong interaction by the linking of two (or more) knots where each knot formed by a gauge field gives an elementary particle. This means that we use links to model the strong interactions among the elementary particles where each elementary particle is in a knot form. This model of strong interaction is a nonabelian effect because the effect of linking can only be achieved by nonabelian gauge group such as $SU(2)$ or $SU(3)$ (In contrast to mesons modeled by $SU(2)$ we shall model baryons by $SU(3)$). It follows that electron can not feel this strong effect because electron is only interacted by the abelian gauge group $U(1)$.  

As a simple example of linking let us consider the Hopf link depicted in Fig.5. For the Hopf link we can show that the linking gives a $R$-matrix which is independent of the $R$-matrices of the two knots (for the two elementary particles) \cite{Ng}. This $R$- matrix which is a new degree of freedom gives the strong interaction between the two elementary particles formed by the two linked knots (In the above sections we show that for the $K^0$ and $K^+$ mesons the strange degree of freedom is from a $R$-matrix denoted by $R_s$ which is from a type of linking with linking number $0$. This type of linking is suitable to model the strong interaction of the associated production of strange particles \cite{Pai}\cite{Roc}\cite{Ada}-\cite{Rol}). 

For simplicity let us use diagrams in Fig.4 and Fig.5  to describe the two types of strong interaction between two elementary  particles (For simplicity in these diagrams we use a circle to mean a nontrivial knot such as the knot ${\bf 4_1}$). The diagrams of Fig.4 is similar to the usual Feymann diagram which gives the residual force of strong interaction described by the Yukawa theory with $e$ as the charge. The diagram of Fig.5 is a Hopf link which also gives strong interaction between two elementary  particles. This interaction is strong and of short range which is shorter than the rasidual force because it is a linking effect. 

This interaction  also gives attractive effect between two elementary particles because it is a linking effect and thus it can replace the existing theory of attractive nuclear force which is supposed given by the exchange of the $\sigma$ meson whose existence  has not yet been as clearly confirmed as the other mesons \cite{PDA}-\cite{Ng1}. 

This interaction is also asymptotically free because it is a linking effect that the  knots representing the elematary particles are free to move except that they are linked together. This gives the property of asymptotic freedom of the strong interaction \cite{Gro}\cite{Pol}. 

Then when the up or down quarks represented by complex numbers $z_i$ forming a complex vector $Z$ is attached to a quantum knot $W(K)$ to form a meson (or a baryon) we have that these up or down quarks are in strong interaction in the knot form; as described in the above sections. Then when the $z_i$ are slowly separated (but are still attached to $W(K)$ at different proper times (or at different positions  of $W(K)$) we have that the up or down quarks are in weak interaction; as described in the above sections (In the following of this section we shall  estimate the strength of weak interaction). Then since $W(K)$ is a knot it is closed and thus the real parts of the interaction propagators of the form $\log (u_i-u_j)$ in the formation of $W(K)$ cancel each other, as shown in the sections on the formation of $W(K)$. These real parts of the interaction propagators of the form $\log (u_i-u_j)$ are for the strength of the interaction of the up and down quarks represented by the components of the complex vector $Z$ attached to $W(K)$. Thus in this case the up or down quarks are basically asymptotically free. This thus gives the property of asymptotic freedom of the strong interaction when the up or down quarks are in a knot form of strong interaction. This also shows that the strong and weak interactions are closedly related.

Then when the components $z_i$ are further apart (but are stll attached to $W(K)$) we have that the mass $M$ of the elematary particle is decreased to give energy for this further apart (This decreasing of mass can be obtained by reducing the winding number $n_2$ of $W(K)$). In this case the strength $\frac{e^2}{\frac{M^2}{m^2}}$ between the quarks represented by the components $z_i$ increases (while $z_i$ are stll attached to $W(K)$). This thus gives both the phenomenon of asymptotic freedom and the phenomenon of quark confinement of strong interaction.

Then we notice that this knot modeling of mesons (or baryons) is based on the representation $W(K)Z$. 
Thus to separate the components of $Z$ by strong interaction the quantum knot $W(K)$ must be separated. Then the way to separate $W(K)$ is that the closed loop $W(K)$ becomes several closed loops connected together and these closed loops as quantum knots $W(K_i)$ are then separated to act on quarks to form mesons. Since these resulting quantum knots $W(K_i)$ are matrices that they must act on complex vectors $Z_i$ whose components represent quarks to form  $W(K_i)Z_i$ and that as matrices they cannot act on a single complex number $z$ representing a single quark. Thus the result of strong decay of a meson based on the knot model
$W(K)Z$ is a bunch of mesons formed by $W(K_i)Z_i$ and there are no free quarks separated from the meson. This also gives the phenomenon of quark confinement of strong interaction.
 
This thus shows that this knot modeling of strong interaction gives the basic properties of strong interaction.

\begin{figure}[hbt]
\centering
\includegraphics[scale=0.8]{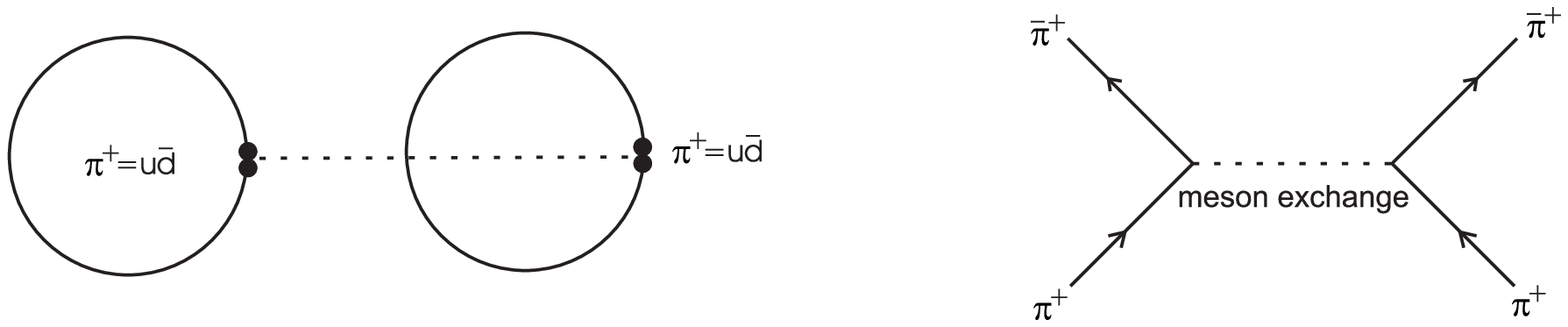}

\begin{minipage}[t]{6cm} 
\centerline{Fig.4a }
Yukawa residual force of strong interaction for $\pi\pi$ scattering ($\pi$ may be replaced by $p$ or $n$). 
\end{minipage}
\begin{minipage}{3cm} 
\hspace*{3cm}
\end{minipage}
\begin{minipage}[t]{5cm}
\begin{center}
Fig.4b  \\
Feynman diagram for Fig.4a 
\end{center}
\end{minipage}
\end{figure}

\begin{figure}[hbt]
\centering
\includegraphics[scale=0.8]{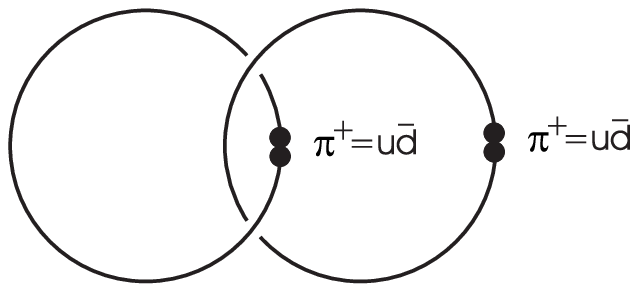} 

\begin{center}Fig.5 \\
Strong interaction in link form ($\pi^+$ may be replaced by $p$ or $n$).
\end{center}
\end{figure}

\begin{figure}[hbt]
\centering
\includegraphics[scale=0.8]{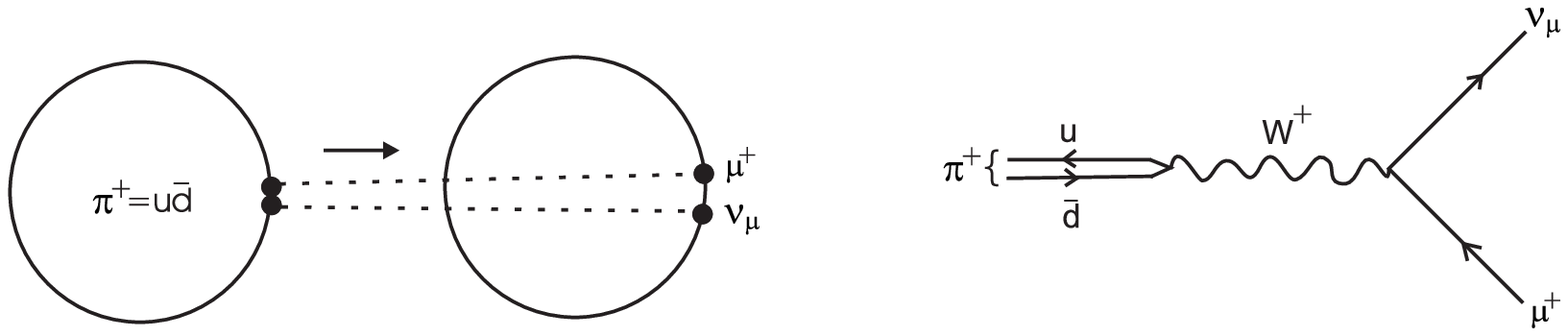}

\indent\begin{minipage}[t]{6cm} 
\begin{center}Fig.6a \\
Weak decay $\pi^+ \to \mu^+ + \nu_{\mu}$ \end{center}
\end{minipage}
\begin{minipage}{2cm} 
\hspace*{2cm}
\end{minipage}
\begin{minipage}[t]{5cm}
\begin{center}Fig.6b \\
Feynman diagram for Fig.6a 
\end{center}
\end{minipage}

\end{figure}

Let us then consider the weak interaction. First from the residual force of strong interaction depicted in Fig.4 we have the strength $\frac{e^2}{\frac{M^2}{m^2}}$. Then when the neutrino (transformed from the $d$ quark) is deviated from $\mu^+$ (trnsformed from the $u$ quark) we have a decay. We want to show that this decay is roughly of the strength of the weak interaction and thus this decay is the decay of the weak interaction. When the $d$ quark is deviated from the $u$ quark  from the residual force of strong interaction we have the strength:
 \begin{equation}
\frac{e^2}{\frac{M^2}{m^2}}\frac{e^2}{\frac{M^2}{m^2}}
 \label{strength2}
\end{equation}
where the two strengths $\frac{e^2}{\frac{M^2}{m^2}}$ are from the residual forces of strong interaction  for the interaction of $u$ and $\mu^+$ and for the interaction of $d$ and $\nu_{\mu}$ respectively (This strength (\ref{strength2})  is in a product form because it is for the amplitude of the appearing of both the interaction of $u$ and $\mu^+$ and the interaction of $d$ and $\nu_{\mu}$). 

Further during this weak decay the $\mu^+$ (which is similar to the electron) comes out  from the $\pi^+$ and the mass of $\pi^+$ is transformed to the mass of $\mu^+$. Thus one of the above two interactions involving the $\mu^+$ is more accurate with the 
strength $\frac{e^2}{\frac{M_{\mu}^2}{m^2}}$ (where $M_{\mu}$ denotes the mass of the $\mu^+$) instead of the strength $\frac{e^2}{\frac{M^2}{m^2}}$.
Thus the strength of this weak  decay is roughly given by:
\begin{equation}
\frac{e^2}{\frac{M^2}{m^2}}\frac{e^2}{\frac{M_{\mu}^2}{m^2}} \sim 
\frac{10^{-2}}{100^2}\frac{10^{-2}}{100^2} Mev^{-2}
=10^{-12} Mev^{-2}
\label{strength3}
\end{equation}
We see that this is roughly the strength of the Fermi coupling constant of weak interaction. This shows that this model of decay of $\pi^+$ is possible to describle the weak decay of $\pi^+$. This also shows that we can deduce the strength of weak interaction from the strength of the usual Yukawa theory of strong interaction and we have that the strong and weak interactions are closedly related.

Now with the Fermi coupling constant of weak interaction as similar to the Yukawa theory of the exchange of $\pi$ mesons we may describe the weak interaction by the exchange of the $W^{\pm}$ and $Z^0$ particles of the standard model of weak and electromagnetic interaction \cite{Wei}\cite{Gla}\cite{Sal}\cite{Mar}.
In Fig.6a and Fig.6b we give the relation of this model of decay of $\pi^+$ with the standard model of weak and electromagnetic interaction. 

\section{Knot Model of Gauge Particles $W^{\pm}$ and $Z^0$ }

During the process of the above weak interaction (\ref{sw4})
 in section \ref{sec100} (depicted in Fig.6) we have that the components $z_i, i=1,2$ representing the up and down quarks (or the $\mu^+$ and $\nu_{\mu}$) are separated but are still attached to the quantum knot for representing the $\pi^+$ meson. At this intermediate state with the up and down quarks (or the $\mu^+$ and $\nu_{\mu}$) being separated the quantum knot and the attached components $z_i$ no longer representing the $\pi^+$ meson but is as a new particle which is identified as the gauge particle $W^{+}$.

Similarly for the weak decay of the $\pi^-$ meson the corresponding intermediate state with the up and down quarks being separated the quantum knot of the $\pi^-$ meson and the attached components $z_i, i=1,2$ is as a new particle which is identified as the gauge particle $W^{-}$.

Then the mass of the gauge particle $W^{\pm}$ can be roughly determined by the analysis in the above section (or by the usual methods of determining the mass of $W^{\pm}$ in the literature of particle physics \cite{Wei}\cite{Gla}\cite{Sal}\cite{Mar}).
From the strength (\ref{strength3}) in the above section we have that because of the separation of the up and down quarks (or the $\mu^+$ and $\nu_{\mu}$) the mass of $W^{\pm}$ is heavier than $\pi^{\pm}$ and is roughly given by the value 
$\frac{MM_{\mu}}{me} Gev \sim 86 Gev$ where $M=135 Mev$ and $M_{\mu}=105 Mev$. This roughly agrees with the experimental value $80.41 \pm 0.18  Gev$ of $W^{\pm}$. Then when the components $z_i$ are further apart we have that the masses $M$ and $M_{\mu}$ decrease, as the case of asymptotic freedom of strong interaction (This also shows that strong and weak interactions are closedly related). In this case the mass of $W^{\pm}$ decreases and thus is more close to the experimental value $80.41 \pm 0.18  Gev$ of $W^{\pm}$.

On the other hand for the following weak interaction \cite{Wei}\cite{Gla}\cite{Sal}\cite{Mar}:
\begin{equation}
e^+ + \nu_{e} \to e^+ + \nu_{e}
\label{moun1a}
\end{equation}
the corresponding intermediate state with the $e^+$ and  $\nu_{e}$ being separated and represented by the corresponding quantum knot and the attached components $z_i, i=1,2$ (representing the $e^+$ and  $\nu_{e}$ respectively) is as a new particle which is identified as the gauge particle $Z^0$ where the charge of the intermediate state is carried by the current of $e^+$. This is just the same as for the above weak decay with the gauge particle $W^{+}$. 
The difference between $W^{+}$ and $Z^0$ is that $W^{+}$ is with the two dot-lines in Fig.6a which are as the propargator for the weak decay by $W^{+}$ while $Z^0$ is not with 
these two dot-lines in Fig.6a and the propagator for the weak interaction (\ref{moun1a}) is the quantum knot of $Z^0$ connecting the attached components $z_i, i=1,2$.

Thus for the weak interaction (\ref{moun1a}) the two currents formed by $e^+$ and  $\nu_{e}$ respectively are intercted by the quantum knot $W({\bf 4_1})$ of $Z^0$ such that these two currents are separated and are connected (and interacted) by 
the quantum knot $W({\bf 4_1})$ of $Z^0$ (The current of $\nu_{e}$ is as a neutral current \cite{Has}). 

Now since the quantum knot of $Z^0$ is closed we have that the interaction of the $e^+$ and  $\nu_{e}$ currents by the quantum knot of $Z^0$ is asymptotically free; as similar to the property of asymptotic freedom of strong interaction.
Thus the weak interaction (\ref{moun1a}) by the gauge particle $Z^0$
is weaker than the weak interaction by $W^{+}$ and this causes the mass of $Z^0$ to be heavier than the mass of $W^{+}$. Also we notice that the weak interaction by $Z^0$ is without the propagator (represented by the dot-lines in Fig.6a) which gives the appearing of the $\mu^{+}$ whose mass $M_{\mu}$ is transformed from the mass $M$ of $\pi^0$ and is less than the mass $M$ of $\pi^0$.

Thus as similar to the case of the gauge particle $W^{\pm}$ from the knot model of the gauge particle $Z^0$ we have that the mass of $Z^0$ can be derived from the strength 
(\ref{strength2}) and
is roughly given by the value $\frac{M^2}{me} Gev \sim 100 Gev$. This roughly agrees with the experimental value $91.1884 \pm 0.0022  Gev$ of $Z^0$. Then when the components $z_i$ are further apart we have that the mass $M$  decreases, as the case of asymptotic freedom of strong interaction (This also shows that strong and weak interactions are closedly related). In this case the mass of $Z^0$ decreases and thus is more close to the experimental value $91.1884 \pm 0.0022  Gev$ of $Z^0$.

\section{Knot Model of Octet of Baryons}\label{sec13a}

In this section we give knot models of the basic octect of baryons. As the case of mesons these knot model are as strong interaction for forming baryons.
We show that the $SU(3)\otimes U(1)$ gauge symmetry can give knot models of baryons.
We have the following  generators $T^k=\lambda^k$ of $SU(3)$:
\begin{equation}
\begin{array}{cccc}
\lambda^1=\left(\begin{array}{ccc}

          0&1&0 \\

          1&0&0 \\
          0&0&0 \end{array}\right)
&       
\lambda^2=\left(\begin{array}{ccc}
          0&-i&0 \\
          i&0&0 \\
          0&0&0 \end{array}\right)
&
\lambda^3=\left(\begin{array}{ccc}
          1&0&0 \\
          0&-1&0 \\
          0&0&0 \end{array}\right) 
&
\lambda^4=\left(\begin{array}{ccc}
          0&0&1 \\
          0&0&0 \\
          1&0&0 \end{array}\right) \\

\lambda^5=\left(\begin{array}{ccc}
          0&0&-i \\
          0&0& 0 \\
          i&0& 0 \end{array}\right) 
&
\lambda^6=\left(\begin{array}{ccc}
          0&0&0 \\
          0&0&1 \\
          0&1&0 \end{array}\right) 
&
\lambda^7=\left(\begin{array}{ccc}
          0&0& 0 \\
          0&0&-i \\
          0&i& 0 \end{array}\right) 
&
\lambda^8= \frac{1}{\sqrt 3}\left(\begin{array}{ccc}
          1&0& 0 \\
         0&1& 0 \\
          0&0&-2 \end{array}\right) 
\end{array}
\label{lambda}
\end{equation}

Let us first consider the proton. 
Let $W_{p}(K_{p})$ denote the generalized Wilson loop of the proton $p$ where $K_{p}$ denotes
a knot to be specified and this loop is on $SU(3)$. 
Then we have that the following knot model of the proton $p$: 
\begin{equation}
W_{p}(K_{p})Z
\label{proton1a}
\end{equation}
where $Z=(z_1,z_2,z_3,z_1,z_2,z_3,z_1,z_2,z_3)^T$. We shall choose the knot $K_{p}={\bf 6_2}$ for both the proton and the neutron. This prime knot ${\bf 6_2}$ is assigned with the prime number $13$ as its power index \cite{Ng}.

Let a gauge fixing be such that $A^0={\sqrt 3}A^8$ where $A^0$ denotes the electromagnetic field and $A^8$ is the gauge field associated with $\lambda^8$.
From this gauge fixing we have the following generator:
\begin{equation}
\lambda_{p}:=\frac13 I_3 +\frac{1}{\sqrt 3}\lambda^8= \frac13
\left(\begin{array}{ccc}
          2&0& 0 \\
         0&2& 0 \\
          0&0&-1 \end{array}\right) \\
\label{protong}
\end{equation}
where $I_3$ denotes the three dimensional identity and $A^0$ is associated with $I_3$ for the $U(1)$ gauge group. This generator shows the $uud$ structure of proton where the up quark $u$ represented by $z_1$ and $z_2$ is with charge $\frac23 e$ and the down quark $d$ represented by $z_3$ is with charge $-\frac13 e$. 
This agrees with the usual quark model.

Similar to the knot models of mesons we have:
 \begin{equation}
W_{p}(K_{p})=R_{p}A_{p}
\label{proton1a1}
\end{equation}
where $R_{p}$ denotes a $R$-matrix of $SU(3)$ and $A_{p}$ denotes the initial operator of $W_{p}(K_{p})$. 
Similar to the knot models of mesons we have that $R_{p}$ is constructed from the mass operator of proton, denoted by $M_p$, which is to be constructed from the $\lambda$ matrices.

Similar to the mass operators of mesons from the generator $\lambda_{p}$ we have that
the mass operator of proton
is given by:
\begin{equation}
\begin{array}{rl}
M_p &:=  \lambda_{p}\otimes \lambda_{p} + \sum_{a\neq 8} \lambda^a\otimes \lambda^a \\
& \\
 & =\left(\begin{array}{ccccccccc}
1+\frac49&0       &0       & 0      &0        &0       &0       &0       &0 \\
   0     &-\frac59&0       & 2      &0        &0       &0       &0       &0 \\
   0     &0       &-\frac29& 0      &0        &0       &2       &0       &0 \\
   0     &2       &0       &-\frac59&0        &0       &0       &0       &0 \\
   0     &0       &0       &0       &1+\frac49&0       &0       &0       &0 \\
   0     &0       &0       &0       &0        &-\frac29&0       &2       &0 \\

   0     &0       &2       &0       &0        &0       &-\frac29&0       &0  \\
   0     &0       &0       &0       &0        &2       &0       &-\frac29&0     \\
   0     &0       &0       &0       &0        &0       &0       &0       &\frac19\\
   \end{array}\right) \\
\end{array}
\label{proton}
\end{equation}

Let us then compute the mass of proton. 
We have that the mass operator $M_p$ has $3$ positive eigenvalues: $ \frac{13}{9}, \frac1{9}, \frac{16}{9}$ with multipicities $3$, $1$ and $2$ respectively. These positive eigenvalues are considered as  eigenvalues for mass and energy.  
It can be checked  that the corresponding eigenmatrices are given by:
\begin{equation}
\begin{array}{rl}
v_1=\left ( \begin{array}{ccccccccc}
             1 & 0 & 0 & 0 & 0 & 0 & 0 & 0 & 0\\
             0 & 0 & 0 & 0 & 0 & 0 & 0 & 0 & 0\\
             0 & 0 & 0 & 0 & 0 & 0 & 0 & 0 & 0\\
             0 & 0 & 0 & 0 & 0 & 0 & 0 & 0 & 0\\ 
             0 & 0 & 0 & 0 & 0 & 0 & 0 & 0 & 0\\
             0 & 0 & 0 & 0 & 0 & 0 & 0 & 0 & 0\\
             0 & 0 & 0 & 0 & 0 & 0 & 0 & 0 & 0\\
             0 & 0 & 0 & 0 & 0 & 0 & 0 & 0 & 0\\
             0 & 0 & 0 & 0 & 0 & 0 & 0 & 0 & 0
             \end{array}\right),

&
v_2=\left ( \begin{array}{ccccccccc}
             0 & 0 & 0 & 0 & 0 & 0 & 0 & 0 & 0\\
             0 & 0 & 0 & 0 & 0 & 0 & 0 & 0 & 0\\
             0 & 0 & 0 & 0 & 0 & 0 & 0 & 0 & 0\\
             0 & 0 & 0 & 0 & 0 & 0 & 0 & 0 & 0\\ 
             0 & 0 & 0 & 0 & 1 & 0 & 0 & 0 & 0\\
             0 & 0 & 0 & 0 & 0 & 0 & 0 & 0 & 0\\
             0 & 0 & 0 & 0 & 0 & 0 & 0 & 0 & 0\\
             0 & 0 & 0 & 0 & 0 & 0 & 0 & 0 & 0\\
             0 & 0 & 0 & 0 & 0 & 0 & 0 & 0 & 0
             \end{array}\right),
\\
& 
\\
v_3= \left ( \begin{array}{ccccccccc}
             0 & 0 & 0 & 0 & 0 & 0 & 0 & 0 & 0\\
             0 & 0 & 0 & 0 & 0 & 0 & 0 & 0 & 0\\
             0 & 0 & 0 & 0 & 0 & 0 & 0 & 0 & 0\\
             0 & 0 & 0 & 0 & 0 & 0 & 0 & 0 & 0\\ 
             0 & 0 & 0 & 0 & 0 & 0 & 0 & 0 & 0\\
             0 & 0 & 0 & 0 & 0 & 0 & 0 & 0 & 0\\
             0 & 0 & 0 & 0 & 0 & 0 & 0 & 0 & 0\\
             0 & 0 & 0 & 0 & 0 & 0 & 0 & 0 & 0\\
             0 & 0 & 0 & 0 & 0 & 0 & 0 & 0 & 1
             \end{array}\right),

&
v_4=\left ( \begin{array}{ccccccccc}
             0 & 0       & 0 & 0       & 0 & 0 & 0 & 0 & 0\\
             0 & \frac12 & 0 & \frac12 & 0 & 0 & 0 & 0 & 0\\
             0 & 0       & 0 & 0       & 0 & 0 & 0 & 0 & 0\\
             0 & \frac12 & 0 & \frac12 & 0 & 0 & 0 & 0 & 0\\ 
             0 & 0       & 0 & 0       & 1 & 0 & 0 & 0 & 0\\
             0 & 0       & 0 & 0       & 0 & 0 & 0 & 0 & 0\\
             0 & 0       & 0 & 0       & 0 & 0 & 0 & 0 & 0\\
             0 & 0       & 0 & 0       & 0 & 0 & 0 & 0 & 0\\
             0 & 0       & 0 & 0       & 0 & 0 & 0 & 0 & 0
             \end{array}\right),

\\
\mbox{} & \mbox{}
\\
v_5=\left ( \begin{array}{ccccccccc}
             0 & 0       & 0       & 0       & 0 & 0 & 0       & 0 & 0\\
             0 & 0       & 0       & 0       & 0 & 0 & 0       & 0 & 0\\
             0 & 0       & \frac12 & 0       & 0 & 0 & \frac12 & 0 & 0\\
             0 & 0       & 0       & 0       & 0 & 0 & 0       & 0 & 0\\ 
             0 & 0       & 0       & 0       & 0 & 0 & 0       & 0 & 0\\
             0 & 0       & 0       & 0       & 0 & 0 & 0       & 0 & 0\\
             0 & 0       & \frac12 & 0       & 0 & 0 & \frac12 & 0 & 0\\
             0 & 0       & 0       & 0       & 0 & 0 & 0       & 0 & 0\\
             0 & 0       & 0       & 0       & 0 & 0 & 0       & 0 & 0
             \end{array}\right),

&
v_6=\left ( \begin{array}{ccccccccc}
             0 & 0       & 0       & 0       & 0 & 0       & 0 & 0       & 0\\
             0 & 0       & 0       & 0       & 0 & 0       & 0 & 0       & 0\\
             0 & 0       & 0       & 0       & 0 & 0       & 0 & 0       & 0\\
             0 & 0       & 0       & 0       & 0 & 0       & 0 & 0       & 0\\ 
             0 & 0       & 0       & 0       & 0 & \frac12 & 0 & \frac12 & 0\\
             0 & 0       & 0       & 0       & 0 & 0       & 0 & 0       & 0\\
             0 & 0       & 0       & 0       & 0 & 0       & 0 & 0       & 0\\
             0 & 0       & 0       & 0       & 0 & \frac12 & 0 & \frac12 & 0\\
             0 & 0       & 0       & 0       & 0 & 0       & 0 & 0       & 0
             \end{array}\right)
\end{array}             
\label{casimir}
\end{equation} 
are eigenmatrices of positive eigenvalues: $ \frac{13}{9},\frac1{9},  \frac{16}{9}$ respectively where $v_1, v_2, v_4$ are eigenmatrices of $ \frac{13}{9}$; $v_3$ is eigenmatrix of $ \frac{1}{9}$; and $v_5, v_6$ are eigenmatrices of $ \frac{16}{9}$.

Let the initial operator $A_{p}$ of $W_{p}(K_{p})$ be given by:
\begin{equation}
A_{p}:= (v_1b_1 + v_{41}b_4+ v_{51}b_5 )+ (v_2b_2 + v_{42}b_2+ v_{61}b_6 )+(v_3b_3 + v_{52}b_5+ v_{62}b_6 )  
\label{op1}
\end{equation}
where $b_i, i=1,...,6$ are independent initial operators for $W_{p}(K_{p})$.
With this initial operator $A_{p}$ we have that the effect of negative eigenvalues are eliminated.

Let the winding number of the $R$-matrix be $2\cdot 9=18$. Let the knot $K_{p}$ be 
${\bf 6_2}$ which is with power index $13$.
Then by using the knot $K_{p}={\bf 6_2}$ we have that the mass of proton is given by:
\begin{equation}
13\cdot[\frac{13}{9}\cdot 3 + \frac19 + \frac{16}{9}\cdot 2]
2\cdot 9 \cdot m= 13\cdot 8\cdot 9 Mev=13\cdot 72 Mev=936 Mev
\label{proton2}
\end{equation}
where $m=0.5 Mev$ is as the mass of electron. 
This agrees with the observed mass $938 Mev$ of the proton.
 
Let us then consider the neutron $n$.
Let $W_{n}(K_{n})$ denote the generalized Wilson loop of the neutron $n$ where we choose the knot $K_{n}=K_{p}$. 
Then we have the following knot model of the neutron $n$: 
\begin{equation}
W_{n}(K_{n})Z
\label{proton12}
\end{equation}
where $Z=(z_1,z_2,z_3,z_1,z_2,z_3,z_1,z_2,z_3)^T$. 

Let us then consider the mass operator of the neutron $n$. Similar to the proton we have that $n$ is with the mass operator $M_{n}$ given by:
\begin{equation}
\begin{array}{rl}
M_{n} & := \lambda^8 \otimes \lambda^8 +c^2\lambda^3 \otimes \lambda^3+
\sum_{a\neq 3,8} \lambda^a \otimes \lambda^a \\
& \\
& =\left(\begin{array}{ccccccccc}
\frac13+c^2&0          &0       & 0         &0          &0       &0       &0       &0 \\
   0       &\frac13-c^2&0       & 2         &0          &0       &0       &0       &0 \\
   0       &0          &-\frac23& 0         &0          &0       &2       &0       &0 \\
   0       &2          &0       &\frac13-c^2&0          &0       &0       &0       &0 \\
   0       &0          &0       &0          &\frac13+c^2&0       &0       &0       &0 \\
   0       &0          &0       &0          &0          &-\frac23&0       &2       &0 \\

   0       &0          &2       &0          &0          &0       &-\frac23&0       &0  \\
   0       &0          &0       &0          &0          &2       &0       &-\frac23&0     \\
   0       &0          &0       &0          &0          &0       &0       &0       &\frac43 
   \end{array}\right) \\
\end{array}
\label{neutron2}
\end{equation}
where the matrix $\lambda^8$ representing that the neutron $n$ is of the form $ddu$ where the down quark $d$ is with charge $-\frac{e}{3}$ and the up quark $u$ is with charge $\frac{2e}{3}$; and the constant $c^2 >1$ is closed to $1$ and as a deviation from $1$. Comparing to $c^2= 1 $ for the proton this constant $c^2\neq 1 $ represents the existence of the structure matrix $\lambda^3 \otimes \lambda^3$ of $SU(3)$.

The positive eigenvalues of $M_{n}$ are $\frac13+c^2, 2+\frac13-c^2,\frac43$ with multiplicities $2,1$ and $3$ respectively ; and with the eigenmatrices given by the above $v_i, i=1,...,6$.
When $c^2= 1 $ this Casimir operator has only one positive eigenvalues $\frac43$  with
multiplicity $6$. Since $\frac43$ is positive it is considered as an eigenvalue for mass and energy. 

Let the winding number of the $R$-matrix be $2\cdot 9=18$ and the knot $K_n=K_p$ be 
${\bf 6_2}$ which is with power index $13$ as the case of proton. 
We let the initial operator $A_n$ of $W_n(K_n)$ be just $A_p$.

Then when $c^2=1 $ we have that the mass of $n$ is given by:
\begin{equation}
13[\frac43\cdot 6]\cdot 2\cdot 9\cdot m =13\cdot 72 Mev=936 Mev
\label{neutron3a}
\end{equation}
We notice that this is exactly the computed mass of the proton. To split the mass of the proton and the neutron let us then choose $c^2>1 $ such that $\frac43\cdot 3 +(\frac13+c^2)\cdot 2+ (2+\frac13-c^2)=7+c^2=8+\frac1{81}$ (We shall show that this $c^2= 1+\frac1{81}$ is determined by a quantum condition). This spliting of mass represents the structure matrix of $SU(3)$. 
Then  the mass of $n$ is given by:
\begin{equation}
13[\frac43\cdot 3 +(\frac13+c^2)\cdot 2+ (2+\frac13-c^2)]\cdot 2\cdot 9\cdot m Mev
=13 \cdot (8+\frac1{81})\cdot 9 Mev=937.5 Mev
\label{neutron3b}
\end{equation}
This computed mass of the neutron $n$ agrees with the observed mass $939.5 Mev$ of the neutron $n$.

Let us then consider the baryon $\Lambda$ which consists of a singlet.
Since $\Lambda$ is with a strange degree of freedom we have that $\Lambda$ is modeled by:
\begin{equation}
R_s^{n_s}W_{\Lambda}(K_{\Lambda})Z
\label{lambda1a}
\end{equation}
where as the case of the $K^0$ and $K^+$ mesons the matrix $R_s^{n_s}$ is for the strange degree of freedom \cite{Ng1}; and $Z=(z_1,z_2,z_3,z_1,z_2,z_3,z_1,z_2,z_3)^T$. We shall choose 
the knot $K_{\Lambda}= {\bf 7_1}$ for the $\Lambda$. This knot is with the index number $19$. 

Let us then consider the mass operator of $\Lambda$. Similar to the neutron $n$ we have that $\Lambda$ is modeled such that it is with the mass operator $M_{\Lambda}$ given by:
\begin{equation}
\begin{array}{rl}
M_{\Lambda} & := \lambda^8 \otimes \lambda^8 +\lambda^3 \otimes \lambda^3+
c^2\sum_{a\neq 3,8} \lambda^a \otimes \lambda^a \\
&\\
& = \left(\begin{array}{ccccccccc}
\frac43&0       &0       &0       &0      &0       &0       &0       &0 \\
   0   &-\frac23&0       &2c^2    &0      &0       &0       &0       &0 \\
   0   &0       &-\frac23&0       &0      &0       &2c^2    &0       &0 \\
   0   &2c^2    &0       &-\frac23&0      &0       &0       &0       &0 \\
   0   &0       &0       &0       &\frac43&0       &0       &0       &0 \\
   0   &0       &0       &0       &0      &-\frac23&0       &2c^2    &0 \\

   0   &0       &2c^2    &0       &0      &0       &-\frac23&0       &0  \\
   0   &0       &0       &0       &0      &2c^2    &0       &-\frac23&0     \\
   0   &0       &0       &0       &0      &0       &0       &0       &\frac43 
   \end{array}\right) 
\end{array}
\label{lambda30}
\end{equation}
where the matrix $\lambda^8$ representing that the $\Lambda$ is of the form $dsu$ where the down quark $d$ and the strange quark $s$ are with charge $-\frac{e}{3}$ respectively; and the up quark $u$ is with charge $\frac{2e}{3}$; and the constant $c^2 <1$ is closed to $1$ and as a deviation from $1$. Comparing to $c^2= 1 $ for the proton this constant 
$c^2\neq 1 $ represents the existence of the structure matrix $\sum_{a\neq 3,8} \lambda^a \otimes \lambda^a$ of $SU(3)$.

The positive eigenvalues of $M_{\Lambda}$ are $\frac43, 2c^2-\frac23$ which are with multiplicity $3$ respectively; and the eigenmatrices are again the $v_i, i=1,...,6$.

 Let us write the initial operator $A_{\Lambda}$ of $W_{\Lambda}(K_{\Lambda})$ in the form:
\begin{equation}
A_{\Lambda}:= (v_1b_1 + v_{41}b_4+ v_{51}b_5 )+ (v_2b_2 + v_{42}b_2+ v_{61}b_6 )+(v_3b_3 + v_{52}b_5+ v_{62}b_6 )  
\label{lambda2}
\end{equation}
where each eigenmatrix $v_i, i=4,5,6$ is written as a sum of two eigenmatrices:
$v_i=v_{i1}+v_{i2}, i=4,5,6$ by the following summation:
\begin{equation}
\left(\begin{array}{cc}
          \frac12&\frac12 \\
          \frac12&\frac12 \\
          \end{array}\right) \\
          =\left(\begin{array}{cc}
          \frac12&0 \\
          \frac12&0 \\
          \end{array}\right)
          +\left(\begin{array}{cc}
            0&\frac12 \\
            0&\frac12 \\
          \end{array}\right)
\label{eigenmatrix}
\end{equation}
Then the $R_s^{n_s}$ matrix for $\Lambda$ is defined by:
\begin{equation}
R_s^{n_s}A_{\Lambda}:= R_s^{n_s}[v_1b_1 + v_{41}b_4]+ v_{51}b_5 + (v_2b_2 + v_{42}b_2+ v_{61}b_6 )+(v_3b_3 + v_{52}b_5+ v_{62}b_6 ) 
\label{lambda3}
\end{equation}
This means that $R_s^{n_s}$ only acts on $v_1b_1 + v_{41}b_4$ (On the other hand the matrix $R_{SU(3)}$ of $W_{\Lambda}(K_{\Lambda})$ acts on the whole $A_{\Lambda}$). 

Let $R_s^{n_s}$ act on $v_1$ giving the mass $8 Mev$. Then $R_s^{n_s}$ acts on $v_{41}$ giving the mass $4 Mev$. Thus $R_s^{n_s}$ acts on $v_1b_1 + v_{41}b_4$ giving the mass $12 Mev$ which is due to the strange quark $s$. Thus the mass of $\Lambda$ is given by:
\begin{equation}
19[\frac43\cdot 3 +(2c^2-\frac23)\cdot 3]\cdot 2\cdot 9\cdot m +12 Mev=19\cdot 58 +12 Mev= 1114 Mev
\label{lambda4}
\end{equation}
where we choose $c^2$ by requiring that $c^2 <1$ is as the smallest deviation from $1$ such that the expression $[\frac43\cdot 3 +(2c^2-\frac23)\cdot 3]\cdot 9=[2+6c^2]\cdot 9$ is an integer. From this integer condition we have that $c^2$ is determined by the equation $[2+6c^2]\cdot 9=58$ which gives $c^2=\frac{40}{54}$. We remark that this integer condition is from the property of quantum knots that the total winding of a quantum knot is an integer which is as the quantum phenomeon of energy. Thus this integer condition can be reinterpreted as a quantum condition
such that $[\frac43\cdot 3 +(2c^2-\frac23)\cdot 3]\cdot 9=[2+6c^2]\cdot 9=58$. This agrees with the observed mass $1115 Mev$ of $\Lambda$.

Let us then consider other particles of the basic octet of baryons.
Consider first the three $\Sigma$ baryons. 
Since $\Sigma^0$ is with a strange degree of freedom we have that $\Sigma^0$ is modeled by:
\begin{equation}
R_s^{n_s}W_{\Sigma^0}(K_{\Sigma})Z
\label{sigma12}
\end{equation}
where as the case of the $K^0$ and $K^+$ mesons the matrix $R_s^{n_s}$ is for the strange degree of freedom; and $Z=(z_1,z_2,z_3,z_1,z_2,z_3,z_1,z_2,z_3)^T$. We shall choose 
the knot $K_{\Sigma}=K_{\Lambda}= {\bf 7_1}$ for the three $\Sigma$ particles. This prime knot is with the prime index number $19$.

Since $\Sigma^0$ is similar to $\Lambda$ we let the mass operator of $\Sigma^0$ be the following operator which is similar to the mass operator of $\Lambda$:
\begin{equation}
\begin{array}{rl}
M_{\Sigma^0}& := 
\lambda^8\otimes\lambda^8 +\sum_{a=1}^3\lambda^a\otimes \lambda^a +
c^2\sum_{a=4}^7 \lambda^a\otimes \lambda^a \\
& \\
& =\left(\begin{array}{ccccccccc}
\frac43&0       &0       &0       &0      &0       &0       &0       &0 \\
   0   &-\frac23&0       &2    &0      &0       &0       &0       &0 \\
   0   &0       &-\frac23&0       &0      &0       &2c^2    &0       &0 \\
   0   &2    &0       &-\frac23&0      &0       &0       &0       &0 \\
   0   &0       &0       &0       &\frac43&0       &0       &0       &0 \\
   0   &0       &0       &0       &0      &-\frac23&0       &2c^2    &0 \\

   0   &0       &2c^2    &0       &0      &0       &-\frac23&0       &0  \\
   0   &0       &0       &0       &0      &2c^2    &0       &-\frac23&0     \\
   0   &0       &0       &0       &0      &0       &0       &0       &\frac43 
   \end{array}\right)
\end{array} 
\label{sigma}
\end{equation}
where the matrix $\lambda^8$ representing that the $\Sigma^0$ is of the form $dsu$ where the down quark $d$ and the strange quark $s$ are with charge $-\frac{e}{3}$ respectively; and the up quark $u$ is with charge $\frac{2e}{3}$; and the constant $c^2 <1$ is closed to $1$ and as a deviation from $1$. Comparing to $c^2= 1 $ for the proton this constant 
$c^2\neq 1 $ represents the existence of the structure matrix $\sum_{a=4}^7 \lambda^a \otimes \lambda^a$ of $SU(3)$.

The positive eigenvalues of $M_{\Sigma^0}$ are $\frac43, 2c^2-\frac23$ with multiplicities $4$ and $2$ respectively; and the eigenmatrices are again the $v_i, i=1,...,6$.

Let $A_{\Sigma^0}=A_{\Lambda}$.
Then the $R_s^{n_s}$ matrix for $\Sigma^0$ is defined by:
\begin{equation}
R_s^{n_s}A_{\Sigma^0}:= R_s^{n_s}[v_1b_1 + v_{41}b_4]+ v_{51}b_5 + (v_2b_2 + v_{42}b_2+ v_{61}b_6 )+(v_3b_3 + v_{52}b_5+ v_{62}b_6 ) 
\label{sigma3}
\end{equation}
This means that $R_s^{n_s}$ only acts on $v_1b_1 + v_{41}b_4$ (On the other hand the $R$-matrix of $W_{\Sigma^0}(K_{\Sigma})$ acts on the whole $A_{\Sigma^0}$). We notice that this definition for $\Sigma^0$ is the same as the $\Lambda$. 

As the case of $\Lambda$ we let $R_s^{n_s}$ act on $v_1$ giving the mass $8 Mev$. Then $R_s^{n_s}$ act on $v_{41}$ giving the mass $4 Mev$. Thus $R_s^{n_s}$ acts on $v_1b_1 + v_{41}b_4$ giving the mass $12 Mev$ which is due to the strange quark $s$. Thus the mass of $\Sigma^0$ is given by:
\begin{equation}
19[\frac43\cdot 4 +(2c^2-\frac23)\cdot 2]\cdot 2\cdot 9\cdot m +12 Mev=19\cdot 62 +12 Mev= 1190 Mev
\label{sigma4}
\end{equation}
where we choose $c^2<1$ such that $[\frac43\cdot 4 +(2c^2-\frac23)\cdot 2]\cdot 9=62$. This agrees with the observed mass $1192 Mev$ of $\Sigma^0$.

Let us then consider $\Sigma^+$. Similar to $\Sigma^0$ we model $\Sigma^+$ by:
\begin{equation}
R_s^{n_s}W_{\Sigma^+}(K_{\Sigma})Z
\label{sigma12a}
\end{equation}
where as the case of the $K^0$ and $K^+$ mesons the matrix $R_s^{n_s}$ is for the strange degree of freedom; and $Z=(z_1,z_2,z_3,z_1,z_2,z_3,z_1,z_2,z_3)^T$. We again choose 
the knot $K_{\Sigma}=K_{\Lambda}= {\bf 7_1}$ for the $\Sigma^+$. Then we need to specify the mass operator of $\Sigma^+$ for $W_{\Sigma^+}(K_{\Sigma})$.

Since $\Sigma^+$ is similar to $\Sigma^0$ and proton we let the mass operator of $\Sigma^+$ be the following operator which is similar to the mass operators of proton and $\Sigma^0$:
\begin{equation}
\begin{array}{rl}
M_{\Sigma^+}& :=  \lambda_{p}\otimes \lambda_{p} + 
\sum_{a=1}^3\lambda^a\otimes \lambda^a +
c^2\sum_{a=4}^7 \lambda^a\otimes \lambda^a \\
&\\
& =\left(\begin{array}{ccccccccc}
1+\frac49&0       &0       & 0      &0        &0       &0       &0       &0 \\
   0     &-\frac59&0       & 2      &0        &0       &0       &0       &0 \\
   0     &0       &-\frac29& 0      &0        &0       &2c^2    &0       &0 \\
   0     &2      &0        &-\frac59&0        &0       &0       &0       &0 \\
   0     &0      &0        &0       &1+\frac49&0       &0       &0       &0 \\
   0     &0      &0        &0       &0        &-\frac29&0       &2c^2    &0 \\

   0     &0      &2c^2     &0       &0        &0       &-\frac29&0       &0  \\
   0     &0      &0        &0       &0        &2c^2    &0       &-\frac29&0     \\
   0     &0      &0        &0       &0        &0       &0       &0       &\frac19
   \end{array}\right)
   \end{array} 
\label{proton20}
\end{equation}
where the matrix $\lambda_{p}$ representing that the $\Sigma^+$ is of the form $dds$ where the down quark $d$ and the strange quark $s$ are with charge $-\frac{e}{3}$ respectively; and the constant $c^2 <1$ is closed to $1$ and as a deviation from $1$. Comparing to $c^2= 1 $ for other particles such as the proton this constant $c^2\neq 1 $ represents the existence of the structure matrix $\sum_{a=4}^7 \lambda^a \otimes \lambda^a$ of $SU(3)$.

The positive eigenvalues of $M_{\Sigma^+}$ are $1+\frac49, 2c^2-\frac29, \frac19$ which are with multiplicities $3$, $2$ and $1$ respectively; and the eigenmatrices are again the $v_i, i=1,...,6$.

Let the initial operator $A_{\Sigma^+}$ of $W_{\Sigma^+}(K_{\Sigma^+})$ be again of the form $A_{\Lambda}$.
Then the $R_s^{n_s}$ matrix for $\Sigma^+$ is defined by
\begin{equation}
R_s^{n_s}A_{\Sigma^+}:= R_s^{n_s}[v_1b_1] + v_{41}b_4+ v_{51}b_5 + (v_2b_2 + v_{42}b_2+ v_{61}b_6 )+(v_3b_3 + v_{52}b_5+ v_{62}b_6 ) 
\label{sigma31}
\end{equation}
This means that $R_s^{n_s}$ only acts on $v_1b_1 $. 

As the cases of $\Lambda$ and $\Sigma^0$ we let $R_s^{n_s}$ act on $v_1$ giving the mass $8 Mev$.  Thus $R_s^{n_s}$ acts on $v_1b_1 $ giving the mass $8 Mev$ which is due to the strange quark $s$ (We notice the difference between $\Sigma^0$ and $\Sigma^+$). Thus the mass of $\Sigma^+$ is given by:
\begin{equation}
19[(1+\frac49)\cdot 3+(2c^2-\frac29)\cdot 2+\frac19]\cdot 2\cdot 9\cdot m +8 Mev
=19\cdot 62 +8 Mev= 1186 Mev
\label{sigma51}
\end{equation}
where we choose $c^2<1$ such that $[(1+\frac49)\cdot 3+(2c^2-\frac29)\cdot 2+\frac19]\cdot 9=62$. This agrees with the observed mass $1189 Mev$ of $\Sigma^+$.

Let us then consider $\Sigma^-$. Similar to $\Sigma^-$ we model $\Sigma^+$ by:
\begin{equation}
R_s^{n_s}W_{\Sigma^-}(K_{\Sigma})Z
\label{sigma123a}
\end{equation}
where as the case of the $K^0$ and $K^+$ mesons the matrix $R_s^{n_s}$ is for the strange degree of freedom; and $Z=(z_1,z_2,z_3,z_1,z_2,z_3,z_1,z_2,z_3)^T$. We again choose 
the knot $K_{\Sigma}=K_{\Lambda}= {\bf 7_1}$ for the $\Sigma^+$. Then we need to specify the mass operator of $\Sigma^-$ for $W_{\Sigma^-}(K_{\Sigma})$.

Since the structure of $W_{\Sigma^-}(K_{\Sigma})$ of $\Sigma^-$ is the same as the $W_{\Sigma^-}(K_{\Sigma})$ of $\Sigma^+$ except that $\lambda_{p}$ is replaced by $-\lambda_{p}$ we have that the mass operator $M_{\Sigma^-}$ of $\Sigma^-$ is equal to 
$M_{\Sigma^+}$ in (\ref{proton20}).

Then we again let $A_{\Sigma^-}$ be of the form $A_{\Lambda}$. Then as a difference from $\Sigma^+$ the
$R_s^{n_s}$ matrix for $\Sigma^-$ is defined by:
\begin{equation}
R_s^{n_s}A_{\Sigma^-}:= R_s^{n_s}[v_1b_1 + v_{41}b_4+ v_{51}b_5] + (v_2b_2 + v_{42}b_2+ v_{61}b_6 )+(v_3b_3 + v_{52}b_5+ v_{62}b_6 ) 
\label{sigma32}
\end{equation}
This means that $R_s^{n_s}$ acts on $v_1b_1 + v_{41}b_4+ v_{51}b_5$ . Thus $R_s^{n_s}$ acts on $v_1b_1+ v_{41}b_4+ v_{51}b_5 $ giving the mass $8+4+4 Mev$ which is due to the strange quark $s$ (We notice the difference among the three $\Sigma$ particles). Thus the mass of $\Sigma^-$ is given by:
\begin{equation}
19[(1+\frac49)\cdot 3+(2c^2-\frac29)\cdot 2+\frac19]\cdot 2\cdot 9\cdot m +16 Mev
=19\cdot 62 +16 Mev= 1194 Mev
\label{sigma52}
\end{equation}
where we choose $c^2$ such that $[(1+\frac49)\cdot 3+(2c^2-\frac29)\cdot 2+\frac19]\cdot 9=62$ (This $c^2<1$ represents the existence of the structure matrix $\sum_{a=4}^7 \lambda^a \otimes \lambda^a$ of $SU(3)$). This agrees with the observed mass $1196 Mev$ of $\Sigma^-$.

Let us then consider $\Xi^0$. Similar to the $\Sigma^0$ particles we model $\Xi^0$ by:
\begin{equation}
R_s^{n_s}W_{\Xi^0}(K_{\Xi})Z
\label{sigma123}
\end{equation}
where as the case of the $K^0$ and $K^+$ mesons the matrix $R_s^{n_s}$ is for the strange degree of freedom; and $Z=(z_1,z_2,z_3,z_1,z_2,z_3,z_1,z_2,z_3)^T$. We again choose 
the knot $K_{\Xi}=K_{\Lambda}= {\bf 7_1}$ for the two $\Xi$ particles. Then we need to specify the mass operator of $\Xi^0$ for $W_{\Xi^0}(K_{\Xi})$.

Since the structure of $\Xi^0$ is similar to $\Sigma^0$ we let the mass operator of $\Xi^0$ be the following operator which is similar to the mass operator of $\Sigma^0$:
\begin{equation}
\begin{array}{rl}
M_{\Xi^0}& := 
\lambda^8\otimes\lambda^8 +\sum_{a=1}^5\lambda^a\otimes \lambda^a +
c^2\sum_{a=6}^7 \lambda^a\otimes \lambda^a \\
& \\
& =\left(\begin{array}{ccccccccc}
\frac43&0       &0       &0       &0      &0       &0       &0       &0 \\
   0   &-\frac23&0       &2       &0      &0       &0       &0       &0 \\
   0   &0       &-\frac23&0       &0      &0       &2       &0       &0 \\
   0   &2       &0       &-\frac23&0      &0       &0       &0       &0 \\
   0   &0       &0       &0       &\frac43&0       &0       &0       &0 \\
   0   &0       &0       &0       &0      &-\frac23&0       &2c^2    &0 \\

   0   &0       &2       &0       &0      &0       &-\frac23&0       &0  \\
   0   &0       &0       &0       &0      &2c^2    &0       &-\frac23&0     \\
   0   &0       &0       &0       &0      &0       &0       &0       &\frac43 
   \end{array}\right)
   \end{array} 
\label{xi33}
\end{equation}
where the matrix $\lambda^8$ representing that the $\Xi^0$ is of the form $ssu$ where the strange quark $s$ are with charge $-\frac{e}{3}$ respectively; and the up quark $u$ is with charge $\frac{2e}{3}$; and the constant $c^2 <1$ is closed to $1$ and as a deviation from $1$. Comparing to $c^2= 1 $ for other particles such as the proton this constant 
$c^2\neq 1 $ represents the existence of the structure matrix $\sum_{a=6}^7 \lambda^a \otimes \lambda^a$ of $SU(3)$.

The positive eigenvalues of $M_{\Xi^0}$ are $\frac43, 2c^2-\frac23$ with multiplicities $5$ and $1$ respectively; and the eigenmatrices are again the $v_i, i=1,...,6$.

Then again we let the initial operator $A_{\Xi^0}$ of $W_{\Xi^0}(K_{\Xi^0})$ be of the form $A_{\Lambda}$. Then as a difference from $\Sigma^0$ and for the two strange degree of freedom of $\Xi^0$ the
$R_s^{n_s}$ matrix for $\Xi^0$ is defined by:
\begin{equation}
R_s^{n_s}A_{\Xi^0}:= R_s^{n_s}[v_1b_1 + v_{41}b_4]+ v_{51}b_5 + R_s^{n_s}[v_2b_2 + v_{42}b_2]+ v_{61}b_6 +(v_3b_3 + v_{52}b_5+ v_{62}b_6 ) 
\label{sigma32a}
\end{equation}
This means that $R_s^{n_s}$ is acted on $[v_1b_1 + v_{41}b_4]+ [v_2b_2 + v_{42}b_2]$ . Thus $R_s^{n_s}$ acts on $[v_1b_1 + v_{41}b_4]+ [v_2b_2 + v_{42}b_2]$ giving the mass $8+4+4+8=24 Mev$ which is due to the two strange quarks $s$ where $\Xi^0$ is modeled as $ssu$. Thus the mass of $\Xi^0$ is given by:
\begin{equation}
19[\frac43\cdot 5 +(2c^2-\frac23)]\cdot 2\cdot 9\cdot m +24 Mev= 19\cdot 68 +24 Mev= 1316 Mev
\label{sigma52a}
\end{equation}
where we choose $c^2<1$ such that $[\frac43\cdot 5 +(2c^2-\frac23)]\cdot 9=68$. This agrees with the observed mass $1317 Mev$ of $\Xi^0$.

Let us then consider $\Xi^-$. Similar to the $\Xi^0$ particles we model $\Xi^-$ by:
\begin{equation}
R_s^{n_s}W_{\Xi^-}(K_{\Xi})Z
\label{xi23}
\end{equation}
where as the case of the $K^0$ and $K^+$ mesons the matrix $R_s^{n_s}$ is for the strange degree of freedom; and $Z=(z_1,z_2,z_3,z_1,z_2,z_3,z_1,z_2,z_3)^T$. We again choose 
the knot $K_{\Xi}=K_{\Lambda}= {\bf 7_1}$ for $\Xi^-$. Then we need to specify the mass operator of $\Xi^-$ for $W_{\Xi^-}(K_{\Xi})$.

Since $\Xi^-$ is similar to $\Sigma^-$ and $\Xi^0$ we let the mass operator of $\Xi^-$ be the following operator which is similar to the mass operators of $\Sigma^-$ and $\Xi^0$:
\begin{equation}
\begin{array}{rl}
M_{\Xi^-}& :=  \lambda_{p}\otimes \lambda_{p} + 
\sum_{a=1}^5 \lambda^a\otimes \lambda^a +
c^2\sum_{a=6}^7 \lambda^a\otimes \lambda^a \\
& \\
& =\left(\begin{array}{ccccccccc}
1+\frac49&0       &0       & 0      &0        &0       &0       &0       &0 \\
   0     &-\frac59&0       & 2      &0        &0       &0       &0       &0 \\
   0     &0       &-\frac29& 0      &0        &0       &2       &0       &0 \\
   0     &2      &0        &-\frac59&0        &0       &0       &0       &0 \\
   0     &0      &0        &0       &1+\frac49&0       &0       &0       &0 \\
   0     &0      &0        &0       &0        &-\frac29&0       &2c^2    &0 \\

   0     &0      &2        &0       &0        &0       &-\frac29&0       &0  \\
   0     &0      &0        &0       &0        &2c^2    &0       &-\frac29&0     \\
   0     &0      &0        &0       &0        &0       &0       &0       &\frac19
   \end{array}\right) 
\end{array}
\label{xi20}
\end{equation}
where the matrix $\lambda_{p}$ representing that the $\Xi^-$ is of the form $dss$ where the down quark $d$ and the strange quark $s$ are with charge $-\frac{e}{3}$ respectively;  
and the constant $c^2 <1$ is closed to $1$ and as a deviation from $1$. Comparing to $c^2= 1 $ for other particles such as the proton this constant $c^2\neq 1 $ represents the existence of the structure matrix $\sum_{a=6}^7 \lambda^a \otimes \lambda^a$ of $SU(3)$.

The positive eigenvalues of $M_{\Xi^-}$ are $1+\frac49, 2c^2-\frac29, \frac19$ which are with multiplicities $4$, $1$ and $1$ respectively; and the eigenmatrices are again the $v_i, i=1,...,6$.

Then again we let the initial operator $A_{\Xi^-}$ of $W_{\Xi^0}(K_{\Xi^0})$ be of the form $A_{\Lambda}$. Then as a difference from $\Sigma^-$ and for the two strange degree of freedom of $\Xi^-$ the
$R_s^{n_s}$ matrix for $\Xi^-$ is defined by:
\begin{equation}
R_s^{n_s}A_{\Xi^-}:= R_s^{n_s}[v_1b_1 + v_{41}b_4 + v_{51}b_5] + 
R_s^{n_s}[v_2b_2 + v_{42}b_2+ v_{61}b_6] +(v_3b_3 + v_{52}b_5+ v_{62}b_6 ) 
\label{sigma33}
\end{equation}
This means that $R_s^{n_s}$ acts on $[v_1b_1 + v_{41}b_4+ v_{51}b_5]+ [v_2b_2 + v_{42}b_2+ v_{61}b_6]$ . Thus $R_s^{n_s}$ acts on $[v_1b_1 + v_{41}b_4+ v_{51}b_5]+ [v_2b_2 + v_{42}b_2+ v_{61}b_6]$ giving the mass $8+4+4+8+4+4=32 Mev$ which is due to the two strange quarks $s$ where $\Xi^-$ is modeled as $ssd$ (We notice the difference of the effects of the strange degree of freedom of $\Xi^-$ and $\Xi^0$). Thus the mass of $\Xi^-$ is given by:
\begin{equation}
19[(1+\frac49)\cdot 4+(2c^2-\frac29)+\frac19]\cdot 2\cdot 9\cdot m +32 Mev=19\cdot 68 +32 Mev= 1324 Mev
\label{sigma53}
\end{equation}
where we choose $c^2<1$ such that $[(1+\frac49)\cdot 4+(2c^2-\frac29)+\frac19]\cdot 9=68$. This agrees with the observed mass $1321 Mev$ of $\Xi^-$.

\section{Knot Model of Decuplet of Baryons}\label{sec16b}

Let us then consider the decuplet family of baryons. Let us first consider
the $\Delta^{-}$ particle. 
Since $\Delta^{-}$ is a $ddd$-baryon we model $\Delta^{-}$ by:
\begin{equation}
W_{\Delta^{-}}(K_{\Delta^{-}})Z
\label{delta23}
\end{equation}
where the matrix $R_s^{n_s}$ for the strange degree of freedom does not appear since $\Delta^{-}$ is modeled as a $ddd$-baryon; and $Z=(z_1,z_2,z_3,z_1,z_2,z_3,z_1,z_2,z_3)^T$. We choose 
the knot $K_{\Delta^{-}}={\bf 7_1}$ which is the knot for the members of the octet of baryons (except the proton and the neutron). Then we need to specify the mass operator of $\Delta^{-}$ for $W_{\Delta^{-}}(K_{\Delta^{-}})$.

To this end let us recall the mass operator of $\pi^+$. For the mass operator of $\pi^+$ we have that the charge of $\pi^+$ is from the charge matrix $\frac{2}3\cdot \frac{1}2 I_2$ where $I_2$ denotes the two dimensional identity matrix and the trace  of the matrix $\frac{1}2 I_2$ is $1$ which represents the unit charge of $\pi^+$. Thus a factor $\frac{2}3$ is multiplied to the matrix $\frac{1}2 I_2$ which represents the unit charge of $\pi^+$ to form the charge matrix $\frac{2}3\cdot \frac{1}2 I_2$ of $\pi^+$. 

Let us then consider $\Delta^-$. We have that $\Delta^-$ is a baryon of the $ddd$-form. Thus the charge matrix of $\Delta^-$ must be a matrix proportional to the three dimensional identity matrix $I_3$. Thus in analogy to the charge matrix $\frac{2}3\cdot \frac{1}2 I_2$ of $\pi^+$ we have that the charge matrix of $\Delta^-$ is of the form $\frac{2}3\cdot \frac{-1}3 I_3$. Thus 
we let the mass operator of $\Delta^-$ be the following operator:
\begin{equation}
\begin{array}{rl}
M_{\Delta^{-}}& := \frac{-2}9 I_3\otimes\frac{-2}9 I_3
+c^2\lambda^3\otimes \lambda^3+ \sum_{a\neq3,8}\lambda^a\otimes \lambda^a \\
& \\
& =\left(\begin{array}{ccccccccc}
c^2+\frac{4}{81}&0               &0           &0               &0               &0           &0           &0       &0 \\
   0            &\frac{4}{81}-c^2&0           &2               &0               &0           &0           &0       &0 \\
   0            &0               &\frac{4}{81}&0               &0               &0           &2           &0       &0 \\
   0            &2               &0           &\frac{4}{81}-c^2&0               &0           &0           &0       &0 \\
   0            &0               &0           &0               &c^2+\frac{4}{81}&0           &0           &0       &0 \\
   0            &0               &0           &0               &0               &\frac{4}{81}&0           &2       &0 \\

   0            &0               &2           &0               &0               &0           &\frac{4}{81}&0       &0\\
   0            &0               &0           &0               &0               &2           &0       &\frac{4}{81}&0\\
   0            &0               &0           &0               &0               &0           &0     &0    &\frac{4}{81} 
   \end{array}\right)
\end{array} 
\label{omega21}
\end{equation}
where the charge matrix $\frac{-2}9 I_3$ representing that the $\Delta^{-}$ is of the $ddd$-form and is with one charge; and the constant $c^2 <1$ is as a small deviation from $1$. Comparing to $c^2= 1 $ for other particles such as the proton this constant $c^2\neq 1 $ represents the existence of the structure matrix $\lambda^3 \otimes \lambda^3$ of $SU(3)$. 

The positive eigenvalues of $\Delta^{-}$ are $c^2+\frac{4}{81}, 2+\frac{4}{81}-c^2,2+\frac{4}{81}, \frac{4}{81}$ with multiplicities $2,1,2$ and $1$ respectively ; and with the eigenmatrices given by the above $v_i, i=1,...,6$.

We let the initial operator $A_{\Delta^{-}}$ of $W_{\Delta^{-}}(K_{\Delta^{-}})$ be of the form $A_{\Lambda}$. 
  Then the mass of $\Delta^{-}$ is given by:
\begin{equation}
19[(c^2+\frac{4}{81})\cdot 2 +(2+\frac{4}{81}-c^2)+(2+\frac{4}{81})\cdot 2+\frac{4}{81}]\cdot 2\cdot 9\cdot m Mev
= 19\cdot 65 Mev= 1235 Mev
\label{omega211}
\end{equation}
where we  choose $c^2<1$ by requiring that $c^2 <1$ is as the smallest deviation from $1$ such that the expression $[(c^2+\frac{4}{81})\cdot 2 +(2+\frac{4}{81}-c^2)+(2+\frac{4}{81})\cdot 2+\frac{4}{81}]\cdot 9=[6+\frac{24}{81}+c^2]\cdot 9 $ is an integer. From this integer condition we have that $c^2$ is determined by the equation $[6+\frac{24}{81}+c^2]\cdot 9 =65$ which gives $c^2=\frac{75}{81}$. We remark that this integer condition is from the property of quantum knots that the total winding of a quantum knot is an integer which is as the quantum phenomeon of energy. Thus this integer condition can be reinterpreted as a quantum condition.

Then we notice that the observed mass of $\Delta^{0}$ is $1233.6\pm 5 Mev$ \cite{PDA}\cite{PDA2}\cite{PDA1}\cite{PDA4}. Thus this computed mass $1235 Mev$ of $\Delta^{-}$ agrees with the observed mass $1233.6\pm 5 Mev$ of $\Delta^{0}$. Further from this computed mass of $\Delta^{-}$ we can predict that the mass of $\Delta^{-}$ is greater than the mass of $\Delta^{0}$ and further we predict that the mass of $\Delta^{-}>$ the mass of $\Delta^{0}>$ the mass of $\Delta^{+}>$ the mass of $\Delta^{++}$. From this condition  let us compute the masses of $\Delta^{0}$, $\Delta^{+}$ and $\Delta^{++}$.

Let us first consider
the $\Delta^{0}$ particle. 
 Similar to  
 the neutron $n$ we model $\Delta^{0}$ by:
\begin{equation}
W_{\Delta^{0}}(K_{\Delta^{0}})Z
\label{delta111}
\end{equation}
where the matrix $R_s^{n_s}$ for the strange degree of freedom does not appear since $\Delta^{0}$ is modeled as a $udd$-baryon; and $Z=(z_1,z_2,z_3,z_1,z_2,z_3,z_1,z_2,z_3)^T$. We choose 
the knot $K_{\Delta^{0}}=K_{\Delta^{-}}={\bf 7_1}$ which is the knot for the members of the basic octet of baryons (except the proton and the neutron). Then we need to specify the mass operator of $\Delta^{0}$ for $W_{\Delta^{0}}(K_{\Delta^{0}})$.

We let the mass operator of $\Delta^{0}$ be the following operator which is similar to the mass operator of the neutron $n$:
\begin{equation}
\begin{array}{rl}
M_{\Delta^{0}}& :=\lambda^8 \otimes \lambda^8 +c^2\lambda^3 \otimes \lambda^3+
\sum_{a\neq 3,8} \lambda^a \otimes \lambda^a \\
& \\
& =\left(\begin{array}{ccccccccc}
\frac13+c^2&0          &0       & 0         &0          &0       &0       &0       &0 \\
   0       &\frac13-c^2&0       & 2         &0          &0       &0       &0       &0 \\
   0       &0          &-\frac23& 0         &0          &0       &2       &0       &0 \\
   0       &2          &0       &\frac13-c^2&0          &0       &0       &0       &0 \\
   0       &0          &0       &0          &\frac13+c^2&0       &0       &0       &0 \\
   0       &0          &0       &0          &0          &-\frac23&0       &2       &0 \\

   0       &0          &2       &0          &0          &0       &-\frac23&0       &0  \\
   0       &0          &0       &0          &0          &2       &0       &-\frac23&0     \\
   0       &0          &0       &0          &0          &0       &0       &0       &\frac43 
   \end{array}\right) \\
\end{array}
\label{neutron211}
\end{equation}
where the matrix $\lambda^8$ representing that the $\Delta^{0}$ is of the $ddu$-form; and the constant $c^2 <1$ is as a deviation from $1$. Comparing to $c^2= 1 $ for the proton this constant $c^2\neq 1 $ represents the existence of the structure matrix $\lambda^3 \otimes \lambda^3$ of $SU(3)$.

The positive eigenvalues of $\Delta^{0}$ are $\frac13+c^2, 2+\frac13-c^2,\frac43$ with multiplicities $2,1$ and $3$ respectively ; and with the eigenmatrices given by the above $v_i, i=1,...,6$.

Then again we let the initial operator $A_{\Delta^{0}}$ of $W_{\Delta^{0}}(K_{\Delta^{0}})$ be of the form $A_{\Lambda}$.  Thus the mass of $\Delta^{0}$ is given by:
\begin{equation}
19[(c^2+\frac13)\cdot 2 +(2+\frac13-c^2)+\frac43\cdot3]\cdot 2\cdot 9\cdot m Mev
=19[7+c^2]\cdot 9 Mev
\label{delta235}
\end{equation}
where we shall choose $c^2<1$ to satisfy some quantum conditions.

Let us then consider
the $\Delta^{+}$ particle. 
 Similar to the proton $p$ we model $\Delta^{+}$ by:
\begin{equation}
W_{\Delta^{+}}(K_{\Delta^{+}})Z
\label{delta231}
\end{equation}
where the matrix $R_s^{n_s}$ for the strange degree of freedom does not appear since $\Delta^{+}$ is modeled as a $uud$-baryon; and $Z=(z_1,z_2,z_3,z_1,z_2,z_3,z_1,z_2,z_3)^T$. We choose 
the knot $K_{\Delta^{+}}=K_{\Delta^{-}}={\bf 7_1}$ which is the knot for the members of the octet of baryons (except the proton and the neutron). Then we need to specify the mass operator of $\Delta^{+}$ for $W_{\Delta^{+}}(K_{\Delta^{+}})$.

We let the mass operator of $\Delta^{+}$ be the following operator which is similar to the mass operator of the proton $p$:
\begin{equation}
\begin{array}{rl}
M_{\Delta^{+}}& := 
\lambda_p\otimes \lambda_p +c^2\lambda^3\otimes \lambda^3+ \sum_{a\neq3,8}\lambda^a\otimes \lambda^a \\
& \\
& =\left(\begin{array}{ccccccccc}
c^2+\frac49&0          &0       &0          &0          &0       &0       &0       &0 \\
   0       &\frac49-c^2&0       &2          &0          &0       &0       &0       &0 \\
   0       &0          &-\frac29&0          &0          &0       &2       &0       &0 \\
   0       &2          &0       &\frac49-c^2&0          &0       &0       &0       &0 \\
   0       &0          &0       &0          &c^2+\frac49&0       &0       &0       &0 \\
   0       &0          &0       &0          &0          &-\frac29&0       &2       &0 \\

   0       &0          &2       &0          &0          &0       &-\frac29&0       &0  \\
   0       &0          &0       &0          &0          &2       &0       &-\frac29&0     \\
   0       &0          &0       &0          &0          &0       &0       &0       &\frac19 
   \end{array}\right)
\end{array} 
\label{delta333}
\end{equation}
where the matrix $\lambda_p$ representing that the $\Delta^{+}$ is of the $uud$-form; and the constant $c^2 <1$ is as a deviation from $1$. Comparing to $c^2= 1 $ for other particles such as the proton this constant $c^2\neq 1 $ represents the existence of the structure matrix $\lambda^3 \otimes \lambda^3$ of $SU(3)$.

The positive eigenvalues of $M_{\Delta^{+}}$ are $c^2+\frac49, 2+\frac49-c^2, 2-\frac29, \frac19$ with multiplicities $2$, $1$, $2$ and $1$ respectively; and the eigenmatrices are again the $v_i, i=1,...,6$.

Then again we let the initial operator $A_{\Delta^{+}}$ of $W_{\Delta^{+}}(K_{\Delta^{+}})$ be of the form $A_{\Lambda}$.  Thus the mass of $\Delta^{+}$ is given by:
\begin{equation}
19[(c^2+\frac49)\cdot 2 +(2+\frac49-c^2)+(2-\frac29)\cdot 2+\frac19]\cdot 2\cdot 9\cdot m Mev
=19[7+c^2]\cdot 9 Mev
\label{delta233}
\end{equation}
where we shall choose $c^2<1$ to satisfy some quantum conditions.

Let us then consider
the $\Delta^{++}$ particle. Since $\Delta^{++}$ is a $uuu$-baryon
 as similar to the $\Delta^-$ particle we model $\Delta^{++}$ by:
\begin{equation}
W_{\Delta^{++}}(K_{\Delta^{++}})Z
\label{delta23a}
\end{equation}
where the matrix $R_s^{n_s}$ for the strange degree of freedom does not appear since $\Delta^{++}$ is modeled as a $uuu$-baryon; and $Z=(z_1,z_2,z_3,z_1,z_2,z_3,z_1,z_2,z_3)^T$. We choose 
the knot $K_{\Delta^{++}}=K_{\Delta^{-}}={\bf 7_1}$ which is the knot for the members of the basic octet of baryons (except the proton and the neutron). Then we need to specify the mass operator of $\Delta^{++}$ for $W_{\Delta^{++}}(K_{\Delta^{++}})$.

Since $\Delta^{++}$ is of the $uuu$-form which is similar to the $ddd$-form of $\Delta^-$ and is with two charges
we let the mass operator of $\Delta^{++}$ be the following operator: 
\begin{equation}
\begin{array}{rl}
M_{\Delta^{++}}& :=
\frac{4}9 I_3\otimes\frac{4}9 I_3 + c^2\lambda^3\otimes \lambda^3+ 
\sum_{a\neq3,8}\lambda^a\otimes \lambda^a \\
& \\
& =\left(\begin{array}{ccccccccc}
c^2+\frac{16}{81}&0                &0            &0            &0            &0            &0           &0       &0 \\
   0             &\frac{16}{81}-c^2&0            &2            &0            &0            &0           &0       &0 \\
   0             &0                &\frac{16}{81}&0            &0            &0            &2           &0       &0 \\
   0             &2               &0            &\frac{16}{81}-c^2&0            &0         &0           &0    &0\\
   0             &0               &0            &0            &\frac{16}{81}+c^2&0         &0           &0    &0\\
   0             &0               &0            &0            &0            &\frac{16}{81}&0            &2       &0\\

   0             &0               &2            &0            &0            &0            &\frac{16}{81}&0       &0 \\
   0             &0               &0            &0            &0            &2            &0         &\frac{16}{81}&0\\
   0             &0               &0            &0            &0            &0            &0           &0&\frac{16}{81} 
\end{array}\right)
\end{array}   
\label{delta112}
\end{equation}
where the charge matrix $2\cdot\frac{2}9 I_3$ representing that the $\Delta^{++}$ is of the $uuu$-form and is with two charges where the up quark $u$ is with charge $\frac{2e}{3}$; and the constant $c^2 <1$ is as a deviation from $1$. Comparing to $c^2= 1 $ for other particles such as the proton this constant $c^2\neq 1 $ represents the existence of the structure matrix $\lambda^3 \otimes \lambda^3$ of $SU(3)$. 

The positive eigenvalues of $\Delta^{++}$ are $c^2+\frac{16}{81}, 2+\frac{16}{81}-c^2,2+\frac{16}{81}, \frac{16}{81}$ with multiplicities $2,1,2$ and $1$ respectively ; and with the eigenmatrices given by the above $v_i, i=1,...,6$.

Then again we let the initial operator $A_{\Delta^{++}}$ of $W_{\Delta^{++}}(K_{\Delta^{++}})$ be of the form $A_{\Lambda}$.  Thus the mass of $\Delta^{++}$ is given by:
\begin{equation}
19[(c^2+\frac{16}{81})\cdot 2 +(2+\frac{16}{81}-c^2)+(2+\frac{16}{81})\cdot 2+\frac{16}{81}]\cdot 2\cdot 9\cdot m Mev
=19[7+\frac{15}{81}+c^2]\cdot 9
\label{delta222}
\end{equation}
where we shall choose $c^2<1$ to satisfy some quantum conditions.
 
 Now from $[7+\frac{15}{81}+c^2]\cdot 9$ in (\ref{delta222}) for $\Delta^{++}$ and $[7+\frac{18}{81}]\cdot 9=65$ for $\Delta^{-}$ we set a quantum condition that the number $[7+\frac{17}{81}]\cdot 9$ is for $\Delta^{0}$, the number $[7+\frac{16}{81}]\cdot 9$ is for $\Delta^{+}$ and the number $[7+\frac{15}{81}]\cdot 9$ is for $\Delta^{++}$. This quantum condition determines that $c^2=\frac{17}{81}$ for $\Delta^{0}$, $c^2=\frac{16}{81}$ for $\Delta^{+}$ and $c^2=0$ for $\Delta^{++}$. 
 
 From this quantum condition we have that the mass of $\Delta^{0}$ is given by:
 \begin{equation}
19[7+\frac{17}{81}]\cdot 9=1233.1 Mev
\label{delta2a}
\end{equation}
This agrees with the observed mass $1233.6\pm 5 Mev$ of $\Delta^{0}$ \cite{PDA}\cite{PDA2}\cite{PDA1}\cite{PDA4}.

 Then from this quantum condition we have that the mass of $\Delta^{+}$ is given by:
 \begin{equation}
19[7+\frac{16}{81}]\cdot 9=1231.2 Mev
\label{delta2b}
\end{equation}
This agrees with the observed mass $1231.8 Mev$ of $\Delta^{+}$ \cite{PDA}\cite{PDA2}\cite{PDA1}\cite{PDA4}.
 
Then from this quantum condition we have that the mass of $\Delta^{++}$ is given by:
 \begin{equation}
19[7+\frac{15}{81}]\cdot 9=1229.3 Mev
\label{delta2c}
\end{equation}
This agrees with the observed mass $1230.9\pm 3 Mev$ of $\Delta^{++}$ \cite{PDA}\cite{PDA2}\cite{PDA1}\cite{PDA4}.
 
Further we notice that there is a mass gap 
$1233.1-1231.2=1.9 Mev$ between the computed mass $1233.1$ of $\Delta^{0}$ and the computed mass $1231.2$ of $\Delta^{+}$. Then we let the above quantum condition be applied to the neutron and the proton such that $c^2=1+\frac1{81}$ for the neutron. From this quantum condition we have that the computed mass gap of the neutron and the proton is $937.5-936=1.5 Mev$. This computed mass gap $1.5 Mev$ agrees with the observed mass gap of the neutron and the proton.
 
 Let us then consider the $\Sigma(1385)$ in the decuplet of basic baryons. This baryon is similar to the $\Sigma$  baryons of the above octet. Thus its mass operator is of the same form as the corresponding $\Sigma$ baryon in the above octet with the difference that we choose $c^2\geq 1$ such that the proportional winding number is $72$. Also we define the $R_s^{n_s}A_{\Sigma}$ of the strange degree of freedom such that the effect of the strange degree of freedom is given by the number $16$. Also the prime knot $K_{\Sigma(1385)}={\bf 7_1}$ is with the assigned number $19$. Then we have that the computed mass of $\Sigma(1385)$ is given by ${\bf 19}\times 72+16= 1384 Mev$. This agrees with the observed mass $1385 Mev$ of $\Sigma(1385)$.
 
 Let us then consider the $\Xi(1530)$ in the decuplet of basic baryons. This baryon is similar to the $\Xi$ baryons of the above octet. Thus its mass operator is of the same form as the corresponding $\Xi$ baryon in the above octet with the difference that we choose $c^2\geq 1$ such that the proportional winding number is $79$. Also we define the $R_s^{n_s}A_{\Xi}$ of the strange degree of freedom such that the effect of the strange degree of freedom is given by the number $32$. Also the prime knot $K_{\Xi(1530)}={\bf 7_1}$ is with the assigned number $19$. Then we have that the computed mass of $\Xi(1530)$ is given by ${\bf 19}\times 79+32= 1533 Mev$. This agrees with the observed mass $1530 Mev$ of $\Xi(1530)$.

  Let us then consider
the $\Omega^-=\Omega(1672)$ particle. 
Similar to the baryons of the basic octet with a strange quark we model $\Omega^-$ by:
\begin{equation}
R_s^{n_s}W_{\Omega^-}(K_{\Omega^-})Z
\label{omega123}
\end{equation}
where as the case of the $K^0$ and $K^+$ mesons the matrix $R_s^{n_s}$ is for the strange degree of freedom; and $Z=(z_1,z_2,z_3,z_1,z_2,z_3,z_1,z_2,z_3)^T$. We choose 
the knot $K_{\Omega^-}={\bf 7_2}$ which is assigned with the prime number $23$ \cite{Ng}. Then we need to specify the mass operator of $\Omega^-$ for $W_{\Omega^-}(K_{\Omega^-})$.

Since $\Omega^-$ is a baryon of the $sss$-form which is similar to the $ddd$-form of $\Delta^-$
we have that the mass operator of $\Omega^-$ is the following operator which is of the same form as the mass operator of $\Delta^-$:
\begin{equation}
\begin{array}{rl}
M_{\Omega^{-}} & := \frac{-2}9 I_3\otimes\frac{-2}9 I_3
+c^2\lambda^3\otimes \lambda^3+ \sum_{a\neq3,8}\lambda^a\otimes \lambda^a \\
& \\
& =\left(\begin{array}{ccccccccc}
c^2+\frac{4}{81}&0               &0           &0               &0               &0           &0           &0       &0 \\
   0            &\frac{4}{81}-c^2&0           &2               &0               &0           &0           &0       &0 \\
   0            &0               &\frac{4}{81}&0               &0               &0           &2           &0       &0 \\
   0            &2               &0           &\frac{4}{81}-c^2&0               &0           &0           &0       &0 \\
   0            &0               &0           &0               &c^2+\frac{4}{81}&0           &0           &0       &0 \\
   0            &0               &0           &0               &0               &\frac{4}{81}&0           &2       &0 \\

   0            &0               &2           &0               &0               &0           &\frac{4}{81}&0       &0\\
   0            &0               &0           &0               &0               &2           &0       &\frac{4}{81}&0\\
   0            &0               &0           &0               &0               &0           &0     &0    &\frac{4}{81} 
   \end{array}\right)
\end{array} 
\label{omega21a}
\end{equation}
where the charge matrix $\frac{-2}9 I_3$ representing that the $\Omega^{-}$ is of the $sss$-form and is with one charge; and by symmetry to the $\Delta^-$ we have $c^2 >1$ for this $\Omega^{-}$. 
Comparing to $c^2= 1 $ for other particles such as the proton this constant $c^2\neq 1 $ represents the existence of the structure matrix $\lambda^3 \otimes \lambda^3$ of $SU(3)$. 

The positive eigenvalues of $\Omega^{-}$ are $c^2+\frac{4}{81}, 2+\frac{4}{81}-c^2,2+\frac{4}{81}, \frac{4}{81}$ with multiplicities $2,1,2$ and $1$ respectively ; and with the eigenmatrices given by the above $v_i, i=1,...,6$.

Then again we let the initial operator $A_{\Omega^-}$ of $W_{\Omega^-}(K_{\Omega^-})$ be of the form $A_{\Lambda}$. Then for the three strange degrees of freedom of $\Omega^-$ the
$R_s^{n_s}$ matrix for $\Omega^-$ is defined by:
\begin{equation}
R_s^{n_s}A_{\Omega^-}:= R_s^{n_s}[v_1b_1] + v_{41}b_4+ v_{51}b_5 + v_2b_2 +R_s^{n_s}[ v_{42}b_4]+ v_{61}b_6 +v_3b_3 + v_{52}b_5+ R_s^{n_s}[v_{62}b_6 ] 
\label{omega32a}
\end{equation}
This means that $R_s^{n_s}$ is acted on $[v_1b_1 + v_{42}b_4+  v_{62}b_6]$ where we let $v_1b_1$, $v_{42}b_4$ and $v_{62}b_6$ represent the three strange degrees of freedom of $\Omega^-$.  Then $R_s^{n_s}$ acts on $[v_1b_1 + v_{42}b_4+  v_{62}b_6]$ giving the mass $8+4+4=16 Mev$ which is due to the three strange quarks $s$ of $\Omega^-$ .

  Thus the mass of $\Omega^{-}$ is given by:
\begin{equation}
23[(c^2+\frac{4}{81})\cdot 2 +(2+\frac{4}{81}-c^2)+(2+\frac{4}{81})\cdot 2+\frac{4}{81}]\cdot 2\cdot 9\cdot m +16 Mev
= 23\cdot 72 +16 Mev= 1672 Mev
\label{omega222}
\end{equation}
where we choose $c^2>1$ such that $c^2$ satisfies the quantum condition $[(c^2+\frac{4}{81})\cdot 2 +(2+\frac{4}{81}-c^2)+(2+\frac{4}{81})\cdot 2+\frac{4}{81}]\cdot 9=[6+\frac{24}{81}+c^2]\cdot 9=72$. This quantum condition determines $c^2$ such that $c^2$ is the smallest deviation from $1$ and gives $c^2=2-\frac{24}{81}$. 
 This agrees with the observed mass $1672 Mev$ of $\Omega^{-}$ \cite{PDA}\cite{PDA2}\cite{PDA1}\cite{PDA4}.
 
   Let us then consider the singlet $\Lambda(1405)$. This baryon is similar to the $\Lambda^0$ baryon of the above octet. Thus its mass operator is of the same form as the corresponding $\Lambda^0$ baryon in the above octet with the difference that we choose $c^2\geq 1$ such that the proportional winding number is $73$. Also we define the $R_s^{n_s}A_{\Lambda}$ of the strange degree of freedom such that the effect of the strange degree of freedom is given by the number $16$. Also the prime knot $K_{\Lambda(1405)}={\bf 7_1}$ is with the assigned number $19$. Then we have that the computed mass of $\Lambda(1405)$ is given by ${\bf 19}\times 73+16= 1403 Mev$. This agrees with the observed mass $1405 Mev$ of $\Lambda(1405)$.

  In summary we have the following classification table of baryons of the basic octet and decuplet:
\begin{displaymath}
\begin{array}{*{4}{|c}|} \hline
\multicolumn{4}{|c|}{\mbox{Octet}}
\\ \hline
   N(939) & \Lambda(1115)& \Sigma(1193)& \Xi(1317)
\\
  {\bf 13}\times 72= 936 &{\bf 19}\times 58+12=1114 &{\bf 19}\times 62+12=1190&{\bf 19}\times 68+24=1316 
\\ \hline
\end{array}
\end{displaymath}

\begin{displaymath}
\begin{array}{*{4}{|c}|} \hline
\multicolumn{4}{|c|}{\mbox{Decuplet}}
\\ \hline
\Delta(1232)& \Sigma^*=\Sigma(1385)& \Xi^*=\Xi(1530)& \Omega^- =\Omega(1672)\\
{\bf 19}\times 65=1235&{\bf 19}\times 72+16=1384&{\bf 19}\times 79+32=1533&{\bf 23}\times 72+16=1672 
 \\ \hline
 \end{array}
\end{displaymath}

\begin{displaymath}
\begin{array}{*{1}{|c}|} \hline
\multicolumn{1}{|c|}{\mbox{Singlet}}
\\ \hline
 \Lambda^* = \Lambda(1405)\\
  {\bf 19}\times 73+16=1403
\\ \hline
\end{array}
\end{displaymath}
where  the prime knot for modeling the above baryons are given by the following classification table of prime knots by prime numbers

\section{Conclusion}\label{sec16a}

In this paper we have established a quantum knot model of the two nonets of pseudoscalar and vector mesons $\pi^0,\pi^+, K^0, K^+, \eta, \rho, \omega, K^*, \eta^{\prime},\phi$ and the two scalar mesons $a_0(980)$ and $f_0(980)$. This knot model gives the required strong interaction  for the formation of these mesons. For the $K^0$ and $K^+$ mesons we have introduced the strange degree of freedom which is from the $R$-matrix of $SU(2)$ in knot theory for braiding (and we denote this $R$-matrix by $R_s$) and this $R$-matrix is a linking effect. This linking effect gives the strong interaction between two elementary particles and gives the associated production property for strong interactions such as the interaction 
$\pi^- +p\to K^0 + \Lambda$. We have computed the masses of $K^0$ and $K^+$ and show that the mass of $K^+$ is less than $K^0$ even though $K^+$ is with charge. This computation agrees with the experimental masses of $K^0$ and $K^+$.

We have also established a knot model of the basic octet and decuplet of baryons including the proton $p$, the neutron $n$, and the baryons $\Lambda, \Sigma, \Xi, \Delta, \Lambda^*, \Sigma^*, \Xi^*,\Omega^-$. This knot model gives the strong interaction  for the formation of these baryons.
For the  baryons $\Lambda, \Sigma, \Xi, \Lambda^*, \Sigma^*, \Xi^*, \Omega^-$  we have introduced the pure strange degree of freedom which is from the $R_s$ matrix of representing knots on $SU(3)$ and this $R_s$ matrix is a linking effect of knots. This linking effect gives the strong interaction between two elementary particles and gives the associated production property for strong interactions producing particles with the strange quark $s$ such as the interaction 
$\pi^- +p\to K^0 + \Lambda$. We have computed the masses of $\Sigma^0$ and $\Sigma^+$ and show that the mass of $\Sigma^+$ is less than $\Sigma^0$ even though $\Sigma^+$ is with charge. This computation agrees with the experimental masses of $\Sigma^0$ and $\Sigma^+$.

We show that the proton $p$ and the neutron $n$ are modeled by the prime knot ${\bf 6_2}$ assigned with the prime number $13$. The baryon $\Omega^-$ constructed with the strange quark $s$ is modeled by the prime knot ${\bf 7_2}$ assigned with the prime number $23$. Then all other baryons (including all baryons with the strange quark $s$) in the basic octet and decuplet of baryons are modeled by the prime knot ${\bf 7_1}$ assigned with the prime number $19$.
This shows that the family ${\bf 7_{(\cdot)}}$ of prime knots with seven crossings (such as the prime knots ${\bf 7_1}$ and ${\bf 7_2}$) is a family for modeling baryons with the strange quark $s$. From this we can predict that the family ${\bf 8_{(\cdot)}}$ of prime knots with eight crossings  is a family for modeling baryons with the charm quark $c$. Then we can predict that the family ${\bf 9_{(\cdot)}}$ of prime knots with nine crossings  is a family for modeling baryons with the bottom quark $b$; and we can predict that the family ${\bf 10_{(\cdot)}}$ of prime knots with ten crossings  is a family for modeling baryons with the top quark $t$; and so on.

This quantum knot model  includes the usual quark model and thus agrees with the usual quark model.
As an example in the knot model of $\pi^+$ we give the required $\frac23 e$ charge to the up quark $u$ and give the required $\frac13 e$ charge to the antidown quark $\overline{d}$.

We give a mass mechanism for the generation of the masses of the mesons and baryons. This mechanism of generating mass supersedes the
conventional mechanism of generating mass through the Higgs particles and
makes hypothesizing the existence of the Higgs particles unnecessary. This
perhaps explains why we cannot physically find such Higgs particles.

We then have a mass formula to compute the masses of other mesons of the two nonets of pseudoscalar and vector mesons and the two scalar mesons $a_0(980)$ and $f_0(980)$ 
and the  basic octet and decuplet of baryons
where we show that the mesons and baryons are modeled by prime knots which are indexed by prime numbers. This computation gives an evidence that mesons and baryons are modeled as knots. We also shows that the phenomeon of the generations of quarks is derived from the properties of knots. This is also an evidence that elementary particles are modeled as knots.

We have established a quantum knot modeling of weak interaction and strong interaction and the $CP$ violation. We show that the weak and the strong interactions are closedly related in the knot form. This gives a unification of the electromagnetic, the weak and the strong interactions. In this unification the common interaction charge is the electric charge $e$ (or the bare electric charge $e_0$).
 
We show that the strong interaction of quarks for forming mesons and baryons is just the quantum knot modeling of mesons and baryons. In this knot modeling of the strong interaction the quantum knot $W(K)$ is as the gluon which acts on a complex vector $Z$ whose components $z_i$ represent quarks to form the mesons and baryons. Then when the components $z_i$ are separated at different proper times (or at different positions of $W(K)$) while they are still attached to $W(K)$ we have the weak interaction of quarks and leptons represented by the components $z_i$.

We show that the asymptotic freedom of strong interaction and the weak interaction are closedly relatd where the separation of the components $z_i$ representing quarks and leptons while these $z_i$ attaching to the quantum knot $W(K)$ gives the asymptotic freedom of strong interaction and also gives the weak interaction. Thus the strong and the weak interaction are closedly relatd.

In the knot modeling of the weak decay of $\pi^+$ we show that by a gauge fixing the $\overline{d}$ quark becomes massless and is identified as the neutrino $\nu_{\mu}$ and the $u$ quark is transformed to the muon lepton $\mu$.  This shows that we can indirectly observe free quarks in the sense that quarks become leptons while we also show that quarks are confined by the strong interaction in the knot form. Thus the strong and the weak interactions are closedly relatd.

We show that the strength of weak interaction is as a product of two strengths of the strong interaction of the Yukawa form. This also shows that the strong and weak interactions are closedly related. From this relation of the strong and weak interactions and the knot modeling of the gauge particles $W^{\pm}$ and $Z^0$ of weak interaction we derive an estimate of the masses of $W^{\pm}$ and $Z^0$. This estimate roughly agrees with the experimantal value of the masses of $W^{\pm}$ and $Z^0$. 

From the knot modeling of the strong and weak interactions we derive the $P$, $C$ and $CP$ violations of the weak interaction. We show that the $\pi$ mesons modeled by the prime knot ${\bf 4_1}$ gives the $P$, $C$ and $CP$ violations of the weak interaction. These violations are due to that the prime knot ${\bf 4_1}$ is an amphichiral knot. 


 {\bf Remarks}.

1. Finkelstein published a different approach to modeling elementary particles by knots. He proposed knot models of leptons and quarks by adding the quantum group $SU_q(2)$ to the gauge group $SU(2)\times U(1)$ of the standard electroweak theory \cite{Fin1}. In this paper, we use the gauge group $SU(2)\times U(1)$ directly to construct a quantum knot theory from which we derived knot models of mesons. Knot models of quarks are derived as components of mesons. Then under weak interaction of mesons we also obtain knot models of leptons. As an example of
the weak interaction $\pi^+\to \mu^+ + \nu_{\mu}$ and the knot model of the meson $\pi^+$, we constructed knot models of the leptons $\mu^+$ and $\nu_{\mu}$.

2. We notice that by using high performce computing Faddeev and Niemi showed the existence of knot phenomena of gauge field \cite{Fad2}.

3. We notice that our knot model of quark confinement is of some similarity to the bag model of quark confinement \cite{Cho}-\cite{Mor}. In our knot model of mesons (and baryons) quarks are confined by knots while in the bag model of strong interaction quarks are confined by bags.

\end{document}